\documentclass{PoS}

\usepackage{epsfig}
\epsfclipon
\usepackage{multicol}
\usepackage{epsfig,bm}
\usepackage{amssymb,amsmath}

\newcommand{\roughly}[1]{\mathrel{\raise.3ex\hbox{$#1$\kern-0.85em
\lower1ex\hbox{$\sim$}}}}

\newcommand{\lsim}{\roughly<}
\newcommand{\gsim}{\roughly>}

\def\be{\begin{equation}}
\def\beq\begin{equation}
\def\ee{\end{equation}}
\def\bea{\begin{eqnarray}}
\def\eea{\end{eqnarray}}

\def\pref#1{(\ref{#1})}

\def\beq{\begin{equation}}
\def\eeq{\end{equation}}
\def\beqa{\begin{eqnarray}}
\def\eeqa{\end{eqnarray}}

\def\cX{{\cal X}}
\def\cB{{\cal B}}

\newcommand{\bmat}{\left(\begin{array}}
\newcommand{\emat}{\end{array}\right)}

\def\yzero{\smash{\hbox{$y\kern-4pt\raise1pt\hbox{${}^\circ$}$}}}

\def\-{\hphantom{-}}

\def\s2{\frac{1}{2}}

\def\IF{\relax{\rm I\kern-.18em F}}
\def\II{\relax{\rm I\kern-.18em I}}
\def\IP{\relax{\rm I\kern-.18em P}}
\def\IC{\relax{\rm I\kern-.48em C}}
\def\IR{\relax{\rm I\kern-.18em R}}
\def\IK{\relax{\rm I\kern-.20em K}}
\def\IM{\relax{\rm I\kern-.25em M}}

\def\Dsl{\,\raise.15ex\hbox{/}\mkern-13.5mu D} 
\def \one{\relax{\rm 1\kern-.26em I}}

\def\exd{{\rm d}}

\def\bfk{{\bf k}}

\def\bfn{{\bf n}}

\def\bfr{{\bf r}}

\def\bfx{{\bf x}}

\def\nott#1{\setbox0=\hbox{$#1$}                
   \dimen0=\wd0                                 
   \setbox1=\hbox{/} \dimen1=\wd1               
   \ifdim\dimen0>\dimen1                        
      \rlap{\hbox to \dimen0{\hfil/\hfil}}      
      #1                                        
   \else                                        
      \rlap{\hbox to \dimen1{\hfil$#1$\hfil}}   
      /                                         
   \fi}                                         %

\def\nn{\nonumber}

\newcommand{\cE}{{\cal E}}

\newcommand{\cL}{{\cal L}}
\newcommand{\cM}{{\cal M}}

\newcommand{\cW}{{\cal W}}

\newcommand{\cV}{{\cal V}}
\newcommand{\cT}{{\cal T}}

\newcounter{oldcounter}
\addtocounter{equation}{1}
\title{Lectures on Cosmic Inflation\\and its Potential
Stringy Realizations}

\ShortTitle{Lectures on String Inflation}

\author{C.P. Burgess\\

Department of Physics \& Astronomy, McMaster University,\\
\qquad 1280 Main Street West, Hamilton, Ontario, Canada, L8S 4M1.\\
Perimeter Institute for Theoretical Physics,\\ \qquad 31 Caroline
Street North, Waterloo, Ontario, Canada, N2L 2Y5. }

\date{}

\abstract{These notes present a brief introduction to Hot Big Bang
cosmology and Cosmic Inflation, together with a selection of some
recent attempts to embed inflation into string theory. They
provide a partial description of lectures presented in courses at
Dubrovnik in August 2006, at CERN in January 2007 and at Carg\`ese
in August 2007. They are aimed at graduate students with a working
knowledge of quantum field theory, but who are unfamiliar with the
details of cosmology or of string theory.}

\FullConference{School on Particle Physics, Gravity and Cosmology\\
        21 August - 2 September 2006 \\
        Dubrovnik, Croatia\\
        $\phantom{XX}$\\
        Winter School on Constituents, Fundamental Forces
        and Symmetries of the Universe\\
        15 January - 19 January 2007\\
        CERN, Switzerland\\
        $\phantom{XX}$\\
        School on Cosmology and Particle Physics
        Beyond the Standard Models\\
        30 July - 11 August 2007 \\
        Carg\`ese, France}

\begin{document}



The Hot Big Bang model of cosmology has recently been tested with
unprecedented redundancy and precision, and has emerged all the
stronger for having done so. The redundancy of these tests gives
confidence that the basic picture --- the expansion of an initial
hot primordial soup --- is basically right. Their precision allows
a detailed inference of the model's parameters, including the
first-ever survey of the energy content of the Universe as a
whole.

Although the Big Bang works well, it does so only provided that
the Universe is started off in a particular way. The theory of
Cosmic Inflation \cite{Inflation} was invented in order to try to
explain these initial conditions, by postulating a much earlier
epoch during which the Universe expanded increasingly rapidly with
time. Remarkably, this proposal turns out also to give a good
explanation for the properties of the temperature fluctuations
that were later seen in the Cosmic Microwave Background Radiation
(CMBR) --- the residual radiation left over from the first epoch
when the universe became transparent to photons, due to its
cooling enough to allow ordinary matter to become dominated by
neutral atoms.

Because inflation likely takes place at temperatures much higher
than any ever seen in the lab on Earth, its study necessarily
involves making assumptions about what kinds of physics are
involved at such high energies. This, together with the
observational successes, has stimulated a variety of attempts to
try to find inflationary configurations within string theory,
which remains our best candidate for the physics relevant to such
high energies. These notes are meant as a brief introduction to
inflationary cosmology and its potential stringy realizations,
aimed at an audience of graduate students in particle physics.

\section{Hot Big Bang Cosmology}

We start with a description of the geometry of spacetime on which
all of the subsequent sections rely, together with a telegraphic
summary of the essentials of the Hot Big Bang model. (More details
can be found in one of the following excellent books
\cite{CosmoBooks,Mukhanov}.) The key underlying assumption in this
section is that the universe is homogeneous and isotropic when
seen on the largest distance scales. Until relatively recently
this assertion about the homogeneity and isotropy of the universe
was an assumption, often called the {\sl Cosmological Principle}.
More recently it has become possible to put this assertion on an
observational footing, based on large-scale surveys of the
distribution of matter and radiation within the universe we see
around us. Most notable among these is the incredible uniformity
of the observed temperature of the CMBR, for which temperature
fluctuations are observed to be of order $\delta T/T \sim
10^{-5}$.

\subsection{Friedman-Robertson-Walker Cosmology}

In General Relativity the geometry of spacetime is specified by
its metric tensor, which defines the differential distance, $\exd
s^2 = g_{\mu\nu} \, \exd x^\mu \, \exd x^\nu$, associated with
infinitesimal coordinate displacements, $\exd x^\mu$. The most
general 4D geometry which is consistent with isotropy and
homogeneity of its spatial slices is described by the
Robertson-Walker metric:
\bea \label{FRWMetric}
    \exd s^2 &=& - \exd t^2 + a^2(t) \, \left[
    \frac{\exd r^2}{1 - \kappa
    r^2} + r^2 \, \exd\theta^2 + r^2 \sin^2\theta \,
    \exd \phi^2 \right] \nonumber\\
    &=& - \exd t^2 + a^2(t) \, \left[
    \exd \ell^2 + r^2(\ell) \, \exd\theta^2 +
    r^2(\ell) \sin^2\theta \, \exd \phi^2 \right]
    \,,
\eea
where $0 < \theta < \pi$ and $0 \le \phi < 2\pi$ are the usual
angular coordinates on a two-sphere, and we choose ourselves to
lie at the origin, $r = 0$, of the radial coordinate.

Homogeneity and isotropy dictate that the 3-dimensional spatial
slices through this geometry at fixed $t$ are maximally symmetric,
and so are described by the three-valued quantity, $\kappa = 0,
1,-1$. If $\kappa = 1$ then the spatial slices are three-spheres
and $0 < r < 1$; if $\kappa = -1$ they are hyperbolic surfaces and
$0 < r < \infty$; and if $\kappa = 0$ they are flat and again $r$
ranges from zero to infinity. The metric of eq.~\pref{FRWMetric}
follows the standard convention, wherein the freedom to redefine
$r \to \lambda r$ has been used to absorb the radius of curvature
of the spatial metric into the overall scale factor, $a(t)$.

The second form given for the metric in eq.~\pref{FRWMetric}
instead uses the proper distance, $\ell$, (at fixed $t$) as the
radial coordinate, where $\exd \ell = \exd r/(1 - \kappa
r^2)^{1/2}$, and so
\be  \label{rvsell}
    r(\ell) = \left\{ \begin{matrix}
    \sin \ell & \;\;\hbox{if} \quad \kappa = +1 \\
    \ell & \hbox{if}\quad \kappa = 0 \\
    \sinh \ell & \;\;\hbox{if}\quad \kappa = -1
    \end{matrix}
    \right. \,.
\ee

\begin{quote}{\bf Exercise 1:}
Find the rate of change, $V_H = \exd D/\exd t$, of the proper
distance, $D = a \Delta \ell$, from us to another co-moving
observer located on a galaxy at fixed position $(\ell, \theta,
\phi)$. Show that this is given by the Hubble Law: $V_H = H \, D$,
where $H = \dot a/a$ defines the instantaneous Hubble parameter.
\label{Ex:1}
\end{quote}

Detailed observations of many, many galaxies broadly confirm that
galaxies do recede from us in a way that is consistent with the
Hubble law defined in Exercise 1, with a present-day Hubble
parameter of $H_0 \sim 75$ km/sec/Mpc. Strictly speaking, however,
the Hubble law only applies once the peculiar motion due to the
gravitational influence of local matter is removed. But since the
Hubble law implies that the apparent recession due to the
universal expansion becomes more important for more distant
galaxies, in practice peculiar velocities are an important
complication only for the nearest galaxies.

\begin{quote}{\bf Exercise 2:}
For the Robertson-Walker geometry show that if a photon having
wavelength $\lambda_{\rm em}$ is emitted at a time $t_{\rm em}$,
when $a(t_{\rm em}) = a_{\rm em}$, and is received with a
wavelength $\lambda_{\rm obs}$ at a later time $t_{\rm obs}$ for
which $a(t_{\rm obs}) = a_{\rm obs}$, then it experiences a
redshift $z = (a_{\rm obs}/a_{\rm em})-1$, where redshift is
defined by $z \equiv (\lambda_{\rm obs} - \lambda_{\rm
em})/\lambda_{\rm em}$. Notice that this implies that universal
expansion ({\it i.e.} $a_{\rm obs} > a_{\rm em}$) implies $z > 0$,
making the observed wavelength longer (more red) than the emitted
one. \label{Ex:2}
\end{quote}

How the scale factor evolves with time depends on what kind of
matter the universe contains, in a way which is dictated by the
field equations for gravity. Assuming these are given by
Einstein's General Theory of Relativity implies that this
connection between spacetime geometry and universal energy content
is given by
\be \label{EinsteinEqn}
    R_{\mu\nu} - \frac12 \, R \, g_{\mu\nu} = 8 \pi G \,
    T_{\mu\nu} \,,
\ee
where $G$ is Newton's constant, and $R = g^{\mu\nu} R_{\mu\nu}$
where $R_{\mu\nu}$ denotes the Ricci tensor --- a particular
measure of the curvature of spacetime.

The tensor $T_{\mu\nu}$ on the right-hand-side of
eq.~\pref{EinsteinEqn} is the energy-momentum stress tensor of the
universe's matter content, which is locally conserved in the sense
that $\nabla^\mu T_{\mu\nu} = 0$. The most general form for
$T_{\mu\nu}$ consistent with the homogeneity and isotropy of
spacetime has the perfect-fluid form:
\be \label{PerfectFluidTmn}
    T_{\mu\nu} = \begin{pmatrix} \rho & 0 \cr
    0 & p \, g_{ij} \end{pmatrix} \,,
\ee
where $\rho$ is the local energy density and $p$ the local
pressure. The indices $i,j = 1,2,3$ run over the spatial
coordinates (as opposed to the spacetime indices $\mu,\nu =
0,1,2,3$).

Once eq.~\pref{EinsteinEqn} is specialized to the Robertson-Walker
metric, eq.~\pref{FRWMetric}, and to \pref{PerfectFluidTmn}, it
reduces to two independent equations governing the time-evolution
of the scale factor, $a(t)$: the Friedmann equation,
\be \label{FriedmannEqn}
    \left( \frac{\dot a}{a} \right)^2
    + \frac{\kappa}{a^2} = \frac{\rho}{3M_p^2}
    \qquad \hbox{(Friedmann)}\,,
\ee
where $M_p^{-2} \equiv 8\pi G$, and the Raychaudhuri equation,
\be \label{RaychaudhuriEqn}
    \frac{\ddot a}{a} = - \frac{1}{6M_p^2} \Bigl( \rho + 3p \Bigr)
    \qquad \hbox{(Raychaudhuri)}\,.
\ee
It is often useful to trade eq.~\pref{RaychaudhuriEqn} for the
equivalent first-order equation which expresses conservation of
energy:
\be \label{EnergyConservationEqn}
    \frac{\exd}{\exd t} \Bigl( \rho \, a^3 \Bigr)
    = - p \, \frac{\exd}{\exd t} \Bigl( a^3 \Bigr)
    \qquad \hbox{(energy conservation)}\,,
\ee
since eqs.~\pref{FriedmannEqn} and \pref{EnergyConservationEqn}
together imply eq.~\pref{RaychaudhuriEqn}.

\subsection{Universal energy content}

At present, the universe appears to be well-described by a fluid
which contains four independent contributions to its stress
energy,
\be
    T_{\mu\nu} = \sum_{i = 1}^4 T^i_{\mu\nu} \,.
\ee
Furthermore, each component of this fluid appears to exchange
energy and momentum negligibly with the others, so $\nabla^\mu
T^i_{\mu\nu} = 0$, for each $i$. In terms of the corresponding
energy densities, $\rho_i$, and pressures, $p_i$, --- defined for
$T^i_{\mu\nu}$ as in eq.~\pref{PerfectFluidTmn} --- this implies
that each component separately satisfies
eq.~\pref{EnergyConservationEqn}.

For the purposes of cosmology, several important things are known
about the universal stress-energy content.

\medskip\noindent{\bf Total Energy Density:}

\medskip\noindent The best current measurements of the
present-day Hubble scale, $H_0 = (\dot a/a)_0$, together with the
measured overall curvature of space, $\kappa/a_0^2$, taken with
the Friedmann equation, eq.~\pref{FriedmannEqn}, tell us the
present value of the total energy density, $\rho_{\rm tot} =
\sum_i \rho_i$, of the universe. The curvature of space,
$\kappa/a_0^2$, can be inferred from the properties of the
measured temperature fluctuations of the CMBR together with the
measured value of $H_0$, and imply $\kappa/a_0^2$ is presently
consistent with zero ({\it i.e.} a spatially flat universe). Using
this, and the measured value for $H_0$, in eq.~\pref{FriedmannEqn}
then implies
\be \label{TotalRhoValue}
    \rho_{\rm tot} \sim \rho_c = 3 M_p^2 H_0^2 \sim 10^{-29} \,
    \hbox{g/cm}{}^3 \,.
\ee
The Friedmann equation, eq.~\pref{FriedmannEqn}, can then be
rewritten as
\be
    \sum_i \Omega_i = 1 \,,
\ee
where $\Omega_i = \rho_i/\rho_c$ denotes the present-day fraction
of energy density contributed by each fluid component, and the sum
runs over all components.

At present there is good evidence for there being the following
four components to the cosmic fluid:

\medskip\noindent{\bf Radiation:}

\medskip\noindent We see the universe around us is filled with
photons, whose energy density is dominated by the photons of the
CMBR. The pressure and energy density of a gas of photons are
related by the equation of state
\be \label{EoS:Radiation}
    p_{\rm rad} = \frac13 \, \rho_{\rm rad} \,.
\ee
These photons are observed to have a thermal distribution, with
temperature $2.715$ K.

On particle-physics grounds it is also believed that there are
also an almost equally large number of Cosmic Relic Neutrinos
(CRNs), whose masses are small enough to have been relativistic at
least up to very recent epochs of the universe. Furthermore, these
neutrinos are calculated to be thermally distributed, with
temperature $T_\nu \sim 1.9$ K. Since any gas of
weakly-interacting relativistic particles satisfies the equation
of state, eq.~\pref{EoS:Radiation}, these neutrinos are normally
lumped together with the photons into the energy density and
pressure of cosmic radiation.

The observed total energy density of radiation is a small fraction
of the present total energy density,
\be
    \Omega_{\rm rad} = \left(
    \frac{\rho_{\rm rad}}{\rho} \right)_{\rm now} \approx
    8 \times 10^{-5} \,,
\ee
of which roughly $3\times 10^{-5}$ comes from the neutrinos.

\medskip\noindent{\bf Baryons:}

\medskip\noindent The universe also contains ordinary
matter (electrons, nuclei, atoms) in large numbers, whose number
density is normally counted as a contribution to the conserved
density of baryon number (for which neutrons and protons carry
$+1$ unit while electrons carry none). (Although this technically
does not count the electrons, the overall electrical neutrality of
the universe tells us that the number of electrons is the same as
the number of protons.)

Since this kind of matter is non-relativistic, its average kinetic
energy --- {\it i.e.} its pressure --- is smaller than the energy
tied up in its rest mass by an amount of order $v^2/c^2$, and so
its equation of state is
\be
    p_B \approx 0 \,.
\ee
Even though the number density of baryons is numerically much less
numerous than photons, $n_B/n_\gamma \approx 5 \times 10^{-10}$,
their relatively large rest mass implies they make up a larger
component of the present day energy density than does the
radiation:
\be
    \Omega_{B} = \left(
    \frac{\rho_B}{\rho} \right)_{\rm now} \approx
    4\%\,.
\ee
The number of {\it visible} baryons is much smaller than this, but
the total amount of baryons present can nonetheless be determined
because of its influence both on the observed temperature
fluctuations of the CMBR and on the relative abundance of light
nuclei which were formed in the very early universe.

\medskip\noindent{\bf Dark Matter:}

\medskip\noindent Observations of how stars move within
galaxies, how galaxies move within clusters and of how the gravity
of matter as a whole influences galaxy formation and the
temperature fluctuations in the CMBR provide good, consistent
evidence for the existence of a large amount of non-relativistic
matter which gravitates just like ordinary baryons do, also with
an equation of state for non-relativistic matter:
\be
    p_{DM} \approx 0 \,.
\ee
Agreement with observations requires the overall abundance of this
Dark Matter to be
\be
    \Omega_{DM} = \left(
    \frac{\rho_{DM}}{\rho} \right)_{\rm now} \approx
    26\%\,.
\ee

Since both baryons and Dark Matter share the same equation of
state, it is common to lump them together into an overall energy
density of non-relativistic matter,
\be
    \Omega_{\rm m} = \Omega_{B} + \Omega_{DM} \approx 30\% \,.
\ee

\medskip\noindent{\bf Dark Energy:}

\medskip\noindent For the past decade evidence has been
accumulating for the existence of yet another kind of invisible
matter, in addition to the Dark Matter just described. The
existence of this matter is inferred in two different ways.

\begin{figure}
\begin{center}\epsfig{file=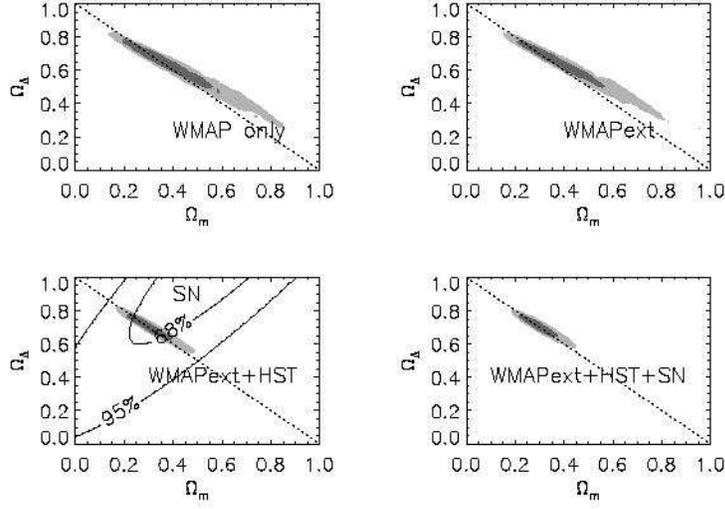,width=4in}
\caption{Current constraints on the relative abundance of Dark
Matter and Dark Energy, as inferred using properties of the CMBR
and measurements of large-scale structure. The diagonal line
corresponds to a universe having total density, $\rho = \rho_c$,
as discussed in the text \cite{WMAPParameters}.}\label{Fig:0.5}
\end{center}
\end{figure}

First, it is clear that the sum of the energy density of the
above-mentioned fluid components does not yet add up to the
observed total energy density, $\rho_c$. (Fig.~\ref{Fig:0.5} shows
the accuracy of this determination obtained using CMBR and
large-scale structure measurements.) This indicates the need for a
missing component --- called `Dark Energy' --- satisfying
\be
    \Omega_{\Lambda} = \left(
    \frac{\rho_{DE}}{\rho} \right)_{\rm now} \approx
    70\%\,.
\ee

Second, detailed tests of the Hubble expansion rate using
supernovae show that the overall expansion rate of the universe,
$H = \dot{a}/a$, appears to be {\it increasing} at present. As
eq.~\pref{RaychaudhuriEqn} shows, this can only happen for
positive energy density, $\rho > 0$, if the total pressure is
sufficiently negative, $p < -\frac13 \rho$. Since this is not true
for any of the fluid components entertained to this point,
something else must exist whose pressure is negative and at
present dominates that of the other forms of matter.

Indeed, present-day understanding of the microscopic laws of
Nature do allow pressure to be negative, and the simplest
candidate is the vacuum itself for which Lorentz invariance
implies its stress energy must satisfy $\langle T_{\mu\nu} \rangle
\propto g_{\mu\nu}$, and so is predicted to have the equation of
state
\be
    p_{DE} \approx - \rho_{DE} \,.
\ee
This equation of state is assumed in what follows for Dark Energy,
and agrees with the present observational bounds, which imply
$p_{DE}/\rho_{DE} < -0.8$. Crucially, the amount of matter having
this equation of state which reproduces the observed acceleration
in the universal expansion is consistent with the energy density
required to ensure $\sum_i \Omega_i = 1$, as required by
measurements of $H_0$ and $\kappa/a_0^2$.

\subsection{Domination by radiation, matter and Dark Energy}

The different equations of state satisfied by radiation,
non-relativistic matter ({\it i.e.} baryons and Dark Matter) and
Dark Energy implies that their relative abundances differed in the
past universe because their energy densities vary differently as
the universe expands.

\subsubsection*{Dependence of $\rho$ on $a$}

Notice that each of the above equations of state implies that the
ratio $w_i = p_i/\rho_i$ is time-independent, with
\be
    w_{\rm rad} = \frac13 \,, \qquad
    w_{\rm m} = 0 \quad \hbox{and} \quad
    w_{DE} = -1 \,,
\ee
and using this allows eq.~\pref{EnergyConservationEqn} to be
integrated to give
\be
    \rho_i = \rho_{i0} \left( \frac{a_0}{a} \right)^{\alpha_i} \,,
\ee
where $\alpha_i = 3(1 + w_i)$, and so
\be
    \alpha_{\rm rad} = 4 \,, \qquad
    \alpha_{\rm m} = 3 \quad \hbox{and} \quad
    \alpha_{DE} = 0 \,.
\ee

\begin{figure}
\begin{center}\epsfig{file=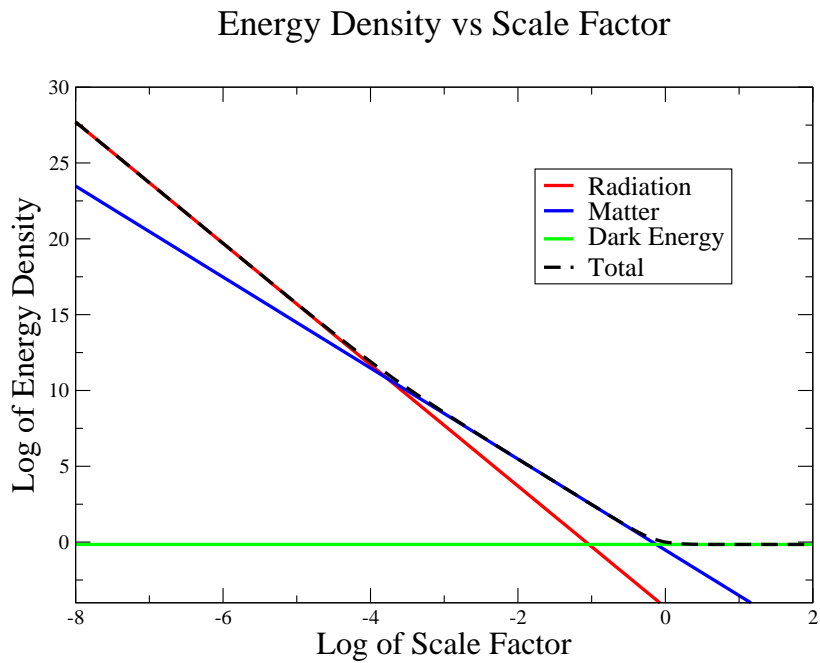,width=5in}
\caption{The energy density of radiation, non-relativistic matter
and Dark Energy as a function of the universal scale factor, in
units for which $\rho = a = 1$ at present.}\label{Fig:1}
\end{center}
\end{figure}

Combining these results shows how the total energy density evolves
with time given an initial density, $\rho_0$, which is divided
into an initial fraction, $f_i = \rho_{i0}/\rho_0$, of radiation
(rad), non-relativistic matter ($m$) and Dark Energy($DE$):
\be
    \rho(a) = \rho_0 \left[ f_{DE}
    + f_{m} \left( \frac{a_0}{a} \right)^{3}
    + f_{\rm rad} \left( \frac{a_0}{a} \right)^{4}
    \right] \,.
\ee
Because each term in the sum varies so differently with time, the
history of the universe breaks up into epochs during each of which
one term or another dominates, and so controls the overall change
of $\rho(a)$, as shown in Fig.~\pref{Fig:1}.

\begin{quote} {\bf Exercise 3:} Given the present-day abundances
of radiation and matter, and using the relation $a_0/a = 1+z$ (see
Problem 2) between redshift and scale factor, show that the epoch
where the energy density in radiation equals that of
non-relativistic matter occurs at redshift $z_{\rm eq} \approx
3600$. Show that if Dark Matter did not exist (so baryons were the
only non-relativistic matter), then the epoch of radiation-matter
equality would have instead occurred much later, at $z_{\rm
Beq}\approx 480$. \label{Ex:3}
\end{quote}

Notice in particular that the contribution to the Friedmann
equation, eq.~\pref{FriedmannEqn}, of the `curvature term',
$\kappa/a^2$, falls more quickly than does $\rho_{DE}$ (which does
not fall at all), but more slowly than $\rho_{\rm rad}$ and
$\rho_m$. Since present-day measurements are consistent with
$\kappa/a^2 \approx 0$, it follows that curvature becomes less and
less important the further back into the past we look.

\subsubsection*{Dependence of $a$ on $t$}

The dependence of $a$ on $t$ (and so also of $H$ on $a$), may be
obtained from the Friedmann equation, eq.~\pref{FriedmannEqn},
using the above expression for $\rho(a)$. Setting $\kappa = 0$,
this implies
\be \label{avstGeneral}
    (t - t_0) =  \sqrt{3} \, M_p \int_{a_0}^a
    \frac{\exd \hat{a}}{\hat{a}\sqrt{\rho(\hat{a})}} \,,
\ee
In general the right-hand side involves elliptic integrals,
however it takes a simple form whenever $\rho(a)$ is dominated by
one component of the cosmic fluid (as it almost always is). In
this instance we have $\rho(a) \approx \rho_0 (a_0/a)^\alpha$,
where $\alpha = 3(1+w)$, and so eq.~\pref{avstGeneral} is easily
integrated, leading to
\be \label{avstEqn}
    a(t) = a_0 \left( \frac{t}{t_0} \right)^\beta \quad
    \hbox{and so} \quad
    H^{-1}(t) = \frac{t}{\beta} \,,
\ee
where $\beta = 2/\alpha = \frac23(1+w)^{-1}$ for $w \ne -1$, and
so $\beta = \frac23$ when $w = 0$ and $\beta = \frac12$ when $w =
\frac13$. For later purposes, two things are worth remarking here.
First, notice that $a(t)$ grows so quickly that it could have
grown from zero size over a finite time interval. Second, $a(t)$
grows more slowly than does the Hubble length, $H^{-1}(t)$, so
long as $\beta < 1$ ({\it i.e.} for $w > -\frac13$). This is true
in particular for both radiation- and matter-dominated universes.

For the exceptional case $w = -1$ we have $\alpha = 0$ and so
$\rho = \rho_\star$ is constant, so integration gives instead
\be \label{InflationaryExpansionEqn}
    a(t) = a_0 \exp \Bigl[ H_\star(t - t_0) \Bigr] \quad
    \hbox{with} \quad
    H^{-1}(t) = H_\star^{-1} = \left( \frac{3M_p^2}{\rho_\star}
    \right)^{1/2} \,.
\ee
Here $a(t)$ grows more quickly than $H^{-1}(t)$ (which in this
case does not grow at all).

\subsection{Major Events}

The Hot Big Bang model for cosmology assumes the universe was
initially a hot soup of elementary particles, whose temperature
was once at least 10 billion degrees. In broad brush strokes, its
later evolution describes the cooling of this hot soup as the
universe expands, for which conservation of entropy implies (for
relativistic particles)
\be \label{TvsaEqn}
    T(t) = T_0 \left( \frac{a_0}{a(t)} \right) \,.
\ee

For the purposes of later observations, there are two main
consequences of such a cooling:
\begin{itemize}
\item {\bf Reduced Reaction Rates:} Reaction rates in dilute
systems are generically proportional to the number of participants
per unit volume, because the reactants must be able to find one
another before they are able to react. But since these particle
densities fall as the universal volume grows, reaction rates also
fall. This implies that one of the main trends of cosmology is the
falling out of equilibrium of various thermal and chemical
reactions.
\item {\bf Formation of Bound States:} A corollary of the previous
point is the appearance of bound states of particles as the
universe ages. Although the reactions forming bound states can
always occur, at the earliest epochs temperatures are high enough
to ensure that collisions very efficiently destroy these bound
states -- leaving very few to survive in equilibrium conditions.
But inter-particle collisions become less violent as the
temperature falls, so that eventually the reactions of formation
can dominate to leave a population of primordial relict bound
states.
\end{itemize}
At very early epochs phase transitions are also expected to play
an important role in the cosmic evolution, but as yet there is no
direct evidence that such transitions took place.

Most of the observational consequences of the Hot Big Bang revolve
about the detection of such relics, together with the detailed
measuring of their properties. A compressed history of the Hot Big
Bang era then becomes a summary of which relics have been
observed, and when they formed.

\bigskip\noindent{\bf Big Bang Nucleosynthesis} is the earliest
cosmic event -- occurring at a redshift $z_{BBN} \sim 10^{10}$ --
for which we have direct observational evidence. At this time the
temperature first cooled below about 1 MeV, at which point light
nuclei (isotopes of Hydrogen, Helium, Lithium and Beryllium) first
began to accumulate from their constituent protons and neutrons.
Observational evidence for this epoch comes from measuring the
relative abundance of these primordial elements, and comparing the
results with the predictions of nuclear physics. The success of
these comparisons also provides a direct measure of the total
baryon number density at this epoch, because this density strongly
affects the various nuclear reaction rates.

\bigskip\noindent{\bf Radiation - Matter Crossover} is defined as
the epoch when relativistic particles (radiation) stop being the
dominant contribution to the cosmic energy density, passing this
baton to non-relativistic Dark Matter (and baryons). As seen in
Problem 3, this occurs at redshift $z_{\rm eq} \sim 3600$. An
important consequence of this crossover is in the speed with which
gravity can enhance the growth of any initial density
inhomogeneities. These can grow proportional to $a$ during matter
domination, while they only grow logarithmically with $a$ during
radiation domination.

\bigskip\noindent{\bf Recombination} is the epoch where free
nuclei and electrons first combine into neutral atoms, at which
point the universe first becomes transparent to photons having
visible and near-UV wavelengths. For Hydrogen, which dominates the
cosmic baryon abundance, this occurs over a comparatively short
epoch (spread over a redshift interval of a few hundred) around
$z_{\rm rec} \sim 1100$. The CMBR has its origin as the light
which is liberated by the universe's newfound transparency at this
epoch, and so measurements of its temperature fluctuations,
$\delta T/T \sim 10^{-5}$, provide direct information about the
size of primordial density fluctuations in the cosmic environment
at this time.

\bigskip\noindent{\bf Galaxy Formation} occurs once primordial
density fluctuations have been amplified to the point that their
evolution is no longer well-described as linear perturbations.
This picture describes well the observed distribution of galaxies
in the universe, but only given the presence of non-relativistic
Dark Matter. Dark Matter is required since the amplitude of
density fluctuations is known to be very small at the epoch of
recombination, and does not grow strongly until after
radiation-matter crossover (which occurs much later in the absence
of Dark Matter).

\subsection{Special Initial Conditions}

In a nutshell, the previous section describes a simple and
consistent picture of the relatively recent universe, described by
Hot Big Bang cosmology, which is able to account for the many
observations of the overall structure and evolution of the
universe which are now being made. This success comes with some
cost, however. Besides having to postulate the existence of two
new forms of matter -- Dark Matter and Dark Energy --- for which
we have no other evidence outside of cosmology --- it is also
necessary to start the universe off with a special kind of initial
conditions. This section describes these initial conditions,
together with a theoretical framework for their explanation in
terms of the still-earlier history of the nascent universe.

It is common to couch the discussion of the special initial
conditions required by the Big Bang model in terms of
initial-condition `problems', of which there are at least three
types.

\medskip\noindent {\bf The Flatness Problem}

\medskip\noindent
The first problem concerns the spatial flatness of the present-day
universe, which is suggested by observations of the temperature
fluctuations in the CMBR. These observations indicate that the
quantity $\kappa/a^2$ of the Friedmann equation,
eq.~\pref{FriedmannEqn}, is at present consistent with zero. In
order to see why this constitutes a problematic initial condition
it is useful to divide this equation by $H(t)$ to give
\be
    1 + \frac{\kappa}{(a H)^2} = \frac{8 \pi G \rho}{3 H^2}
    \equiv
    \Omega(a) \,.
\ee
Since the product $aH$ decreases with time (during both matter and
radiation domination), this shows that the curvature term becomes
more and more important as time passes.

The problem arises because observations indicate that at present
$\Omega = \Omega_0$ is unity to within about 10\%. But during the
matter-dominated era which is just ending the product $(aH)^2
\propto a^{-1}$ so, using the result of Problem 3, at the point of
radiation-matter equality we must have had
\be
    \Omega(z_{\rm eq}) -1 = \Bigl( \Omega_0 - 1 \Bigr) (1+z_{\rm
    eq})^{-1} = \frac{0.1}{3600} \approx 2.8 \times 10^{-5} \,.
\ee
So if $\Omega_0$ is now within 10\% of unity, then it was within a
few tens of a millionth at the time of radiation-matter equality.

Earlier than this the universe was radiation-dominated, and so
$(aH)^2 \propto a^{-2}$. Since $z_{\rm BBN} \sim 10^{10}$ at the
epoch of Big Bang Nucleosynthesis we have
\be \label{OmegaBBNEqn}
    \Omega(z_{\rm BBN}) -1 = \Bigl[ \Omega(z_{\rm eq}) - 1 \Bigr]
    \left( \frac{1 + z_{\rm eq}}{1+z_{\rm BBN}} \right)^{2}
    = \frac{0.1}{3600} \left( \frac{3600}{10^{10}} \right)^2
    \approx 3.6 \times 10^{-18} \,,
\ee
requiring $\Omega$ to be unity with an accuracy of roughly a part
in $10^{18}$. The further back one goes, the more fantastic the
accuracy with which we must start $\Omega$ near 1 in order to
properly describe the universe as we now see it. One's discomfort
of having the success of a theory hinge so sensitively on the
precise value of an initial condition in this way is known as the
Big Bang's {\sl Flatness Problem}.

\medskip\noindent{\bf The Horizon Problem}

\medskip\noindent
The Big Bang's {\sl Horizon Problem} asks why the initial universe
is so very homogeneous. In particular, the temperature
fluctuations of the CMBR only arise at the level of 1 part in
$10^5$, and the question is why this temperature should be so
incredibly uniform across the sky. Why is this regarded as a
problem? After all, gasses on earth often have a uniform
temperature, and this is usually understood as a consequence of
thermal equilibrium because an initially inhomogeneous temperature
distribution equilibrates by having heat flow between the hot and
cold areas, until the gas is eventually all at the same
temperature.

What makes it odd to see the same temperature in all directions of
the sky in the Hot Big Bang model is that the universe generically
expands so quickly -- {\it c.f.} eq.~\pref{avstEqn} -- that there
has not been enough time for light to travel across the entire sky
to bring the news as to what the common temperature is supposed to
be. For instance, in a radiation-dominated universe $a(t) = a_0
(t/t_0)^{1/2}$ and $H(t) = 1/(2t)$ so the maximum proper distance
that a light signal can travel by the time of recombination,
$t_{\rm rec}$, is
\be
    L_{\rm rec} = a_{\rm rec} \int_0^{t_{\rm rec}}
    \frac{\exd \hat t}{a(\hat t)}
    = 2 t_{\rm rec}
    = \frac{1}{H_{\rm rec}}
    = \frac{1}{H_0} \left( \frac{a_{\rm rec}}{a_0} \right)^{3/2}
    \simeq \frac{1}{H_0} \left( \frac{1}{1100} \right)^{3/2} \,,
\ee
which uses $H \propto a^{-3/2}$ during matter domination (as is
appropriate between recombination and now), and $a_0/a_{\rm rec} =
1+z_{\rm rec} \simeq 1100$.

Evaluating this using $H_0 = 75$ km/sec/Mpc --- or (keeping in
mind our units for which $c=1$), $H_0^{-1} \simeq 13$ Gyr $\simeq
4$ Gpc --- gives $L_{\rm rec} \simeq 0.2$ Mpc. Now the surface of
last scattering for the CMBR at present is at a distance of order
\be
    D_0 = a_0 \int_{t_{\rm rec}}^{t_0} \frac{\exd \hat t}{a(\hat t)}
    = 3 t_0 - 3 t_0^{2/3} t_{\rm rec}^{1/3}
    = \frac{2}{H_0} \left[ 1 - \left( \frac{a_0}{a_{\rm rec}}
    \right) \frac{H_0}{H_{\rm rec}} \right]
    = \frac{2}{H_0} \left[ 1 - \left( \frac{a_{\rm rec}}{a_0}
    \right)^{1/2} \right] \,,
\ee
(using $a \propto t^{2/3}$ and $H \propto a^{-3/2}$) and so $D_0
\simeq 2/H_0 \simeq 8$ Gpc. But due to the intervening expansion
of the universe, the angle subtended by $L_{\rm rec}$ placed at
this distance away (in a spatially-flat geometry) is really
$\theta \simeq L_{\rm rec}/D_{\rm rec}$ where $D_{\rm rec} =
(a_{\rm rec}/a_0) D_0 \simeq 7$ Mpc is its distance {\it at the
time of last scattering}, leading to $\theta \simeq 1^o$. Any two
directions separated by more than this angle (about twice the
angular size of the Moon, seen from Earth) are so far apart that
light had not yet had time to reach one from the other since the
universe's beginning. How could all the directions we see then
have known they were all to equilibrate to the same temperature?
It is very much as if we were to find a very uniform temperature
distribution, {\it immediately} after the explosion of a very
powerful bomb.

\medskip\noindent {\bf A Defect Problem?}

\medskip\noindent
A third problem called the {\sl Defect Problem}\footnote{Sometimes
also known as the {\sl Monopole Problem}.} can arise if one
extrapolates the Big Bang back to times much earlier than the
epoch of Big Bang Nucleosynthesis. Unlike the previous two
problems, whether this problem really arises or not depends on the
kind of physics describing these very short distances and high
energies.

The potential problem arises if the physics of these scales
implies the universe passed through the kind of phase transition
during an earlier epoch, which produces topological defects. These
defects can take the form of very massive particles (possibly
carrying magnetic charges, and so called {\sl magnetic
monopoles}); long thin {\sl cosmic strings}, which could now be
stretched across the visible universe; or two-dimensional {\sl
domain walls} or sheets which cross the universe.

These kinds of objects can be fatal to successful late-time
cosmology, depending on how many of them survive down to the
present epoch. For instance if the defects are monopoles, then
they typically are extremely massive and so behave like
non-relativistic matter. But these can cause problems, depending
on how abundantly they are produced -- typically as much one per
Hubble volume: $n \sim H^3$. For instance, since the energy
density of such particles falls more slowly than does radiation as
the universe expands, it can easily come to dominate the universe
well before the nucleosynthesis epoch. This could cause the
universe to expand (and so cool) too quickly as nuclei were
forming, and so give the wrong abundances of light nuclei. Even if
not sufficiently abundant during BBN, the energy density in relict
defects can be inconsistent with measures of the current energy
density.

This is clearly a much more hypothetical problem than are the
other two, unless you are committed to a particular theory for the
high-energy physics of the very early universe which produces
these types of defects.

\section{Cosmic Inflation}

Cosmic Inflation was initially motivated as a way to understand
how these special initial conditions of the Hot Big Bang model
might be understood as naturally arising from the dynamics of a
much earlier epoch. Quite compellingly, it has been found more
recently also to provide a simple explanation for the origin of
the primordial density fluctuations whose presence seeds both the
observed temperature fluctuations of the CMBR and the formation of
galaxies through gravitational collapse. (For textbook treatments
of inflation, see ref.~\cite{Linde,LiddleLyth,Mukhanov}, and for
recent reviews see ref.~\cite{InflationReviews}.)

\subsection{The Inflationary Paradigm}

The idea of {\sl Cosmic Inflation} is that all three of the above
problems can be solved if the history of the universe were to have
undergone a period of accelerated expansion at some point in its
very distant past. For example, suppose the universe were to
temporarily pass through an epoch during which the dominant
component of the cosmic fluid were to have an approximately
constant energy density, $\rho = M_I^4$, which would require the
equation of state $p = - \rho$. This is the equation of state used
above for the vacuum, but now the value of the energy density is
to be chosen to be much larger, such as $M_I \sim 10^{15}$ GeV.

During any such an epoch we have seen that the Hubble scale
remains constant, $H_I \sim M_I^2/M_p$, and the scale factor grows
exponentially, or inflates, according to
eq.~\pref{InflationaryExpansionEqn}: $a(t) = a_0 \, \exp [ H_I
(t-t_0) ]$. This expansion law implies that the combination $aH$
now {\sl grows} exponentially with time, rather than falling as it
did for matter- or radiation-domination. This last observation
shows why this kind of expansion can solve the flatness, horizon
and defect problems, as we now see.

\medskip\noindent{\bf Flatness Problem:}
Since $aH$ grows exponentially it does not take long for any
initial curvature, $\kappa/(a H)$, to be diluted to extremely
small values. Precisely how much dilution is required? For
example, suppose the universe were radiation dominated all the way
back to an extremely high temperature like $T_M \sim M_I \sim
10^{15}$ GeV. Since $T \propto 1/a$ --- and since light nuclei
form at roughly $T_{BBN} \sim 1$ MeV --- the universe expands by a
factor $a_{BBN}/a_M = T_M/T_{BBN} \sim 10^{18}$ while cooling from
$T_M$ to nucleosynthesis. Since $aH \propto 1/a$ (radiation
domination) during this time it also follows that
$(aH)_M/(aH)_{BBN} \sim 10^{18}$. Comparing with
eq.~\pref{OmegaBBNEqn} shows that the universe must have been very
flat indeed at this early epoch:
\bea
    \Omega(z_M) - 1 &\sim& (\Omega_0(z_{\rm BBN}) - 1) \left[
    \frac{(a H)_{\rm BBN}}{(aH)_M} \right]^2 \nn\\
    &\sim& 3.6 \times 10^{-54} \,
    \left( \frac{10^{15} \; \hbox{GeV}}{T_M}
    \right)^2\,.
\eea

Since $(aH)_t/(aH)_0 = a(t)/a_0 = \exp[H_I (t-t_0)]$ during
exponential expansion, even such a small initial condition would
very easily be explained if the radiation-dominated epoch were
preceded by exponential expansion for a period of time, $\Delta
t$, satisfying
\be
    N_e \equiv H_I \Delta t \gsim \frac12 \ln
    \left(3 \times 10^{53} \right) \simeq 62 \,.
\ee
That is, under these circumstances generic initial conditions get
sucked towards very flat geometries by inflation, with sufficient
flatness arising even in extreme circumstances given about 60
$e$-foldings of inflation.

\medskip\noindent {\bf Horizon Problem:}
This type of accelerated expansion can also solve the horizon
problem because once $aH$ is increasing physical distance scales,
$L(t) = a(t) \ell$, grow more quickly than does the Hubble length,
$H^{-1}(t)$. Modes which were initially shorter than the Hubble
length eventually can be stretched to be larger than the Hubble
scale. The larger the co-moving scale, $\ell$, that is involved,
the earlier it grows larger than the Hubble length during
inflation. This makes it possible to have ordinary causal
processes be stretched during inflationary times to appear at late
times as if they were too far apart to be causally related.

How much inflation is required to make this work? The largest
proper scales presently visible to us are of order $H_0^{-1} \sim
4$ Gpc, and so we focus our attention to scales that are presently
this size, $L(t_0) = a_0 \ell \sim H_0^{-1}$, or $\ell \sim
1/(aH)_0$. Because $aH$ decreases during radiation- and
matter-dominated epochs, such scales satisfied $L(t) > H^{-1}(t)$
at earlier times, with for example
\be \label{ScalesatBBNEqn}
    \frac{L(t_{BBN})}{H_{BBN}^{-1}} = \ell (aH)_{BBN}
    = \left[ \frac{(aH)_{BBN}}{(aH)_{\rm eq}} \right]
    \left[ \frac{(aH)_{\rm eq}}{(aH)_{0}} \right]
    = \left( \frac{a_{\rm eq}}{a_{BBN}} \right)
    \left( \frac{a_{0}}{a_{\rm eq}} \right)^{1/2}
    \simeq 2 \times 10^8 \,,
\ee
at the epoch of nucleosynthesis.

During exponential expansion, however, $L/H^{-1}$ grows and so we
ask how much exponential expansion is required in order to ensure
that this scale also satisfies $L < H^{-1}$ at some earlier time,
$t_{\rm he}$, called the time of {\sl horizon exit}. For times
earlier than this (during or before inflation) causal processes
can be at work to explain things like the present-day uniformity
of the CMB temperature over these scales. (See Figure \ref{Fig:3}
for a sketch of the relative sizes of $L$ and $H^{-1}$, during and
after inflation.)

For simplicity we assume that inflation ends when $t = t_{\rm
end}$ and the universe then makes an immediate transition from an
inflationary epoch, where $\rho = \rho_I = M_I^4$ is approximately
constant, to a radiation-dominated epoch whose initial {\sl reheat
temperature} is also $T \sim M_I$ ({\it i.e.} reheats with perfect
efficiency). In this case at the epoch of horizon exit we have (by
assumption) $L(t_{\rm he}) = \ell a_{\rm he} = H^{-1}_{\rm he}$
and so $\ell = (aH)_0^{-1} = (aH)_{\rm he}^{-1}$. Consequently,
\bea \label{1EqualsIdentityEqn}
    1 &=& \frac{a_0 H_0}{a_{\rm he} H_{\rm he}} = \left(
    \frac{a_{\rm end} H_{\rm end}}{a_{\rm he}H_{\rm he}} \right)
    \left( \frac{a_{\rm eq} H_{\rm eq}}{
    a_{\rm end} H_{\rm end}} \right)
    \left( \frac{a_0 H_0}{a_{\rm eq} H_{\rm eq}} \right)
    \,,
\eea
which we solve for $a_{\rm end}/a_{\rm he} = e^{N_e} =
e^{H_I(t_{\rm end} - t_{\rm he})}$, assuming a constant energy
density during inflation, and so $H_{\rm he} \approx H_{\rm end}$.
Using, as above, $(a_{\rm eq} H_{\rm eq})/(a_0 H_0) = (a_0/a_{\rm
eq})^{1/2} \simeq 60$, and $(a_{\rm eq} H_{\rm eq})/(a_{\rm end}
H_{\rm end}) = a_{\rm end}/a_{\rm eq} = T_{\rm eq}/T_M$ with
$T_{\rm eq} \sim 3$ eV leads to
\be \label{NIFromHorizonEqn}
    N_e \sim \ln \left[(3 \times 10^{23}) \times 60 \right]
    + \ln\left( \frac{T_M}{10^{15} \; \hbox{GeV}} \right)
    \approx 58 + \ln\left( \frac{T_M}{10^{15} \;
    \hbox{GeV}} \right)\,.
\ee

Again we see that roughly 60 $e$-foldings of exponential expansion
can provide a framework for explaining how causal physics might
provide the observed correlations that are observed in the CMBR
over the largest scales. We shall see below that life is even
better than this, because in addition to providing a {\sl
framework} in which a causal understanding of correlations could
be solved, inflation itself can provide the {\sl mechanism} for
explaining these correlations (given an inflationary scale of the
right size).

\medskip\noindent {\bf Defect Problem:} Inflation can also solve
the defect problem --- within theories for which this needs
solving --- for similar reasons. Consider for example monopoles,
which are typically predicted to be produced one per Hubble
volume, $H_f^{-3}$, at the epoch where they are formed.
Consequently their number density at that time would be $n_f \sim
H_f^{3}$. The number density at later times is therefore $n =
H_f^3(a_f/a)^3$ and so the number of defects per Hubble volume at
later times is $N_{\rm def} = n H^{-3} = [(aH)_f/(aH)]^3$. As such
it is clear that this number gets enormously diluted {\sl if} the
monopoles are produced before inflation, because of the enormous
exponential suppression which is then possible for $(aH)_f/(aH)$.

\subsection{Single-Field Models}

So far so good, but the devil is in the details. Obtaining the
benefits of such an exponential expansion requires two things:
$(i)$ some sort of physics which can hang the universe up for a
relatively long period with a vacuum-dominated equation of state,
$p \approx -\rho$; {\sl and} $(ii$) some mechanism for ending this
epoch to allow the later appearance of the radiation-dominated
epoch within which the usual Big Bang cosmology starts. Although a
number of models exist for the kinds of physics which might do
this, none of these models yet seems completely compelling. This
section describes some of the very simplest such models, in order
to see some of their successes and limitations, and to see what
their implications can be for the large-scale structure seen in
the later universe.

No way is known to obtain inflation simply using the known
particles and interactions, and so inflationary models are
characterized by what kind of new physics is invented to describe
the inflationary dynamics. For the vast majority of models this
new physics comes from the dynamics of a scalar field,
$\varphi(x)$, (called the {\sl inflaton}) which can be thought to
be an order parameter characterizing the nature of the vacuum in
the theory which describes the very high energy physics relevant
to inflationary cosmology. Although the field $\varphi$ can in
principle depend on both position and time, inflation turns out
rapidly to smooth out spatial variations, and so it suffices to
study $\varphi = \varphi(t)$.

The simplest such a relativistic order parameter has a dynamics
which is determined by a potential energy, $V(\varphi)$, and
satisfies the following field equation,
\be \label{ScalarFieldEqn}
    \ddot{\varphi} + 3 H \dot{\varphi} + V' = 0 \,,
\ee
where $V' = \exd V/\exd\varphi$. Its gravitational influence is
described by the usual Friedmann and acceleration equations, but
including also a $\varphi$-dependent contribution to the energy
and pressure: $\rho = \rho_{\rm rad} + \rho_{\rm m} +
\rho_\varphi$ and $p = \frac13 \, \rho_{\rm rad} + p_\varphi$,
where $\rho_{\rm rad}$ and $\rho_{\rm m}$ describe the energy
density of relativistic and non-relativistic matter, and
\be \label{Rhoandpvarphi}
    \rho_\varphi = \frac12 \, \dot{\varphi}^2 + V(\varphi) \qquad
    \hbox{and} \qquad
    p_\varphi = \frac12 \, \dot{\varphi}^2 - V(\varphi) \,.
\ee
We imagine the Dark Energy of the modern epoch to correspond to
there being a very small constant term in $V$, which is assumed to
presently dominate.

As is easy to check, with these choices energy conservation for
the $\varphi$ field --- $\dot\rho_\varphi + 3 (\dot{a}/a)
(\rho_\varphi + p_\varphi) = 0$ follows from the field equation,
eq.~\pref{ScalarFieldEqn}, and so $\varphi$ exchanges energy with
the rest of the cosmic ingredients purely through their mutual
gravitational interactions. The $\varphi$ field is not imagined to
be in thermal equilibrium with itself or with the other kinds of
matter, and this is self-consistent because it couples to the
other matter only gravitationally (which is too weak to establish
equilibrium).

\subsubsection*{Slow-Roll Inflation}

We seek a solution to these equations for $\varphi(t)$ for which
the Hubble parameter, $H$, is approximately constant. This is
ensured if the total energy density is dominated by
$\rho_\varphi$, with $\rho_\varphi$ also approximately constant.
Energy conservation then requires the pressure to satisfy
$p_\varphi \approx - \rho_\varphi$. It does not matter here that
$\varphi$ is not in equilibrium, since for $\varphi$ we ask that
this relation between $\rho_\varphi$ and $p_\varphi$ to follow as
a consequence of the field equations and not as an equation of
state. Inspection of eqs.~\pref{Rhoandpvarphi} shows that the
regime of interest is when the $\varphi$ kinetic energy is
negligible compared with its kinetic energy: $\frac12
\dot{\varphi}^2 \ll V(\varphi)$ since then $p_\varphi \approx -
V(\varphi) \approx - \rho_\varphi$. So long as $V(\varphi)$ is
also much larger than any other energy densities, it would
dominate and $H^2 \approx V/(3 M_p^2)$ would then be approximately
constant.

What properties must $V(\varphi)$ satisfy in order to allow such
an extended period of slow rolling? Clearly the field equation
\pref{ScalarFieldEqn} only permits precisely time-independent
solutions, $\varphi = \varphi_0$, at points where the potential is
stationary, $V'(\varphi_0) = 0$. As we now quantify, a sufficient
condition for having a long period of time with $\varphi$ very
slowly moving requires {\sl both} $\dot\varphi$ and $\ddot\varphi$
to remain small for the entire inflationary period, and so
requires both $V'$ and $V''$ to be close to zero for a
sufficiently broad range of $\varphi$.

More specifically, in order to have a prolonged slow roll we must
demand $\ddot\varphi \ll H \dot\varphi$, which allows
eq.~\pref{ScalarFieldEqn} to be approximately written in the
following {\sl slow-roll} approximation
\be \label{SRScalarFieldEqn}
    \dot{\varphi} \approx - \left( \frac{V'}{3H} \right) \,.
\ee
Using this in the condition $\frac12 \dot\varphi^2 \ll V$ shows
$V$ must satisfy $(V')^2/(9 H^2 V) \ll 1$, or
\be \label{epsilonParameterDef}
    \epsilon \equiv \frac12 \left( \frac{M_p V'}{V} \right)^2
    \ll 1 \,.
\ee
A self-consistency condition for using eq.~\pref{SRScalarFieldEqn}
throughout inflation is the requirement that $\ddot\varphi$
remains small. Differentiating eq.~\pref{SRScalarFieldEqn} with
respect to $t$, and using the approximate constancy of $H$ gives
$\ddot\varphi \approx - V'' \dot\varphi/(3H)$. Demanding this
remain small (in absolute value) compared with $3 H \dot\varphi$,
then gives $|V''/(3H)^2|  \ll 1$, or $|\eta| \ll 1$ where
\be \label{etaParameterDef}
    \eta \equiv \frac{M_p^2 \, V''}{V} \,.
\ee
As we shall see, all of the important predictions of single-field
slow-roll inflation for density fluctuations can be expressed in
terms of these two small parameters, $\epsilon$ and $\eta$,
together with the value of the Hubble parameter, $H$, during
inflation.

We have seen that the success of inflation relies on obtaining
sufficient expansion, and so it is convenient to relate the amount
of expansion directly to the distance $\varphi$ traverses in field
space. To this end, rewriting eq.~\pref{SRScalarFieldEqn} in terms
of $\varphi' \equiv \exd \varphi/\exd a$, leads to
\be
    \frac{\exd\varphi}{\exd a} = \frac{\dot{\varphi}}{\dot{a}} =
    -\, \frac{V'}{3a H^2} = - \, \frac{M_p^2 \, V'}{aV} \,,
\ee
which when integrated between the initial value, $\varphi_i$, and
final value, $\varphi_{\rm end}$, implies the universal expansion
during inflation is given by $a_{\rm end}/a_i \equiv \exp(N_I)$,
with
\be \label{NIvsPhiEqn}
    N_I(\varphi_i) = \int_{a_i}^{a_{\rm end}} \frac{\exd a}{a}
    = \int_{\varphi_{\rm end}}^{\varphi_i} \exd
    \varphi \left( \frac{V}{M_p^2 \, V'} \right)
    = \frac{1}{M_p} \int_{\varphi_{\rm end}}^{\varphi_i}
    \frac{\exd \varphi}{\sqrt{2 \epsilon}} \,.
\ee
Since $\varphi_{\rm end}$ can be defined by the point where the
slow-roll parameters are no longer small, this last equation can
be read as defining $\varphi_i(N_I)$, as a function of the desired
number of $e$-foldings. This is most usefully applied to finding
the number of $e$-foldings, $N_e$, between the the epoch of
horizon exit -- as defined below eq.~\pref{1EqualsIdentityEqn} --
and the end of inflation: $N_e \equiv N_I(\varphi_{\rm he})$,
since it is this quantity which is constrained to be large by the
horizon and flatness problems. Notice also that if $\epsilon$ were
approximately constant during inflation, then
eq.~\pref{NIvsPhiEqn} implies that $N_I \approx (\varphi_i -
\varphi_{\rm end})/(\sqrt{2\epsilon} \, M_p)$. In such a case
$\varphi$ must traverse a range larger than $O(M_p)$ between
$\varphi_i$ and $\varphi_{\rm end}$ in order to obtain 60 or more
$e$-foldings, unless $\epsilon \lsim 10^{-4}$.

\subsubsection*{Large- and Small-Field Examples}

Consider, for example, the special case where
\be \label{PotentialDef}
    V = A + \frac12 \, B \, \varphi^2 + \frac14 \, \lambda^2 \,
    \varphi^4 \,,
\ee
and so for which
\be
    V' = B \, \varphi + \lambda^2 \, \varphi^3
    \qquad\hbox{and} \qquad
    V'' = B + 3 \lambda^2 \, \varphi^2 \,.
\ee
There are two examples of slow rolls which arise in this case and
which (for observational purposes) are representative of two of
the main classes of inflationary models.

\medskip\noindent {\bf Large-Field Inflation:}

\medskip\noindent For very large $\varphi$ we have $V \approx
\frac14 \, \lambda^2 \, \varphi^4$, $V' \approx \lambda^2 \,
\varphi^3$ and $V'' \approx 3 \lambda^2 \, \varphi^2$ and so
\be
    \epsilon \approx \frac12 \, \left(
    \frac{4 M_p}{\varphi} \right)^2
    \qquad \hbox{and} \qquad
    \eta \approx \frac{12 M_p^2}{\varphi^2}  \,.
\ee
while the scale for inflation is $M_I^4 \equiv V \approx \frac14
\, \lambda^2 \, \varphi^4$ and so $H_I \approx \lambda \,
\varphi^2/(2 \sqrt3 \, M_p)$. [More generally, for $M_I^4 = V
\approx \frac1n \, \lambda^2 \, \varphi^n$, $V' \approx \lambda^2
\, \varphi^{n-1}$ and $V'' \approx (n-1) \lambda^2 \,
\varphi^{n-2}$ and so $\epsilon \approx \frac12 \, \left(n
M_p/\varphi \right)^2$ and $\eta \approx n(n-1) M_p^2/\varphi^2$,
and the Hubble scale for inflation is $H_I \approx \lambda \,
\varphi^2/(\sqrt{3n} \, M_p)$.]

In this case $\eta \approx \frac32 \, \epsilon > 0$ and both are
small provided $\varphi \gg M_p$ (which is consistent with the
large-$\varphi$ approximation being used). In this regime
$\varphi$ (and so also $V$ and $H$) remains approximately constant
despite there being no stationary point for $V$ at large $\varphi$
because Hubble friction keeps $\varphi$ from sliding down the
potential very quickly. Since $\varphi$ evolves towards smaller
values, eventually slow roll ends once $\eta$ and $\epsilon$
become $O(1)$. Since $\eta > \epsilon$, it is convenient to define
$\varphi_{\rm end}$ by $\eta = \frac34$, which implies
$\varphi_{\rm end} = 4 M_p$.

The number of $e$-foldings between horizon exit and $\varphi_{\rm
end} = 4 M_p$ is given by eq.~\pref{NIvsPhiEqn}, which becomes
\be \label{NIvsPhiEqnLF}
    N_e \equiv N_I(\varphi_{\rm he})
    = \int_{\varphi_{\rm end}}^{\varphi_{\rm he}} \exd
    \varphi \left( \frac{\varphi}{4M_p^2} \right)
    = \frac{\varphi_{\rm he}^2}{8 M_p^2} - 2 \,.
\ee
This shows that obtaining $N_e > 60$ $e$-foldings requires
choosing $\varphi_{\rm he} \gsim 22 \, M_p$.

\medskip\noindent{\bf Small-Field Inflation:}

\medskip\noindent Alternatively, imagine again using the potential
of eq.~\pref{PotentialDef}, but instead assuming $B = - \mu^2 < 0$
and so $V$ has a local maximum at $\varphi = 0$. Sufficiently near
this maximum,
\be \label{SmallPhiConditions}
    \varphi^2 \ll \min \left( \frac{2A}{\mu^2} ,
    \frac{2\mu^2}{\lambda^2}
    \right) \,,
\ee
we have $V \approx A \equiv M_I^4$, $V' \approx -\mu^2 \, \varphi$
and $V'' \approx -\mu^2$. If so, the slow-roll parameters become
\be
    \epsilon \approx \frac12 \left( \frac{\mu^2 M_p\,\varphi}{A}
    \right)^2 \qquad \hbox{and} \qquad
    \eta \approx - \left( \frac{\mu^2 M_p^2}{A} \right) \,.
\ee
In this case $\eta < 0$ and $\epsilon = \frac12 (\eta
\varphi/M_p)^2$. $|\eta|$ is small provided $\mu^2 M_p^2 \ll A$
and since we have assumed $\varphi$ to be small we see that
generically in this case $\epsilon \ll |\eta|$. Again the
slow-roll regime is consistent with the small-$\varphi$
approximation with which we start. The inflationary scale is $V
\approx A = M_I^4$, and so $H = M_I^2/(\sqrt3 \, M_p)$.

Physically, the scalar potential in this case can dominate the
energy density because there is always an unstable solution to the
equations of motion corresponding to sitting with $\varphi$
precisely at rest at the local maximum, where $V' = 0$. Solutions
near this static solution can therefore be very slow if they start
sufficiently close to the maximum, or if the maximum is
sufficiently shallow. As we see below, only the second of these
two options provides a bona fide inflationary model.

Since $\eta$ is constant, the end of inflation occurs once either
$\epsilon$ becomes $O(1)$ or once the small-$\varphi$ conditions,
eq.~\pref{SmallPhiConditions}, break down. Since $\epsilon = O(1)$
requires $\varphi = O(M_p/|\eta|)$, it is well outside of the
assumed small-field regime and so it is the failure of
eq.~\pref{SmallPhiConditions} which kicks in first: $\varphi_{\rm
end}^2 \sim \min(2A/\mu^2,2\mu^2/\lambda^2)$. The total number of
$e$-foldings after $\varphi = \varphi_{\rm he}$, becomes in this
case
\be \label{NIvsPhiEqnSF}
    N_e \equiv N_I(\varphi_{\rm he})
    = \int_{\varphi_{\rm end}}^{\varphi_{\rm he}} \exd
    \varphi \left( \frac{A}{M_p^2 \, B \varphi} \right)
    = \frac{A}{M_p^2 \mu^2} \ln \left(
    \frac{\varphi_{\rm end}}{\varphi_{\rm he}} \right)
    = \frac{1}{|\eta|} \ln \left(
    \frac{\varphi_{\rm end}}{\varphi_{\rm he}} \right)
    \,.
\ee
Since this only depends logarithmically on $\varphi_{\rm
end}/\varphi_{\rm he}$, obtaining $N_e \gsim 60$ generically
requires $|\eta| \lsim 1/60 = 0.017$. Taking, for instance, $A =
M_I^4$ with $M_I = 10^{14}$ GeV, then implies $\mu \lsim M_I^2/M_p
= 10^{10}$ GeV.

Another way to make $N_e$ large would be to take $\varphi_{\rm he}
\to 0$, since in this limit $N_e \to \infty$ corresponding to the
solution which sits at the top of the maximum for an indefinitely
long period. At first sight this choice seems attractive because
it appears always to be possible, regardless of how steeply the
potential falls away from this maximum. However, in reality the
inflaton field is subject to fluctuations, such as due to quantum
vacuum fluctuations which arise because the scalar-field
Hamiltonian --- for which the vacuum is an eigenstate -- does not
commute with the field, $\phi$, itself. $\varphi$ is only a
classical approximation to $\langle \phi \rangle$, but in an
exponentially-expanding universe the fluctuations about this value
turn out to be of order $\delta \varphi \sim H_I$. Generically,
then, we can only choose $\varphi_{\rm he} = 0$ to within an
accuracy $\delta \varphi \sim H_I$, and so should restrict
$\varphi_{\rm he} \gsim H_I$. For the potential of current
interest this implies $\varphi_{\rm he} \gsim H_I \sim M_I^2/M_p$
and so since $\varphi_{\rm end} \sim
\hbox{min}(M_I^2/\mu,\mu/\lambda)$, we have $\varphi_{\rm
end}/\varphi_{\rm he} \lsim M_p/\mu$ or $\varphi_{\rm
end}/\varphi_{\rm he} \lsim \mu M_p/(\lambda M_I^2) \sim
\sqrt{|\eta|}/\lambda$, showing that large values for $N_e$ really
do require $|\eta|$ to be small.

\subsubsection*{Consistency of the Approximations}

It is important for any inflationary model to ask whether the
choices made for inflation are consistent with approximations
which are made when writing down a scalar potential. There are
three important criteria which must be satisfied.

\begin{enumerate}
\item {\bf Perturbation theory:} Analyzing the dynamics of
$\varphi$ as a classical field (rather than a quantum one) assumes
the semi-classical approximation. For instance, the validity of
this is justified in the case studied above when $\lambda \ll 1$
and $\varphi^2 \lsim |B|/\lambda^2$.
\item {\bf Quantum Gravity:} Neglect of the complications of
quantum gravity require that no energy densities should ever be
allowed to be greater than Planck density. That is, $\frac12 \,
\dot{\varphi}^2 \ll M_p^4$ and $V \ll M_p^4$. In the example above
this implies choosing $A \ll M_p^4$, $\varphi/M_p \ll M_p/\mu$ and
$\varphi/M_p \ll \lambda^{-1/2}$. Note this can permit the
large-$\varphi$ regime, $\varphi \gg M_p$, provided $\lambda$ and
$\mu/M_p$ are sufficiently small.
\item {\bf High-Energy Corrections to $V$:} Typically the
integrating out of higher-energy physics generates corrections to
the shape of $V$, with the contributions due to physics at mass
scale $M$ generically contributing terms of order $\delta V \sim
\varphi^k/M^{k-4}$ for all possible choices for $k$. If the
success of the inflationary model depends on the particular form
for $V$ it is therefore necessary to understand why these
corrections are not present or important in the example of
interest. For small-field inflation it is the absence of terms
with $k \le 4$ which require explanation, since these are not
automatically suppressed by powers of $\varphi/M$. Since $M$ is
typically smaller than $M_p$, large-field inflation is sensitive
to a potentially enormous range of $k$'s, which is to say that it
must be understood why these corrections do not change the
large-field form of the potential.
\end{enumerate}

\subsection{Primordial Density Fluctuations}

One of the successes of the Hot Big Bang is its description of the
origins of galaxies, which are understood as the final result of
the gravitational amplification of what were initially very small
density inhomogeneities. This picture of structure formation very
successfully describes the observed distribution of galaxies, as
well as how this distribution correlates with the observed small
temperature fluctuations of the CMBR. The structure-formation
picture assumes the existence of initially small primordial
fluctuations about the homogeneous universe, and its success
depends on assumptions made about their detailed properties. These
are simply taken as an initial condition of the Big Bang Model,
with no attempt made to understand their origin.

Although originally motivated as a solution to the horizon and
flatness problems, a bonus for inflationary models is their
subsequent success in predicting the properties of the primordial
fluctuations which the Hot Big Bang requires. This prediction
describes the fluctuations as being due to ordinary microscopic
quantum fluctuations of the inflaton field, $\delta\varphi$, and
the metric, $\delta g_{\mu\nu}$, which become stretched up to
cosmologically interesting scales by the inflationary expansion of
the universe. This section provides a heuristic description of
these fluctuations before quoting the final results which follow
from more sophisticated calculations.

\subsubsection*{Fluctuation Phenomenology}

Before describing what inflation can say about the properties of
primordial fluctuations, first recall how these fluctuations are
characterized. Since the universe seems to be spatially flat, it
is convenient for these purposes to use $\kappa = 0$ for the
background geometry, and to Fourier transform fluctuating
quantities. For instance, writing the fluctuating energy density
in non-relativistic matter as $\rho(\bfr,t) = \rho_{\rm m}(t)[1 +
\delta(\bfr,t)]$, we have
\be
    \delta(\bfr,t) = \int \frac{\exd^3k}{(2\pi)^3}
    \, \delta_k(t) \, \exp[i\bfk \cdot \bfx ]
    = \int \frac{\exd^3k}{(2\pi)^3}
    \, \delta_k(t) \, \exp\left[i(\bfk/a) \cdot \bfr \right]\,,
\ee
where homogeneity and isotropy of the background cosmology implies
$\delta_k(t)$ depends only on $k = |\bfk|$ and $t$. $\bfx$ here
denotes the co-moving coordinate, corresponding to physical
distance $\bfr = a \, \bfx$, so the physical wavelength associated
with co-moving wave-number $k$ is $\lambda = 2\pi a/k$.

A useful statistic for quantifying the galaxy distribution is the
density-density autocorrelation function, $\xi_\rho(\bfr,t)$,
defined by
\be \label{PowerSpectrumDef}
    \xi_\rho(\bfr) = \left\langle \delta(\bfr'+\bfr) \delta(\bfr')
    \right\rangle = \int \frac{\exd^3k}{(2\pi)^3} \, P_\rho(k)
    \, \exp[i(\bfk/a) \cdot \bfr] \,,
\ee
where the average is over all $\bfr'$. This measures how likely it
is to find a density excursion at a physical distance $\bfr$ from
a given density excursion. The integrand in the second equality
defines the power spectrum of density fluctuations, $P_\rho(k)$,
which can be related to $\delta_k$ by $P_\rho(k) \propto |
\delta_k |^2$. A dimensionless measure of the power spectrum is
obtained by performing the angular integration in
eq.~\pref{PowerSpectrumDef}, leading to
\be \label{PowerSpectrumDef2}
    \xi_\rho(\bfr) = \int_0^\infty \frac{\exd k}{k}
    \, \Delta^2_\rho(k)
    \, \frac{\sin(kr/a)}{kr/a} \,,
\ee
where $\Delta^2_\rho = k^3 P_\rho(k)/(2\pi^2)$.

\begin{figure}
\begin{center}\epsfig{file=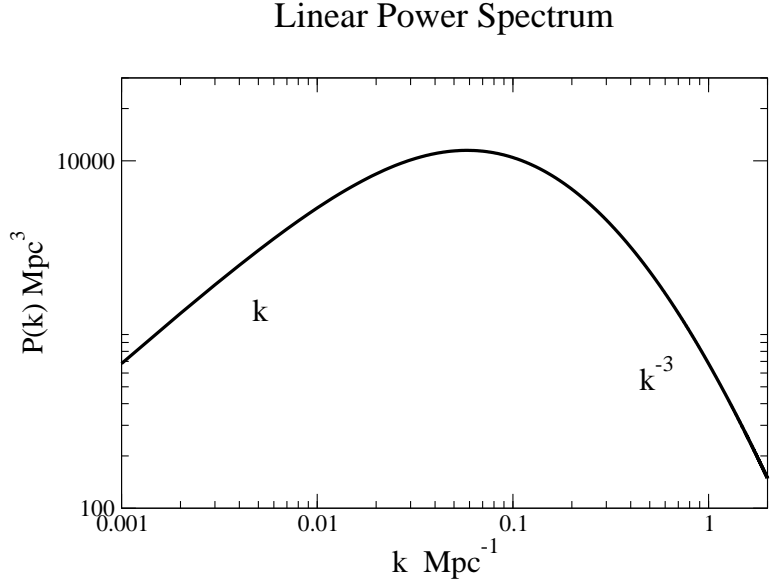,width=4in}
\caption{A sketch of the linear density power spectrum,
$P_\rho(k)$.}\label{Fig:2}
\end{center}
\end{figure}

Theoretically, the evolution of linear perturbations within the
Hot Big Bang allows $P_\rho(k,t)$ to be computed in terms of the
primordial spectrum, $P_\rho^{\,0}(k)$, of initial density
distributions, according to
\be
    P_\rho(k,t_{\rm now}) = P_\rho^{\,0}(k) \,
    \cT(k,t_0,t_{\rm now})
    \,,
\ee
where $\cT(k,t,t')$ is a calculable transfer function. $\cT$ has
the property that it is approximately independent of $k$ for small
$k$, and is proportional to $k^{-4}$ for large $k$, with the
transition between these regimes occurring for co-moving
wave-numbers satisfying $k \simeq (aH)_{\rm eq}$ at the epoch of
radiation-matter equality.

Physically, this form for $\cT$ arises because those modes
satisfying $k > k_{\rm eq} \equiv (aH)_{\rm eq}$ re-enter the
Hubble scale {\it before} radiation-matter equality, while those
with $k < k_{\rm eq}$ do so afterwards. However, density
fluctuations only grow logarithmically with $a$ during radiation
domination, but can grow proportional to $a$ during matter
domination. All other things being equal, one therefore expects
modes with $k < k_{\rm eq}$ to have a $k$-dependent amplitude,
because they grow over the $k$-dependent time interval during
which the universe expands by a factor $a_0/a_k \propto k^2$,
where $a_k$ is defined as the scale factor at the epoch where $aH
= k$, and we use that $aH \propto a^{-1/2}$ during matter
domination to conclude $a_k \propto k^{-2}$. By contrast, modes
with $k > k_{\rm eq}$ all start growing at radiation-matter
equality, and so are amplified by a $k$-independent factor:
$a_0/a_{\rm eq}$. This implies modes with $k > k_{\rm eq}$ are
stunted by an amount proportional to $(k_{\rm eq}/k)^2$ relative
to what one would get by extrapolating from the amplitude of modes
with $k < k_{\rm eq}$, so their contribution to the power spectrum
is suppressed by $\cT \propto 1/k^4$.

\begin{figure}
\begin{center}\epsfig{file=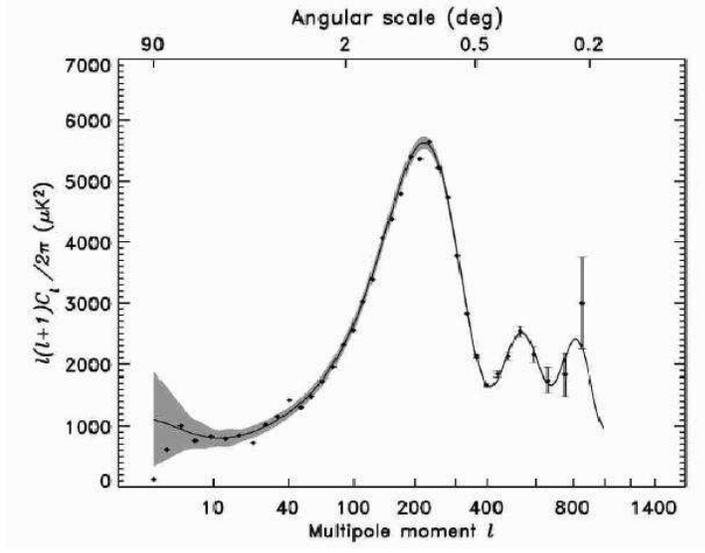,width=4in}
\caption{Legendre coefficients for the CMBR temperature
correlations, as measured by the WMAP collaboration
\cite{WMAP3}.}\label{Fig:4}
\end{center}
\end{figure}

Observationally, $P_\rho(k)$ can be related to the galaxy-galaxy
correlation function, which can be measured from surveys of galaxy
distributions. It can also be used to compute the temperature
fluctuations observed in the CMBR. These measure the correlations
between the temperature deviations seen in two directions, $\bfn$
and $\bfn'$, as a function of their relative direction,
$\cos\theta = \bfn \cdot \bfn'$, with the result averaged over all
possible orientations of these two vectors (for fixed relative
direction, $\theta$) in the sky. The result is conventionally
expressed by expanding in a Legendre series,
\be
    \left\langle \frac{\delta T}{T}(\bfn) \frac{\delta
    T}{T}(\bfn') \right\rangle = \frac{1}{4\pi} \sum_{l =
    0}^\infty (2l+1) C_l P_l(\cos\theta) \,,
\ee
and quoting the measured values for $C_l$. (See Fig.~\ref{Fig:4}
for recent measurements of these coefficients.)

Perturbations to the Dark Matter density, $\delta\rho_{\rm m}$,
are related to $\delta T/T$ because the temperature fluctuations
arise due to the redshift of CMBR photons as they climb out of the
gravitational potential wells that are generated by
$\delta\rho_{\rm m}$ --- a phenomenon called the {\sl Sachs-Wolfe}
effect. (In a matter-dominated universe, the quantity $\Phi +
\delta T/T$ turns out to be a constant along a photon trajectory
\cite{Mukhanov}). Measurements agree well with what is expected
theoretically, provided the primordial power spectrum has a simple
power-law form $P_\rho^0(k) = A k^{n_s}$, with $n_s \simeq 1$.

The choice $n_s = 1$ is called the {\sl Harrison-Zel'dovich} (HZ)
spectrum, and is special because it corresponds to the case where
the dependence of $\Delta_\rho^2$ is approximately scale invariant
for modes which re-enter the horizon during the recent
radiation-dominated universe: $\Delta_\rho^2 \propto k^0$ for $k >
k_{\rm eq}$ (and so $\Delta_\rho^2 \propto k^4$ for $k < k_{\rm
eq}$). It also corresponds to scale-invariant fluctuations for the
Newtonian gravitational potential, $\Phi$ (defined in more detail
for the relativistic case below), when $k < k_{\rm eq}$. To see
why, notice that $\Phi$ is related to $\rho$ by the Poisson
equation --- {\it i.e.} $\nabla^2 \Phi = 4\pi G\rho$ --- and so
the power spectra for $\Phi$ and $\rho$ should be related by
$P_\Phi(k) \propto P_\rho(k)/k^4$. Consequently, if $P_\rho(k)
\propto k^{n_s}$ for small $k$, then $P_\Phi(k) \propto k^{n_s-4}$
and the corresponding dimensionless power spectrum is
$\Delta_\Phi^2(k) \propto k^3 P_\Phi(k) \propto k^{n_s - 1}$,
which is independent of $k$ when $n_s = 1$.

\subsubsection*{Evolution of Primordial Fluctuations}

Since inflation provides the past from which the Hot Big Bang
later evolves, it is natural to try to compute quantities like
$P_\Phi^0(k)$, assuming they arise from this earlier epoch. To
this end it is necessary to follow the evolution of small
fluctuations in the inflaton, $\delta\varphi$, as well as the
metric, $\delta g_{\mu\nu}$, during and after the inflationary
epoch.\footnote{The discussion here follows the excellent
treatment in \cite{Mukhanov}.}

The perturbations of the metric, $\delta g_{\mu\nu}$ come in three
kinds: {\sl scalar}, {\sl vector} and {\sl tensor} fluctuations,
which differ in how they transform under rotations (and so evolve
independently of one another at linear order in the fluctuations).
After transforming to conformal time, $\hat\eta = \int \exd t/a$,
the scalar perturbations may be written
\be
    \delta_S g_{\mu\nu} = a^2 \begin{pmatrix} 2 \phi &&
    \partial_j \cB \\ \partial_i \cB && 2 \psi \, \delta_{ij}
    + \partial_i\partial_j \cE \end{pmatrix} \,,
\ee
while the vector and tensor ones become
\be
    \delta_V g_{\mu\nu} = a^2 \begin{pmatrix} 0 && \cV_j \\
    \cV_i && \partial_i \cW_j + \partial_j \cW_i \end{pmatrix}
    \quad \hbox{and} \quad
    \delta_T g_{\mu\nu} = a^2 \begin{pmatrix} 0 && 0 \\
    0 && h_{ij} \end{pmatrix} \,.
\ee

The freedom to perform infinitesimal coordinate transformations
allows these functions to be changed, so it is useful to define
the following coordinate-invariant combinations:
\bea \label{GaugeInvariantDefs}
    \Phi &=& \phi - \frac{1}{a} \Bigl[ a(\cB - \cE') \Bigr]'
    \,, \qquad
    \Psi = \psi + \frac{a'}{a} (\cB - \cE') \\
    \delta \chi &=& \delta \varphi
    - \varphi' ( \cB - \cE') \,, \quad
    V_i = \cV_i - \cW_i \quad \hbox{and} \quad
    h_{ij}  \,, \nn
\eea
in terms of which all physical inferences can be drawn. Here
primes denote differentiation with respect to conformal time,
$\hat\eta$. Notice that $\Phi$, $\Psi$ and $V_i$ reduce to $\phi$,
$\psi$ and $\cV_i$ in the gauge choice where $\cB = \cE = \cW_i =
0$, and so $\Phi$ is the relativistic generalization of the
Newtonian potential.

\begin{quote}{\bf Exercise 4:}
Show that the combinations given in eqs.~\pref{GaugeInvariantDefs}
are invariant under infinitesimal coordinate transformations:
$\delta \varphi = \xi^\mu \partial_\mu \varphi$ and $\delta
g_{\mu\nu} = \xi^\lambda \partial_\lambda g_{\mu\nu} +
\partial_\mu \xi^\lambda g_{\lambda\nu} + \partial_\nu \xi^\lambda
g_{\lambda\mu}$. \label{Ex:4}
\end{quote}

These functions are evolved forward in time by linearizing the
relevant field equations:
\be
    \Box \varphi - V'(\varphi) = 0 \quad \hbox{and} \quad
    R_{\mu\nu} - \frac12 \,R g_{\mu\nu} = \frac{T_{\mu\nu}}{M_p^2}
    \,,
\ee
and provided we use the invariant stress-energy perturbations,
\bea
    \delta {\cT^0}_0 &=& \delta {T^0}_0 -
    \left[ {t^0}_0 \right]' (\cB - \cE') \,, \nn\\
    \delta {\cT^0}_i &=& \delta {T^0}_i -
    \left[ {t^0}_0 - \frac13 \, {t^k}_k \right]
    \partial_i (\cB - \cE') \,, \\
    \delta {\cT^i}_j &=& \delta {T^i}_j -
    \left[ {t^i}_j \right]' (\cB - \cE') \,, \nn
\eea
(where ${t^\mu}_\nu$ denotes the background stress-energy), the
results can be expressed purely in terms of the gauge-invariant
quantities, eqs.~\pref{GaugeInvariantDefs}.

The equations which result show that in the absence of vector
stress-energy perturbations, the vector perturbation $V_i$ is not
sourced, and decays very rapidly in an expanding universe,
allowing it to be henceforth ignored. Similarly, in the absence of
off-diagonal stress-energy perturbations it is also generic that
$\Psi = \Phi$.

The equations which govern the evolution of tensor modes then
become (after Fourier transforming)
\be \label{TensorModeEqn}
    \ddot h_{ij} + 3 H \, \dot h_{ij}
    + \frac{k^2}{a^2} \, h_{ij} = 0
    \,,
\ee
while the scalar fluctuations similarly reduce to
\bea \label{ScalarModeEqn}
    &&\delta \ddot \chi + 3H \delta \dot \chi + \frac{k^2}{a^2}
    \delta \chi + V''(\varphi) \delta \chi - 4 \dot\varphi
    \,\dot\Phi + 2 V'(\varphi) \,\Phi = 0 \nn\\
    &&\qquad\qquad \hbox{and} \qquad
    \dot \Phi + H \,\Phi = \frac{\dot\varphi}{2M_p^2} \delta \chi
    \,,
\eea
which shows that it is the time-dependence of the background
configurations which forces $\delta \chi$ and $\Phi$ to mix with
one another. The homogeneous background fields in these
expressions themselves satisfy the equations
\be
    \ddot \varphi + 3H \dot \varphi + V'(\varphi) = 0
    \quad\hbox{and}\quad
    3 M_p^2 H^2 = \frac12 \dot\varphi^2 + V(\varphi) \,.
\ee

\begin{figure}
\begin{center}\epsfig{file=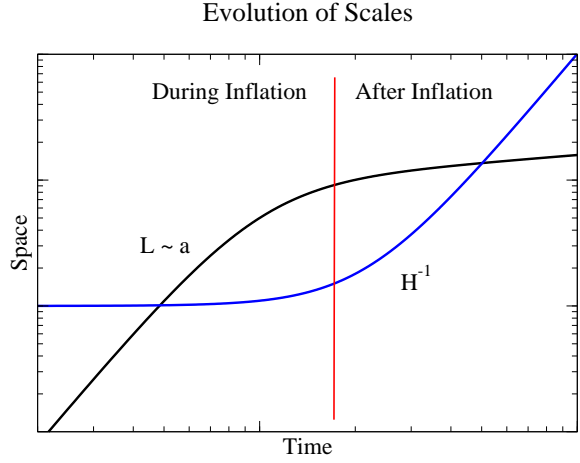,width=3in}
\caption{A sketch of the relative growth of physical scales,
$L(t)$, (in black) and the Hubble length, $H^{-1}$, (in blue)
during and after inflation.}\label{Fig:3}
\end{center}
\end{figure}

\begin{quote}{\bf Exercise 5:}
Derive eq.~\pref{TensorModeEqn}. (Hint: use conformal time,
$\hat\eta = \int \exd t/a$.) \label{Ex:5}
\end{quote}

\subsubsection*{Scalar Perturbations}

The character of the solutions of these equations depends strongly
on the size of $k/a$ relative to $H$, since this dictates the
extent to which the frictional terms can compete with the spatial
derivatives. For instance, an approximate form for the two
independent solutions for $\delta \chi$ that applies when $k/a \gg
H$ is given by damped oscillations
\be \label{DampedOscillationEqn}
    \delta \chi_k \propto \frac{1}{a \sqrt k}
    \exp \left[ \pm i k \int^t
    \frac{\exd t'}{a(t')} \right] \,.
\ee
A similar expression in the limit $k/a \ll H$ is also obtainable
during inflation by using the slow-roll approximation, for which
we neglect $\delta\ddot\chi$, $\ddot \varphi$ and $\dot \Phi$. In
this case the approximate non-decaying solution to
\be \label{SlowRollDE}
    3H \delta\dot\chi + V''(\varphi) \delta \chi +
    2V'(\varphi)\Phi \simeq 0 \quad \hbox{and} \quad
    2M_p^2 H \,\Phi \simeq \dot\varphi \delta \chi \,,
\ee
is given (after Fourier transformation) by
\be \label{SlowRollPerturbationEqn}
    \delta \chi_k \simeq C_k \, \frac{V'(\varphi)}{V(\varphi)}
    \quad\hbox{and} \quad
    \Phi_k \simeq -\frac{C_k}{2} \, \left( \frac{
    V'(\varphi)}{V(\varphi)} \right)^2 \,.
\ee
where $C_k$ is a (potentially $k$-dependent) constant of
integration.

\begin{quote}{\bf Exercise 6:}
Verify that eqs.~\pref{SlowRollPerturbationEqn} satisfy
eqs.~\pref{SlowRollDE}. \label{Ex:6}
\end{quote}

The transition between these two qualitatively different kinds of
behaviour occurs when $k/a \simeq H$. When the product $aH$ is
shrinking (such as during radiation and matter domination) the
condition $k/a = H$ is satisfied for successively smaller values
of $k$ (longer wavelengths) as time goes on. Conversely, when $aH$
grows (as during inflation) it is the larger values of $k$
(shorter wavelengths) which satisfy $k/a = H$ at later times. A
typical mode with wavelength $\lambda = 2\pi a/k$ smaller than the
Hubble length, $H^{-1}$, during inflation is therefore stretched
until it eventually becomes larger than $H^{-1}$, at the epoch of
horizon exit. It continues to grow compared with the Hubble scale
until inflation ends, after which it is $H^{-1}$ which grows
faster than $\lambda$ (see Fig.~\ref{Fig:3}).

During inflation, the modes of interest initially start off with
$k/a \gg H_I$, and any initial oscillations are efficiently damped
by the exponential factor $1/a \propto e^{-H_I t}$ in
eq.~\pref{DampedOscillationEqn}, removing all memory of the
initial configuration. However, eventually $k/a$ falls far enough
that the mode `leaves the horizon' to satisfy $k/a \ll H_I$. At
this point the growing solution,
eq.~\pref{SlowRollPerturbationEqn}, starts to dominate. During
inflation the growth of this solution is slow, because $\delta
\chi_k \propto \sqrt\epsilon$ and $\Phi_k \propto \epsilon$, where
the slow-roll parameter, $\epsilon$, of
eq.~\pref{epsilonParameterDef}, is necessarily small. This
evolution need no longer remain small once inflation ends, but at
this point the slow-roll assumption used to derive the solution,
eq.~\pref{SlowRollPerturbationEqn} breaks down.

\subsubsection*{Source of Fluctuations}

The primordial fluctuation amplitude derived in this way depends
on the integration constants $C_k$, which are themselves set by
the initial conditions for the fluctuation at horizon exit, during
inflation. But why should this amplitude be nonzero given that all
previous evolution is strongly damped, as in
eq.~\pref{DampedOscillationEqn}? The result remains nonzero (and
largely independent of the details of earlier evolution) because
quantum fluctuations in $\delta \chi$ continually replenish the
perturbations long after any initial classical configurations have
damped away.

The starting point for the calculation of the amplitude of scalar
perturbations is the observation that the inflaton and metric
fields whose dynamics we are following are quantum fields, not
classical ones. For instance, for spatially-flat spacetimes the
linearized inflaton field, $\delta\chi$, is described by the
operator
\be
    \delta\chi(x) = \int \frac{\exd^3k}{(2\pi)^3} \Bigl[ c_k \,
    u_k(t) \, e^{i \bfk \cdot \bfr/a} + c_k^* \, u_k^*(t)
    \, e^{-i \bfk \cdot \bfr/a} \Bigr] \,,
\ee
where we expand in a basis of eigenmodes of the scalar field
equation in the background metric, $u_k(t) \, e^{i \bfk \cdot
\bfx}$, labelled by the co-moving momentum $\bfk$. For constant
$H$ the time-dependent mode functions are
\be
    u_k(t) \propto \frac{H}{k^{3/2}} \left( i +
    \frac{k}{aH} \right) \exp \left( \frac{ik}{aH} \right) \,,
\ee
which reduces to the standard flat-space form (up to a
slowly-varying phase), $u_k(t) \propto a^{-1} k^{-1/2} \,
e^{-ikt/a}$, when $k/a \gg H$. The quantities $c_k$ and their
adjoints $c_k^*$ are {\sl annihilation} and {\sl creation
operators}, which define the adiabatic vacuum state, $|\Omega
\rangle$, through the condition $c_k |\Omega \rangle = 0$ (for all
$\bfk$).

The $\delta\chi$ auto-correlation function in this vacuum,
$\langle \delta\chi(x) \delta \chi(x') \rangle$, describes the
quantum fluctuations of the field amplitude in the quantum ground
state. Assuming these quantum fluctuations get decohered sometime
during or after inflation in an as-yet poorly understood way (for
preliminary discussions see ref.~\cite{decoherence}), sometime
between horizon exit and horizon re-entry these quantum
fluctuations eventually become converted into classical
statistical fluctuations of the classical field, $\varphi$, about
its spatial mean, by an amount of order $|\delta \chi_k| \sim
[\langle \delta \chi_k \delta \chi_{-k} \rangle]^{1/2} \propto
|u_k(t)|$. Although the details of this decoherence remain
unclear, for observational purposes all that matters is that the
classical variance of these statistical fluctuations is
well-described by the corresponding quantum auto-correlations -- a
property that is expected to be a good approximation given the
kinds of `squeezed' quantum states which are generated during
inflation \cite{decoherencewodecoherence,Mukhanov}.

Evaluating $\delta\chi_k \sim u_k$ at $t_{\rm he}$ (where $k =
aH$) and equating the result to the fluctuation of
eq.~\pref{SlowRollPerturbationEqn} allows the integration constant
in this equation to be determined
\be
    C_k = u_k(t_{\rm he}) \left( \frac{V}{V'}
    \right)_{\varphi_{\rm he}} \,,
\ee
where both $t_{\rm he}$ and $\varphi_{\rm he} = \varphi(t_{\rm
he})$ implicitly depend on $k$. Using this to compute $\Phi_k$ in
eq.~\pref{SlowRollPerturbationEqn} then gives, near the end of
inflation
\be \label{InflationaryPhiEvolution}
    \Phi_k(t_{\rm end}) = - \frac12 u_k(t_{\rm he})
    \left( \frac{V}{V'}\right)_{\varphi_{\rm he}}
    \left( \frac{V'}{V} \right)^2_{\varphi_{\rm end}}
    = - \epsilon(t_{\rm end})
    \left( \frac{u_k}{\sqrt{2\epsilon} \, M_p}
    \right)_{t_{\rm he}} \,.
\ee
Notice that the factors depending on $t_{\rm end}$ are generically
$O(1)$ if taken at the end of inflation, and do not affect the
$k$-dependence of the result.

\subsubsection*{Post-Inflationary Evolution}

For the case of single-field inflation discussed here, the
subsequent post-inflationary evolution of the fluctuation $\Phi$
--- which is what governs both $\delta \rho_{\rm m}$ and $\delta
T/T$ --- can be solved quite generally (in single-field slow-roll
models), so long as $k/a \ll H$. This is because it can be shown
that when $k \ll aH$ the quantity
\be \label{ConservedCurvatureEqn}
    \zeta = \Phi + \frac23 \left( \frac{\Phi
    + \dot\Phi/H}{1 + w} \right) =
    \frac{1}{3(1+w)} \left[ (5 + 3w) \, \Phi +
    \frac{2\dot\Phi}{H} \right] \,,
\ee
is conserved, $\dot \zeta \simeq 0$, where $w \equiv p/\rho$ is
{\it not} assumed to be constant. This is a very powerful result
because it can be used to evolve fluctuations using $\zeta(t_i) =
\zeta(t_f)$, assuming only that they involve a single scalar
field, and that the modes in question are well outside the
horizon: $k/a\ll H$. Furthermore, although $\dot\Phi$ in general
becomes nonzero at places where $w$ varies strongly with time,
this time dependence quickly damps due to Hubble friction for
modes outside the Hubble scale. We may therefore neglect the
dependence of $\zeta$ on $\dot\Phi$ provided we restrict $t_i$ and
$t_f$ to epochs during which $w$ is roughly constant. This allows
the expression $\zeta(t_i) = \zeta(t_f)$ to be simplified to
\be \label{UsefulConservationEqn}
    \Phi_f = \frac{1+w_f}{1+w_i}
    \left( \frac{5 + 3 w_i}{5+3w_f} \right) \Phi_i  \,,
\ee
where $w_i = w(t_i)$ and $w_f = w(t_f)$, implying in particular
$\Phi_f = \Phi_i$ whenever $w_i = w_f$.

\begin{quote}{\bf Exercise 7:}
Use the conservation of $\zeta$ to show that (when $k/a \to 0$),
$\Phi_{\rm m} = \frac{9}{10} \Phi_{\rm rad}$ for modes evaluated
well before and well after the transition from radiation to matter
domination. \label{Ex:7}
\end{quote}

\begin{quote}{\bf Exercise 8:}
Show that $1 + w \simeq \dot\varphi^2/V \simeq \frac23\, \epsilon$
during single-field slow-roll inflation, and use this with
eq.~\pref{UsefulConservationEqn} to provide an alternate
derivation of eq.~\pref{InflationaryPhiEvolution}. That is, show
that (when $k/a \to 0$), $\Phi_{f}/\Phi_i = \epsilon_f/\epsilon_i$
for times $t_i$ and $t_f$ both well within the inflationary epoch.
\label{Ex:8}
\end{quote}

To infer the value of $\Phi$ in the later Hot Big Bang era we
choose $t_i$ just after horizon exit (where $w_i \simeq -1 +
\frac23 \, \epsilon_{\rm he}$ -- see Exercise 8). $t_f$ is then
chosen in the radiation dominated universe (where $w_f =
\frac13$), either just before horizon re-entry for the mode of
interest, or just before the transition to matter domination,
whichever comes first. Eqs.~\pref{UsefulConservationEqn} and
\pref{InflationaryPhiEvolution} then imply
\be
    \Phi_f \simeq  \left( \frac{6 \,\Phi}{\epsilon} \right)_{\rm he}
    \simeq - \left( \frac{3\sqrt2 \, u_k}{\sqrt\epsilon \, M_p}
    \right)_{\rm he} \,.
\ee
Using this in the definition of the dimensionless power spectrum
for $\Phi$, $\Delta_\Phi^2 = k^3 P_\Phi/(2\pi^2)$, then leads to
\be
    \Delta^2_\Phi(k) \sim k^3 |\Phi_k(t_f)|^2 \sim
    \frac{|k^{3/2} u_k(t_{\rm he})|^2}{ \epsilon(\varphi_{\rm he})
    M_p^2 }
    \sim \left( \frac{H^2}{\epsilon M_p^2}
    \right)_{\varphi_{\rm he}}
    \sim \left( \frac{V}{\epsilon M_p^4}
    \right)_{\varphi_{\rm he}}\,.
\ee
Once the order-unity factors are included from a more detailed
calculation one finds
\be \label{CurvaturePowerSpectrum}
    \Delta^2_\Phi(k) = \frac{k^3
    P_\Phi(k)}{2 \pi^2} = \left(
    \frac{H^2}{8 \pi^2 M_p^2 \, \epsilon} \right)_{\rm
    he} = \left( \frac{V}{24 \pi^2 M_p^4 \,
    \epsilon} \right)_{\rm
    he} \,,
\ee

We see that because it is $V/\epsilon$ which controls the
amplitude of density fluctuations, measurements of this amplitude
provide information about the energy scale which dominates the
universe during inflation. For the purposes of comparison it is
convenient to define \cite{LiddleLyth} the quantity $\delta_H(k)$
by $\delta^{\,2}_H = (4/25) \, \Delta^2_\Phi(k)$, since the
observed amplitude of large-angle temperature fluctuations
requires
\be
    \delta_H(\hat{k}) = 1.91 \times 10^{-5} \,,
\ee
when evaluated at $k = \hat{k} \sim 7.5 a_0 H_0$. In terms of $V$
this implies
\be
    \left( \frac{V}{\epsilon} \right)^{1/4} = 6.6 \times 10^{16}
    \; \hbox{GeV} \,.
\ee
The smaller $\epsilon$ becomes, the smaller a potential energy is
required, and for $\epsilon \sim 0.01$ we have $V \sim 2 \times
10^{15}$ GeV. This is remarkably close to the scale where the
couplings of the three known interactions appear to unify, and may
indicate a connection between inflation and more exotic physics
like the physics of Grand Unification.\footnote{Of course, $V$ can
be much smaller if $\epsilon$ is smaller as well, or if primordial
fluctuations come from another source. For instance generating
primordial fluctuations from TeV scale inflation \cite{EWInfl}
would require $\epsilon \simeq 10^{-55}$.}

\subsubsection*{Spectra}

We now compute in more detail what
eq.~\pref{CurvaturePowerSpectrum} implies for the $k$-dependence
of the primordial fluctuation spectrum. Notice to this end that to
first approximation the size of $\Delta^2(k)$ is set by $H$ and
$\epsilon$ and does not depend explicitly on $k$ at all. This
observation underlies the approximate scale-invariance of the
primordial power spectrum which inflation predicts for the later
universe.

However, inflation does predict a weak $k$-dependence for the
right-hand-side because it must be evaluated for $\varphi =
\varphi_{\rm he}$, defined as the value taken by $\varphi(t)$ at
the epoch $t_{\rm he} = t_{\rm he}(k)$ when the co-moving
wavelength $k$ of interest is just exiting the Hubble length $k =
a(t_{\rm he})H(t_{\rm he})$. It is this $k$-dependence of the
horizon-exit time which introduces small deviations from scale
invariance into the predicted power spectrum.

To quantify this more precisely, recall that in earlier sections a
successful phenomenological parametrization of the density power
spectrum was given by $P_\rho(k) \propto k^{n_s}$, and that this
choice implies the primordial gravitational power spectrum
satisfies $\Delta^2_\Phi = A k^{n_s - 1}$. Deviations from scale
invariance may be computed by evaluating
\be
    n_s - 1 \equiv \left. \frac{\exd \ln \Delta^2_\Phi}{\exd
    \ln k} \right|_{\rm he}\,,
\ee
and using the condition $k = a H$ (and the constancy of $H$ during
inflation) to write $\exd \ln k = H \exd t$. Since the right-hand
side of eq.~\pref{CurvaturePowerSpectrum} depends on $\varphi$, it
is convenient to use the slow-roll equations,
eq.~\pref{SRScalarFieldEqn} to further change variables from $t$
to $\varphi$: $\exd t = - (3H/V') \, \exd \varphi$, and so
\be
    \frac{\exd}{\exd \ln k} = - M_p^2 \left( \frac{V'}{V} \right)
    \, \frac{\exd }{\exd \varphi} \,.
\ee
These expressions allow the derivation of the following relation
between $n_s$ and the slow-roll parameters, $\epsilon$ and $\eta$:
\be \label{nsFormulaEqn}
    n_s - 1 = - 6 \epsilon + 2 \eta \,,
\ee
where the right-hand side is evaluated at $\varphi = \varphi_{\rm
he}$.

Notice that this prediction for the spectral index makes $n_s < 1$
for both the large- and small-field inflation models considered
above. Recall that for large-field models (with $V = \frac1n \,
\lambda^2 \varphi^n$) we had $\epsilon = \frac12 \, n^2
(M_p/\varphi)^2 > 0$ and $\eta = 2(1-1/n)\epsilon$ and so
$-6\epsilon + 2\eta = -(2+4/n)\epsilon < 0$. On the other hand,
for small-field models (where $V = M_I^4 - \mu^2 \varphi^2 +
\cdots$, we had $\eta \le 0$ because we work near $\varphi = 0$,
which is a maximum of $V$. Since in this case $\eta \le 0$ and
$\epsilon \ge 0$, they necessarily both make negative
contributions to $-6\epsilon + 2\eta$.

Observational inferences of $n_s$ from the detailed shape of the
CMBR temperature fluctuation spectrum now give a central value of
$n_s = 0.951 \pm 0.016$, with $n_s = 1$ beginning to be
disfavoured \cite{WMAP3} (assuming no tensor fluctuations -- see
Fig.~\ref{Fig:5} below).

\subsubsection*{Tensor Fluctuations}

A very similar story goes through for the tensor fluctuations that
are generated by quantum fluctuations, although in this case these
fluctuations have not yet been observed. Just like for scalar
fluctuations, for each propagating mode these are generated with
amplitude $H/(2\pi)$, but unlike for scalar modes it is not
necessary for the inflaton to mix with a gravitational mode to
obtain an observable effect, and so the power spectrum for tensor
perturbations does not share the singular factor of $1/\epsilon$.

Similar arguments to those given above then lead to the following
dimensionless tensor power spectrum
\be \label{DeltaSqTSlowRoll}
    \Delta^2_T(k) = \frac{8}{M_p^2} \left( \frac{H}{2 \pi} \right)^2
    = \frac{2V}{3\pi^2 M_p^4}
    \,.
\ee
As expected, this differs from the scalar power spectrum by
depending only on the value of $V$ and not also on the slow-roll
parameter $\epsilon$. Consequently, should both scalar and tensor
modes be measured, a comparison of their amplitudes provides a
direct measure of the slow-roll parameter $\epsilon$. A more
precise version of this comparison can be phrased in terms of a
parameter $r$, which is defined as a ratio of the scalar and
tensor power spectra
\be
    r \equiv \frac{\Delta_T^2}{\Delta_\Phi^2}
    = 16 \, \epsilon \,.
\ee
The failure to detect these perturbations to date places a
relatively weak upper limit: $r < 0.30$ (95\% CL) \cite{WMAP3}, or
$\epsilon < 0.02$.

Once tensor modes are detected, more information can be found from
its power spectrum as a function of $k$. In particular, the tensor
spectral index, $n_T$, is defined by
\be \label{nTinSlowRoll}
    n_T \equiv \frac{\exd \Delta^2_T}{\exd \ln k} = -2 \epsilon
    = -\frac{r}{8}\,,
\ee
where the last equality evaluates the derivative by changing
variables from $k$ to $\varphi$. Again the result is to be
evaluated at the epoch when observable modes leave the horizon
during inflation, $\varphi = \varphi_{\rm he}$.

\subsubsection*{Implications for the CMBR}

\FIGURE[ht]{\epsfig{file=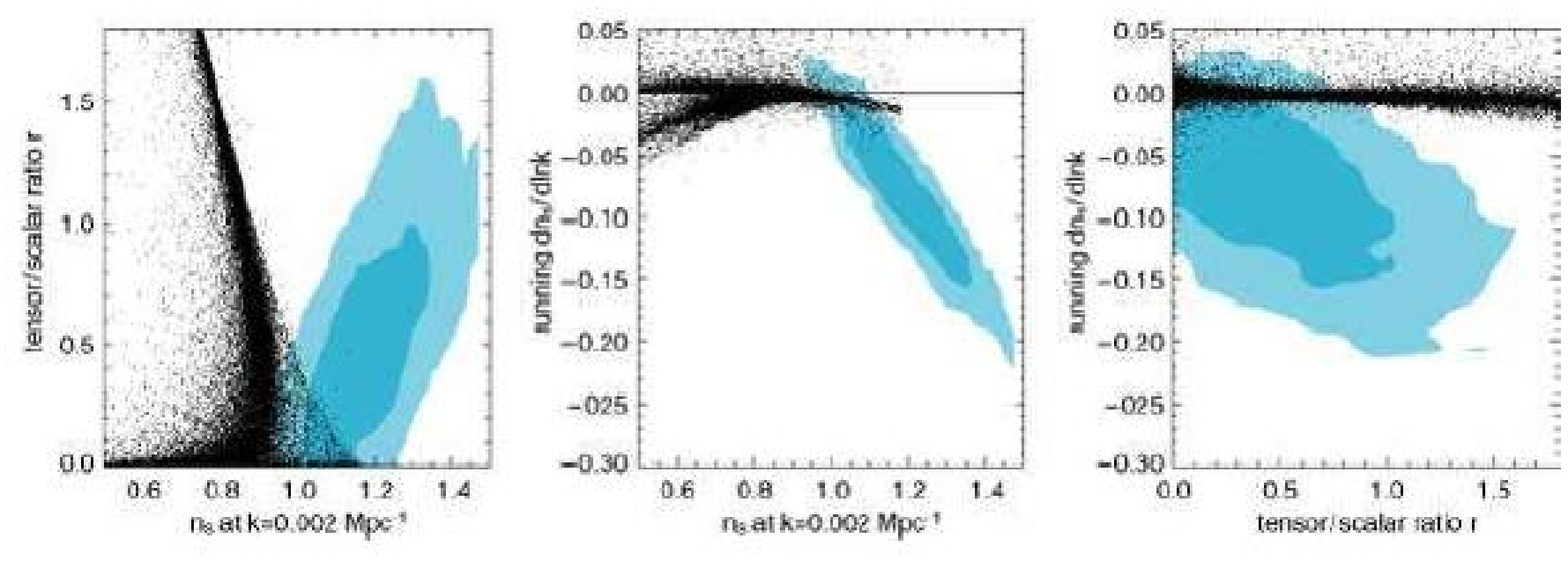,width=10cm}
  \caption{Best fits to the ratio of tensor to scalar fluctuations
  and spectral index from WMAP, compared with the predictions of
  various inflationary models \cite{WMAP3}.}\label{Fig:5}}

In summary, quantum fluctuations generated during slow-roll
inflation provide a natural source for the small temperature
variations visible in the CMBR, which also appear to have seeded
galaxy formation. Furthermore, single-field slow-roll inflation
makes the following detailed, yet successful, predictions for the
form of the primordial fluctuation spectrum which are inferred
from the large-angle properties of temperature fluctuations for
the CMB photons.

\medskip\noindent{\bf Gaussian Fluctuations:} Because inflation
requires such a slow roll, the fluctuations in the inflaton field
are very weakly coupled to one another. This turns out to imply
that the late-time density fluctuations are predicted obey
Gaussian statistics. To date no non-Gaussian correlations have
been seen in the CMBR (more about this below).

\medskip\noindent{\bf In-Phase Perturbations:} The process whereby
fluctuations freeze while they are outside of the Hubble scale,
and then begin to evolve again once liberated by re-entering the
Hubble scale during our much-later epoch, implies these
fluctuations all enter the horizon in phase. Being in phase allows
for the coherent peaks and valleys of the $C_l$'s which are seen
in Fig.~\ref{Fig:4}, and is {\sl not} predicted by many of the
alternative theories of the primordial density fluctuations (such
as their production by cosmic strings or other defects).

\medskip\noindent{\bf Adiabatic Perturbations:} The process of
re-entry of fluctuations, after their having been frozen over long
periods while outside the Hubble scale implies the fluctuations
enter the horizon at rest. This is crucial for determining the $l$
value for the position of the first peak in the CMB spectrum, and
is verified by the observations that this peak occurs at $l
\approx 200$. This prediction need no longer hold if more than one
scalar field is involved in inflation.

\medskip\noindent{\bf Almost Scale Invariant Spectrum:} Inflation
predicts a spectrum of fluctuations which is close to, but not
exactly, scale invariant. For instance if $N_e \sim 60$ implies
$\epsilon \sim |\eta| \sim 1/60$, then the deviation $n_s - 1$
should be a few percent. Current measurements prefer such a
deviation, with an accuracy that is on the verge of excluding an
interesting part of the parameter space of inflationary models.

\medskip\noindent{\bf Scalar to Tensor Ratio:} The same parameters
which determine the scalar fluctuation spectrum also predict the
tensor fluctuation properties. A good test of the theory is
provided once tensor modes are observed, because the tensor and
scalar fluctuations are characterized by 4 observable quantities
(amplitude and spectral index for both scalar and tensor modes),
and the theory predicts these in terms of three parameters: $H$,
$\epsilon$ and $\eta$. The present status of these observational
tests is given in Fig.~\ref{Fig:4}
\cite{WMAPInflation,WMAP3,InflationvsData}.

\subsection{Problems With Inflation?}

The general idea of there being an epoch of accelerated expansion
as a solution for the horizon and flatness problems is very simple
and attractive, and the additional feature that it also accounts
for the primordial spectrum of temperature fluctuations is quite
compelling. Nevertheless some conceptual problems remain with
inflation, and are mostly associated with our ignorance about the
physics which governs the enormously high energies which inflation
could probe. Many of these potential problems can be phrased in
terms of naturalness issues that arise once specific models having
an inflationary dynamics are made (such as the single-field
slow-roll models examined in earlier sections).

Some of the main concerns of this sort are now listed, with an eye
to seeing how the next section's contact with string theory might
help.

\medskip\noindent{\bf Initial Conditions:}
As was seen in the models studied above, inflation tends to arise
only for particular kinds of initial conditions for the fields.
For instance, small-field inflation requires the initial value of
$\varphi$ to be very close to a maximum of the potential, and it
is not clear why the universe should start off in this region. (By
contrast, large-field inflation occurs over a broader range of
initial conditions, but relies on having reliably-calculable
potentials for large field values, $\varphi \gg M_p$.) Relying on
special initial conditions is uncomfortable because inflation was
invented to provide a physical explanation for the origin of the
unnatural initial conditions which are required for the success of
the Hot Big Bang. If we are happy to choose special initial
conditions to obtain inflation, why not instead simply choose the
special conditions required by the Big Bang?

\medskip\noindent{\bf Special Potentials:}
The success of the inflationary models studied above relies on the
potential energy being quite flat, since $V'/V$ and $V''/V$ must
be suppressed to make the slow-roll parameters $\epsilon$ and
$\eta$ sufficiently small. But it is difficult to make such
choices for a scalar potential stable against quantum corrections,
since they are very sensitive to the microscopic particle content
of the theory which underlies the inflationary model. It remains
to be seen if this remains a problem once the best theories we
have for the relevant microscopic physics (like string theory) are
used to try to produce inflation.

\medskip\noindent{\bf Reheating:} Since inflation ruthlessly
`inflates away' any previously existing particles and energy, it
can only precede the Hot Big Bang epoch if it comes with a
mechanism for transferring energy into the heating of the contents
of the observable post-inflationary universe. How is this energy
transfer accomplished? \cite{Reheating}

\medskip\noindent{\bf Predicting in the Multiverse:}
In general causality forbids a completly homogeneous field
evolution, with fields in causally-disconnected regions of
spacetime evolving independent of one another. This means that we
should only imagine the above inflationary picture for $\varphi$
describing one of these regions, with other regions being
described by slightly different initial conditions (and possibly
also different scalar potentials, if the couplings of the inflaton
is related to the values taken of other fields). But since each
region evolves dramatically differently depending on whether it
inflates or not, how does one make predictions in such a diverse
universe? One might expect that inflation exponentially rewards
those parts of the universe that choose the initial conditions
leading to inflation, even if these conditions are comparatively
improbable, because of the exponential growth of the volume of the
region which does so. Does this contain the seeds of a
probabilistic understanding of the properties of the later
universe\cite{lindeinitialcond,SLV}?

\medskip
It remains to be seen how serious each of these problems really
is, but there is considerable motivation to understand them in
some detail given the simplicity of the inflationary understanding
of the large-scale features of the observed CMBR temperature
fluctuations.

\section{Towards String Inflation}

The last section closed with a list of potential problems for
inflation, whose resolution requires an understanding of the
physics at the potentially enormous energies --- possibly as large
as $M_I \lsim 10^{15}$ GeV --- at which inflation can take place.
What guidance can particle physics provide as to what this physics
might be?

Since the energies involved could be not much lower than the
Planck scale, $M_p = (8\pi G)^{-1/2} \sim 10^{18}$ GeV, it is not
unreasonable to look to theories including quantum gravity when
searching for this guidance. At present, the theory which provides
the best-developed and best-motivated framework of quantum gravity
is string theory, making this a natural laboratory for seeking
inflationary dynamics. This section describes some recent progress
along these lines, with several possible inflationary mechanisms
being identified. Since the target audience is not string
theorists, the description will be in broad brush-strokes rather
than fine detail, with an eye towards the broader inflationary
lessons that are being learned.

\subsubsection*{What One Might Hope to Learn}

Before launching into a lengthy technical preamble to building
inflationary scenarios within string theory, it is worth first
stating why one might be interested in doing so in the first
place. (See ref.~\cite{StringInflationReviews} for reviews of
string-based inflation.) After all, present observations can just
barely differentiate amongst the simplest single-field slow-roll
models, so one might reasonably ask why bother building the
inevitably more baroque string models. The thinking is that string
theory potentially can provide new insight into several issues in
inflationary cosmology.

\medskip\noindent{\bf Robustness of Inferences:}
Much of the observational evidence for inflation rests on it being
the source of the primordial fluctuations, but its success in
doing so is largely based on the predictions of very simple
single-field models. But is the single-field approximation too
simple given the many fields which typically arise in fundamental
theories? Even if not, if microscopic physics is being stretched
by inflation up to cosmological distances, can the physics of much
smaller scales be similarly stretched \cite{Transplanckia}, and so
influence inflationary predictions in unexpected ways? If so, then
the observational evidence for inflation would be undermined by
this introduction of an uncontrollable theoretical error into its
predictions \cite{TPIUS}. Such questions can be tested in string
theory, with current evidence supporting the robustness of the
predictions of simple inflationary models \cite{Shenker}.

\medskip\noindent{\bf Validity of Approximations:}
Single field models often rely for their validity on
approximations whose validity cannot be properly established
without better understanding the high-energy limit of the theory.
For instance, for large-field inflationary models successful
inflation relies on fields taking large values, $\varphi \gg M_p$,
and this is also typically required to obtain observably large
primordial tensor fluctuations \cite{lythtensor}. But whether such
large fields make sense depends on properly understanding the
shape of the scalar potential for such large field values. String
theory can shed light on this by providing a physical
interpretation for the inflaton (such as being the distance
between two branes \cite{DvaliTye}), and so can identify upper
limits in its range (such as it not being larger than the size of
the extra dimensions in which the branes move \cite{BBI1}).
Detailed arguments like these have led to the conjecture that
observable primordial tensor fluctuations may be unlikely to be
obtained from string theoretic inflation \cite{SmallTensor}.

\medskip\noindent{\bf Initial Conditions and Naturalness:}
How unusual is inflation? Inflationary models can require
comparatively flat potentials and special initial conditions, but
an understanding of how special these are requires a broader
understanding of the shape of the scalar potential, and of the
likely initial conditions before inflation, which only a
fundamental theory like string theory can ultimately provide.

\medskip\noindent{\bf Reheating:}
As noted above, the energy density which drives inflation must
ultimately get transformed to heat for the later Hot Big Bang.
Just as having a warm house in the winter requires both a good
furnace and good insulation, successful reheating after inflation
requires two things: ($i$) a sufficiently strong coupling between
the inflaton and the ordinary Standard Model particles we now see
around ourselves; and ($ii$) the absence of too strong couplings
between the inflaton and any other, currently unobserved, degrees
of freedom. It is clear that the second part of this question
cannot be properly addressed without knowing the full theory
describing {\it all} the degrees of freedom which are relevant at
the energies available after inflation.

\medskip\noindent{\bf Mind Broadening:}
Simple inflationary models make simplifying assumptions which need
not be true, but which tend to guide our search for models.
Embedding inflation into string theory has already exposed some of
these assumptions, and may yet expose more. For instance, it is
often assumed that the inflaton field remains around after
inflation ends and still appears in the low-energy theory
describing the later Hot Big Bang epoch. However if the inflaton
were the separation between a brane and antibrane which mutually
annihilate at inflation's end \cite{BBI1,BBI2}, then the inflaton
does not even make sense as a field in the later universe.
Similarly, although inflation now seems compelling to us in the
context of field theory, perhaps string theory provides novel
alternative ways \cite{Alternatives} to solve the initial
condition problems which inflation was originally invented to
solve.

\subsection{General Framework}

String theory is much more complicated than the simple inflaton
models discussed above, involving a potentially infinite number of
particle types (string modes), moving in more than 4 dimensions.
The space of vacua which is allowed is only partially understood,
but that part which is already well explored shows that it is
incredibly vast and diverse -- involving many possible vacuum
values for many possible low-energy fields. (See
ref.~\cite{StringBooks} for textbook descriptions of string
theory, and \cite{StringReviews} for useful reviews.)

Part of this complexity can be traced to there being a large
number of scales in string theory, and for inflationary purposes
there are at least three which are very important: the string
scale, $M_s$; the compactification -- or Kaluza-Klein (KK) --
scale(s), $M_c$; and the inflationary scale, $M_I$ (and so also
$H_I \sim M_I^2/M_p$). For strings moving in 10D Minkowski space,
$M_s$ characterizes the mass splitting among generic string modes.
$M_c$ describes the mass splitting within each string mode when it
is placed in a non-trivial background, such as when all but 4 of
the dimensions are compactified. For simple geometries
characterized by a single length scale, $\ell$, ({\it e.g.} a
curvature radius, or a volume, $\cV_n = \ell^n$), the
compactification scale is of order $M_c \sim 1/\ell$. The 4D
Planck mass is not an independent scale because it is calculable
in terms of the others.

Much of what is known in string theory is restricted to the case
$M_c \ll M_s$, since in this case the effective theory describing
energies $E \ll M_s$ is given by a higher-dimensional (usually 10
or 11 dimensional) supergravity. If all but 4 of the dimensions
are compactified at similar scales, then the physics of energies
$E \ll M_c$ is described by some sort of 4D effective theory. The
4D Planck scale is typically of order $M_p \sim g_s^{-1} M_s
(M_s/M_c)^3 \gg M_s$, where $g_s \ll 1$ is the string coupling
(which in string theory is related to the value of one of the
background scalar fields). The field content and symmetries (like
supersymmetry) of this low-energy 4D theory depend on the details
of the kind of higher-dimensional supergravity, and of its
compactification, that is under consideration. In what follows it
is always assumed that $M_c \ll M_s$.

The complexity of an inflationary model in string theory depends
crucially on how large is the inflationary Hubble scale, $H_I \sim
M_I^2/M_p$, compared with both $M_s$ and $M_c$.
\begin{itemize}
\item If $M_s \lsim H_I$ then inflation is an intrinsically
stringy phenomenon. It is stringy because the time-dependence of
the background geometry is sufficient to produce particles having
masses up to $O(H_I)$, and this includes nontrivial string modes
by assumption. In this case inflation can only be convincingly
demonstrated by working with all of the complexity of string
theory.
\item If $M_c \ll H_I \ll M_s$, then inflation can be described
within the effective higher-dimensional field theory, without
requiring all the stringy bells and whistles. However in this
regime all of the extra-dimensional physics is important, and one
is seeking solutions to the full higher-dimensional supergravity
equations.
\item If $H_I \ll M_c \ll M_s$, then inflation can be
intrinsically 4-dimensional, since the energies available to be
pair-produced by the time-dependent geometry are generically not
high enough to excite any of the KK modes associated with the
existence of the extra dimensions.
\end{itemize}

Most of the inflationary models proposed to date\footnote{Here an
inflationary model means one having both an accelerated expansion
{\it and} a mechanism for it to end, and so excludes in particular
higher-dimensional configurations having only accelerated 4D
expansion \cite{AccUniv}.} are formulated within the last of these
categories, with $H_I \ll M_c \ll M_s$, since in this case the
problem reduces to searching for time-dependent inflating
solutions to the effective 4D field equations. Because these
models are being constructed in an explicitly 4D limit, we should
not be surprised to find them to share many features of 4D
inflationary models, and this is indeed what is found. Of more
interest is finding those ways in which inflation differs when the
field theory in which it is found arises as a low energy 4D
effective theory in string theory, and a few of the known examples
of this will be discussed.

\subsection{Multiple Scalars}

Although inflation asks only for one scalar field to be the
inflaton, it is a generic feature of string vacua that their
low-energy limit contains more than one scalar field. This opens
up the possibility that more than one of these fields plays an
inflationary role, and so suggests re-examining slow-roll
inflation in multi-field models.

\subsubsection*{Hybrid Inflation}

A useful starting point for multi-scalar inflationary models is
Hybrid Inflation \cite{HybridInfl}. In its simplest form this
corresponds to the following action for two scalar fields,
$\varphi$ and $\chi$,
\be \label{HybridInflationActionEqn}
    S = - \int \exd^4x \sqrt{-g} \left[ \frac{M_p^2}{2} R
    + \frac12 \partial^\mu \varphi \, \partial_\mu \varphi
    + \partial^\mu \chi \,\partial_\mu \chi + V(\varphi,\chi)
    \right]\,,
\ee
with scalar potential
\be
    V(\varphi,\chi) = \frac{m^2}{2} \varphi^2 +
    \frac{\lambda^2}{4} \varphi^4 + \frac{g^2}{4}
    (\chi^2 - v^2)^2 + \frac{h^2}{4} \varphi^2 \chi^2 \,.
\ee
Here $\lambda$, $g$ and $h$ are dimensionless, real coupling
constants and an additive constant has been chosen to ensure that
$V=0$ when evaluated at the potential's global minimum, which is
situated at $\varphi =0$ and $\chi = v$.

\begin{figure}
\begin{center}\epsfig{file=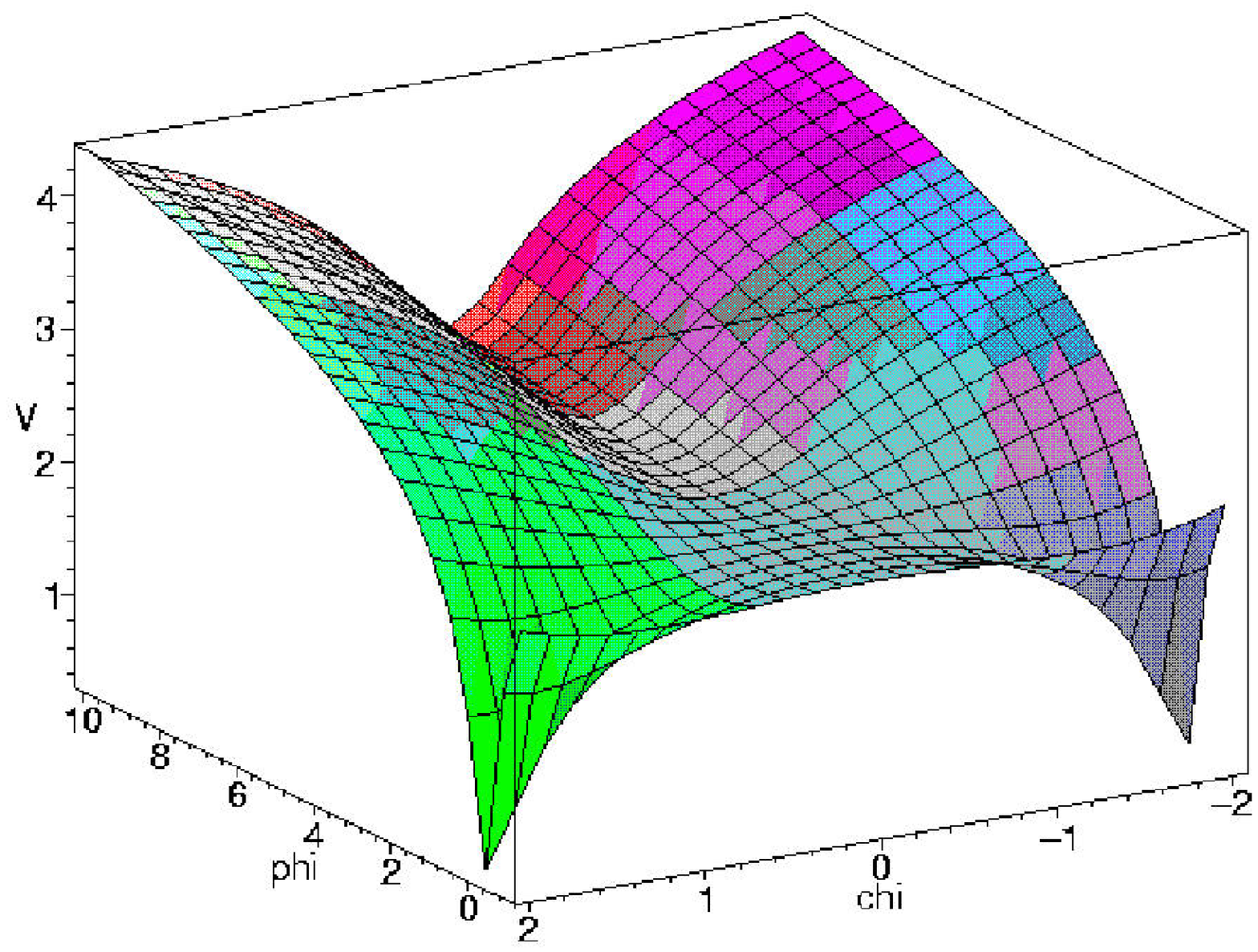,width=3in}
\caption{A sketch of the scalar potential for Hybrid
Inflation.}\label{Fig:6}
\end{center}
\end{figure}

For inflationary purposes our interest is in the case where the
dimensionful constants satisfy $0 \le m \ll gv$, and where
$\varphi$ starts out very large. The derivatives of the potential
are
\be
    V_{,\varphi} = \varphi \left[ m^2 + \lambda^2 \varphi^2 +
    \frac{h^2}{2} \chi^2 \right] \quad\hbox{and}\quad
    V_{,\chi} = \chi \left[ g^2(\chi^2 - v^2) + \frac{h^2}{2}
    \varphi^2 \right]\,,
\ee
and so both vanish at the global minimum ($\varphi = 0$ and $\chi
= v$) as well as at a saddle point at $\varphi = \chi = 0$.
$V_{,\chi}$ vanishes along the entire line $\chi = 0$, along which
the curvature of the potential is given by
\be
    \begin{pmatrix} V_{,\varphi\varphi} & V_{,\varphi\chi} \cr
    V_{,\chi\varphi} & V_{,\chi\chi} \end{pmatrix} =
    \begin{pmatrix} m^2 + 3 \lambda^2 \varphi^2 & 0 \cr
    0 & \frac12 h^2 \varphi^2 - g^2 v^2 \end{pmatrix} \,,
\ee
showing that this line is a trough (local minimum in the $\chi$
direction) if $\varphi > \varphi_\star = \sqrt2 \, gv/h$ ($\sim v$
if $g \sim h$), which gets steeper and steeper the larger
$\varphi$ is. Otherwise, for $\varphi < \varphi_\star$, the line
$\chi = 0$ is a ridge (local maximum in the $\chi$ direction),
which is steepest at the saddle point at $\varphi = 0$. (See
Fig.~\ref{Fig:6} for a sketch of this potential.)

If $\varphi$ starts off initially much bigger than
$\varphi_\star$, with $\chi = 0$, then the potential keeps $\chi$
at zero but allows $\varphi$ to roll towards smaller values.
Furthermore, if $\frac14 g^2 v^4 \gg \frac12 m^2 \varphi^2 +
\frac14 \lambda^2 \varphi^4$ (as is generically true for $\varphi
\sim v$ if $\lambda \ll g$ and $m \ll g v$) then $V(\varphi,\chi =
0) \approx \frac14 g^2 v^4$ is approximately constant during this
roll. Inflation can occur provided the kinetic energy is much
smaller than this constant, which the discussion of earlier
sections shows occurs if the slow roll parameters describing the
motion in the $\varphi$ direction,
\be
    \epsilon = \left[ \frac{M_p \varphi (m^2 + \lambda^2
    \varphi^2)}{g^2 v^4} \right]^2
    \quad\hbox{and} \quad
    \eta = \frac{4 M_p^2 (m^2 + 3\lambda^2 \varphi^2)}{g^2v^4} \,,
\ee
are both small. This provides an inflationary epoch, which lasts
either until the slow roll parameters become too large, or until
$\varphi$ falls below $\varphi_\star$, and so $\chi$ becomes
destabilized away from zero, provoking a fast roll towards the
absolute minimum at $\chi = v$. The condition $\varphi >
\varphi_\star$ would be the first to fail if $\epsilon$ and $\eta$
are small for $\varphi \sim v$, which is true if $m/gv$ and
$\lambda/g$ are both much smaller than $v/M_p$.

This provides an intrinsically two-field inflationary model, where
the second field can play a crucial role in bringing inflation to
an end. The additional parameters available also allow a wide
range for the slow roll parameters at horizon exit, and so allow
examples both with $n_s > 1$ (unlike the previous single-field
models) as well as with $n_s < 1$. For an example with $n_s > 1$,
consider the case where $\lambda \approx 0$, and $h \simeq g$, so
that $\varphi_\star \simeq v$. Taking also $v/M_p = O(\delta)$ and
$m/gv \sim O(\delta^2)$ for some $\delta \ll 1$, the number of
$e$-foldings after horizon exit becomes
\be
    N_e = \frac{1}{M_p} \int_{\varphi_\star}^{\varphi_{\rm he}}
    \frac{\exd \varphi}{\sqrt{2\epsilon}}
    \simeq \frac{g^2 v^4}{\sqrt2 \, m^2 M_p^2} \ln \left(
    \frac{\varphi_{\rm he}}{\varphi_\star} \right) \,,
\ee
which is $O(\delta^{-2})$ even when $\varphi_{\rm he}$ is also of
order $v$. But for $\varphi_{\rm he} \sim v$ we have $\eta =
O(\delta^2) \gg \epsilon = O(\delta^6) > 0$, which implies $n_s >
1$ when used in eq.~\pref{nsFormulaEqn}.

\subsubsection*{General Multi-scalar Models}

Although hybrid inflation shows that multi-field inflationary
models can have interesting properties in their own right, the
form of the action, eq.~\pref{HybridInflationActionEqn}, is not
general enough to capture the generic kinds of scalar dynamics
which emerge in the low-energy limit of string theory.

The most general action describing the low-energy evolution of $N$
real scalar fields, $\phi^a$, is
\be \label{MultiScalarActionEqn}
    S = - \int \exd^4x \sqrt{-g} \left[ \frac{M_p^2}{2} \, R +
    \frac12 \, G_{ab}(\phi) \, \partial^\mu \phi^a \partial_\mu
    \phi^b + V(\phi) \right] \,,
\ee
where $V$ is the scalar potential, and $G_{ab} = G_{ba}$ is a
positive definite symmetric matrix of functions. Notice that there
is no loss in not having a function of $\phi^a$ in front of the
Ricci curvature scalar, such as $\cL \propto \sqrt{-g} \, A(\phi)
R$, because any such term can be removed by performing an
appropriate $\phi$-dependent Weyl re-scaling of the metric:
$g_{\mu\nu} \to A^{-1}(\phi) \, g_{\mu\nu}$. This choice of metric
which makes the Einstein-Hilbert action $\phi^a$-independent is
called the {\sl Einstein Frame}.

One way the action for Hybrid inflation,
eq.~\pref{HybridInflationActionEqn}, differs from
eq.~\pref{MultiScalarActionEqn} is by having $G_{ab} =
\delta_{ab}$, and one might ask whether this can always be
arranged by performing an appropriate redefinition among the
scalar fields. Although this can be done quite generally when only
one scalar field is present, for more than one field it can be
done (as well as ensuring $\partial_a G_{bc} = 0$) only when
evaluated at a specific point, $\phi^a = \phi^a_0$, but not
simultaneously for all $\phi^a$. To see why this is true, notice
that $G_{ab}$ transforms as a rank two tensor under field
redefinitions, $\phi^a \to f^a(\phi)$ (see Exercise 9). Since
$G_{ab}$ is also positive definite, it therefore has a geometrical
interpretation of being a metric on the `target' space, $M$, in
which the $\phi^a$ take their values. As a result, we know that a
change of coordinates can only ensure $G_{ab} = \delta_{ab}$
everywhere if its Riemann tensor, ${R^a}_{bcd}$, vanishes
everywhere. On the other hand, the freedom to arrange
$G_{ab}(\phi_0) = \delta_{ab}$ at any specific point $\phi^a_0$
corresponds to choosing Gaussian normal coordinates at this point.

\begin{quote}{\bf Exercise 9:}
Show that under a field redefinition, $\delta \phi^a =
\xi^a(\phi)$, the action of eq.~\pref{MultiScalarActionEqn}
returns to the same form with $V \to V + \xi^a \partial_a V$ and
$G_{ab} \to G_{ab} + \xi^c \partial_c G_{ab} + G_{ac}
\partial_b \xi^c + G_{cb} \partial_a \xi^c$. This shows
that $V$ transforms as a scalar field, and $G_{ab}$ transforms
like a rank-two tensor.\label{Ex:9}
\end{quote}

The scalar field equations for the action
\pref{MultiScalarActionEqn} are
\be \label{MultiscalarFieldEqn}
    \ddot\phi^a  + \Gamma^a_{bc}(\phi) \dot \phi^b \dot \phi^c
    + 3H \dot\phi^a + G^{ab} V_{,a} = 0
    \,,
\ee
where $V_{,a} = \partial_a V = \partial V/\partial \phi^a$,
$G^{ab}$ is the inverse metric for $G_{ab}$ and $\Gamma^a_{bc} =
\frac12 G^{ad}[\partial_b G_{cd} + \partial_c G_{bd} - \partial_d
G_{bc} ]$ is the Christoffel symbol built from the target-space
metric, $G_{ab}$. These are to be supplemented by the standard
Friedmann (eq.~\pref{FriedmannEqn}) and Raychaudhuri
(eq.~\pref{RaychaudhuriEqn}) equations (or energy conservation,
eq.~\pref{EnergyConservationEqn}), where $p$ and $\rho$ are given
by
\be
    \rho = \frac12 \, G_{ab}(\phi) \dot \phi^a \dot \phi^b +
    V(\phi) \quad \hbox{and} \quad
    p = \frac12 \, G_{ab}(\phi) \dot\phi^a \dot\phi^b - V(\phi)
    \,.
\ee

As before a sufficient condition for inflation is to have $V \gg
\frac12 G_{ab} \dot\phi^a \dot\phi^b$ and approximately constant,
and this is ensured if we may drop both the $\ddot\phi^a$ and
$\Gamma^a_{bc} \dot \phi^b \dot \phi^c$ terms of
eqs.~\pref{MultiscalarFieldEqn}, leading to the slow-roll
equations, $3H \dot\phi^a = - G^{ab} V_{,b}$. These slow-roll
conditions remain good approximations for an appreciable time
provided the multi-scalar generalizations of the slow-roll
parameters are small over a broad enough region. As the Hybrid
inflation example shows, it is important when defining these to be
sure that they measure the derivatives of the potential only along
the steepest direction down the potential, since this is also the
direction of motion if the field starts out close to rest.

Since the gradient, $V_{,a}(\phi)$, of the scalar potential
automatically points in the direction of steepest ascent for the
potential, its negative naturally provides the direction down
which an initially-static configuration starts to roll from any
point, $\phi^a$, in the target space. Consequently, the
generalization of $\epsilon$ which measures the first derivative
of the potential in this direction can be taken to be,
\be \label{MultiEpsilonDef}
    \epsilon = \frac{M_p^2 G^{ab} V_{,a} V_{,b}}{2V^2} \,.
\ee
Notice that because this transforms as a scalar under field
redefinitions, it may be evaluated using any choice of fields and
(unlike the formulae given earlier for single-field inflation) its
use does {\it not} assume the choice $G_{ab}(\phi_0) =
\delta_{ab}$. Furthermore, it agrees with standard multi-field
definitions \cite{LiddleLyth} for $\epsilon$, since it reduces to
these in normal coordinates (for which $G_{ab}(\phi_0) =
\delta_{ab}$).

A multi-scalar generalization of $\eta$ is given by the smallest
of the eigenvalues of the matrix of second derivatives of the
potential, $V_{,ab}(\phi_0)$, since this defines the most unstable
direction (at least in a slow-roll region where $V_{,a}$ is
negligible). (Notice that if this eigenvalue is negative then we
are looking for the negative eigenvalue having the largest
absolute value.) In order to ensure a slow enough evolution for
$\phi^a$ near $\phi^a = \phi^a_0$ it is important to evaluate this
second derivative matrix only {\it after} transforming to
(Gaussian normal) coordinates to ensure that $G_{ab}(\phi_0) =
\delta_{ab}$. Alternatively, this definition can be written in a
way which is equally good when evaluated using an arbitrary choice
of coordinates on the target space, as follows. First define the
eigenvalues, $\lambda$, of the matrix ${N^a}_b(\phi)$, defined by
\be
    {N^a}_b e^b = \lambda e^a, \quad \hbox{with} \quad
    {N^a}_b = \frac{M_p^2 G^{ac} V_{;cb}}{V} \,,
\ee
where $V_{;cb} = V_{,cb} - \Gamma^a_{bc} V_{,a}$ is the covariant
derivative of $V_{,b}$ using the target-space connection
$\Gamma^a_{bc}(\phi)$. Then in an arbitrary coordinate frame $\eta
= \hbox{min} \, \lambda$, minimized over all of the possible
eigenvalues of ${N^a}_b$. This is the appropriate generalization
because as defined $\lambda$ is a scalar under scalar-field
redefinitions, and because it agrees with standard definitions
\cite{LiddleLyth} when evaluated in the canonical Gaussian normal
frame.

\subsubsection*{The special case of K\"ahler metrics}

An important special case of the above discussion is the case
which arises when the scalar fields can be grouped in to complex
fields, $\{\phi^a \} = \{ \phi^i, \overline\phi^{\,
\overline\imath} \}$, where $\overline\phi^{\,\overline\imath}$
denotes the complex conjugate of $\phi^i$. In this case, if the
nonzero components of the metric, $G_{ab}$, locally can be written
$G_{i\overline\jmath} = \partial_i \partial_{\overline\jmath} K$,
for some function $K = K(\phi, \overline\phi)$, then the metric is
called a {\sl K\"ahler} metric, with $K$ being its K\"ahler
potential.

In this case the definition for $\epsilon$ becomes
\cite{realistic}
\be \label{MultiEpsilonDef2}
    \epsilon = \frac{M_p^2 G^{i\overline\jmath} V_{,i}
    V_{,\overline\jmath}}{V^2} \,,
\ee
and $\eta$ is defined in terms of the smallest eigenvalue of the
matrix
\be
    \begin{pmatrix} {N^i}_j & {N^i}_{\overline\jmath} \cr
    {N^{\overline\imath}}_j &
    {N^{\overline\imath}}_{\overline\jmath} \end{pmatrix} \,,
\ee
where
\be \label{KahlerNDefs}
    {N^i}_j = \frac{M_p^2 G^{i\overline k} V_{,\overline{k} j}}{V}
    \quad \hbox{and} \quad
    {N^{\overline\imath}}_j = \frac{M_p^2 G^{\overline\imath k}}{V}
    \Bigl[ V_{,kj} - G^{l\overline{m}} K_{,jk\overline{m}} V_{,l}
    \Bigr] \,,
\ee
while ${N^{\overline\imath}}_{\overline\jmath}$ and
${N^i}_{\overline\jmath}$ are the complex conjugates of these.

\begin{quote}{\bf Exercise 10:}
Derive eqs.~\pref{KahlerNDefs}, by first showing that the only
nonzero Christoffel symbols for a K\"ahler metric are
$\Gamma^i_{jk} = G^{i\overline{m}} K_{,jk\overline{m}}$, and its
complex conjugate, $\Gamma^{\overline\imath}_{\overline\jmath
\overline{k}}$. \label{Ex:10}
\end{quote}

\subsubsection*{Primordial Fluctuations}

The presence of many scalars also changes the kinds of primordial
fluctuations which are possible, because with several scalars
there can be perturbations, $\delta\phi^a$, for which the total
energy density remains unchanged, $\delta\rho = 0$. Any such a
fluctuation is called an `isocurvature' fluctuation, in contrast
to the `adiabatic' fluctuations involving nonzero $\delta \rho$
considered previously.

There are strong observational constraints against the existence
of such isocurvature fluctuations re-entering the Hubble scale
during the Hot Big Bang era. Constraints exist because
isocurvature perturbations at this scale correspond to metric
perturbations which emerge into the sub-Hubble world with a zero
initial amplitude, $\Phi_i = 0$, but nonzero velocity, $\dot\Phi_i
\ne 0$ (in contrast with adiabatic modes, which emerge with
nonzero initial amplitude, $\Phi_i \ne 0$, and initially vanishing
speed, $\dot\Phi_i = 0$). This phase difference is measurable in
the CMBR because it changes the value of $l$ for which the maximum
peak occurs in Fig.~\ref{Fig:4}. Current observations are
consistent with purely adiabatic oscillations at horizon re-entry.

Multi-field inflationary models must therefore either not generate
primordial isocurvature perturbations at all at horizon exit, or
any such primordial perturbations must disappear sometime after
horizon exit but before horizon re-entry. The absence of such
fluctuations must be checked in any specific model
\cite{isocheck}.

Primordial isocurvature modes need not be a problem for an
inflationary model even should they be generated at horizon exit,
however, provided they are subsequently erased before horizon
re-entry. This possibility exists because in the multi-field case
no simple conservation law like eq.~\pref{ConservedCurvatureEqn}
ensures the model-independent survival of perturbed quantities. In
particular, all isocurvature modes are erased if a period of
thermal equilibrium occurs between Hubble exit and re-entry,
because in this case all perturbations are encoded into
temperature fluctuations, whose presence necessarily also perturbs
the energy (and so also the gravitational potential).

\subsection{Moduli and their Stabilization}

We now return to the main development: the description of explicit
inflationary models that are grounded in stringy vacua. By
restricting attention to the case $M_c \ll M_s$, the discussion
can be framed within higher-dimensional field theory.

\subsubsection*{10D Supergravity}

The string solutions about which most is known are those which
preserve some of the supersymmetries of the theory, and the
higher-dimensional field theories which describe their properties
below $M_s$ are supergravities, of which there are several in 10
dimensions. It is the bosonic fields of these supergravities that
are relevant to their classical dynamics, and these always include
the metric, $g_{MN}$, together with its bosonic partners under
supersymmetry: a scalar dilaton, $\phi$, and a rank-2
antisymmetric gauge potential, $B_{MN}$. Other bosonic fields can
also arise, depending on which supergravity is of interest. These
can include gauge potentials, $A^a_M$, for 10D gauge
supermultiplets (where the index `$a$' runs over the generators of
the relevant gauge group), as well as various kinds of $n$th-rank
skew-tensor gauge potentials, $C_{M_1...M_n}$.

In addition to these `bulk' fields, the low-energy supergravity
can also include the positions, $x^M(\sigma^\alpha)$, within 10D
spacetime of each of any D-branes that are allowed for the
supergravity.\footnote{In principle, Type IIA supergravity allows
D0, D2, D4, D6 and D8 branes, while Type IIB supergravity allows
D1, D3, D5, D7 and D9 branes. No D-branes arise at all in
heterotic vacua. 5+1 dimensional surfaces called NS5-branes can
also exist for each of these supergravities.} Here $\sigma^\alpha$
are coordinates on the D-brane world sheet, with $\alpha =
0,1,...,p+1$ running over one time and $p$ space directions for a
D$p$-brane.

The action governing the dynamics of these fields comes as the sum
of brane and bulk terms, $S_{10} = S_{\rm br} + S_B$, where the
bulk action has the generic form
\bea
    S_{B} &=& - \int \exd^{10}x \, \sqrt{-g}\, M_s^8\left[
    \frac{1}{2} R + \frac{1}{2} \partial^M \phi \,
    \partial_M \phi + \frac{1}{6} e^{-\phi} H^{MNP} H_{MNP}
    \right.  \\ &&\qquad \qquad \qquad \qquad \left.
    + \sum_{n} \frac{e^{c_{n} \phi}}{2(n+1)!}
    F_{M_1...M_{n+1}} F^{M_1...M_{n+1}} +
    \cdots \right]\,, \nn
\eea
and $F = \exd C$ is the exterior derivative which corresponds to
the field strength appropriate to each of the skew-tensor gauge
fields. (These sometimes also contain Chern-Simons terms, which in
the above action are rolled into the ellipses.) The number of
fields summed over, and the values of the numerical constants
$c_{n}$, depend on the precise supergravity of interest. For
instance, for the Type IIB supergravity of later interest the bulk
action has one rank-0 potential, $C$, ({\it i.e.} a scalar), no
rank-1 gauge potentials, $C_M$, one additional rank-2 potential,
$C_{MN}$, no rank-3 potentials, $C_{MNP}$ and one rank-4 potential
$C_{MNPQ}$, while the constants are $c_0 = 2$, $c_{2} = 1$ and
$c_{4} = 0$.

The brane action has a similar form,
\be \label{BraneActionEqn}
    S_{\rm br} = \sum_b T_b \int_{\Sigma_b}
    \exd^{p_b + 1} \sigma \, \sqrt{-\gamma} \,
    e^{\lambda_b\phi} \Bigl(1 + \cdots \Bigr)
    + \mu_b \int_{\Sigma_b} \Bigl( \Omega_b + \cdots \Bigr) \,,
\ee
where the sum is over the branes present, and the integral is over
the $(p+1)$-dimensional world-volume of each D$p$-brane. Here
$\lambda_b$ is a known constant, equal to $(p_b-3)/4$ for 10D
supergravity, and the form $\Omega_b$ appearing in the second
integral is either the particular potential, $C_{M_1...M_{p+1}}$,
whose rank is $p + 1$, or the Hodge dual (obtained by contracting
one of the $C$'s with the 10D Levi-Civita tensor,
$\epsilon_{M_1...M_{10}}$) of a form of rank $9-p$. One such a
form exists for each kind of brane allowed by each of the possible
supergravities. The dimensionful constants $T_b$ and $\mu_b$ in
these expressions are proportional to $M_s^{p+1}$ (with known
numerical coefficients). $T_p$ has the physical interpretation of
the brane tension, or energy per unit world-volume. Finally, the
world sheet `metric' appearing in eq.~\pref{BraneActionEqn} is
given by
\be
    \gamma_{\alpha\beta}(\sigma) = \partial_\alpha x^M
    \partial_\beta x^N \left[g_{MN} + B_{MN} + \frac{1}{M_s^2}
    F_{MN}\right] \,,
\ee
where $F_{MN} = \partial_M A_N - \partial_N A_M$ is the $U(1)$
gauge field associated with those open strings both of whose ends
terminate on the brane in question. (A more complicated expression
holds when $N$ branes sit at the same point in spacetime, since
this promotes the gauge group to $U(N)$.)

\subsubsection*{Moduli}

Of particular interest are those string vacua for which only the 4
dimensions of everyday experience are noncompact, and the other 6
dimensions are compactified with a size corresponding to an energy
scale $M_c$. For $M_c \ll M_s$ these correspond to semiclassical
solutions to the corresponding 10D supergravity equations. A
considerable amount is known about these solutions in the case
that the compactification preserves at least one supersymmetry in
4D.

In the absence of branes the supersymmetric solutions have a
metric of the product form \cite{Candelas}
\be \label{ProductMetricEqn}
    \exd s^2 = \eta_{\mu\nu} \, \exd x^\mu \, \exd x^\nu
    + g_{mn}(y) \, \exd y^m \, \exd y^n \,,
\ee
where $x^\mu$ are coordinates for the noncompact 4 dimensions,
$y^m$ label the compact 6 dimensions and $\eta_{\mu\nu}$ is the
usual 4D Minkowski metric. Among other things, $N=1$ supersymmetry
in 4D requires the extra-dimensional metric, $g_{mn}$, to be {\sl
Calabi-Yau} ({\it i.e.} Ricci-flat geometries having $SU(3)$
holonomy). There is generically a many-parameter family of such
metrics which all share the same (fairly complicated) topology,
$g_{mn}(y) = g_{mn}(y;\omega)$, where $\omega_a$ represent the
parameters required to fully describe the geometry.

The parameters required to describe a geometry are known as {\sl
moduli}, and generically arise when solving the Einstein
equations. A simple example of a geometry having moduli is given
by the 2-dimensional torus, which is defined by the condition that
its Riemann curvature vanishes: ${R^m}_{npq} = 0$ in a 2D space
with boundary conditions $y^1 \simeq y^1 + 1$ and $y^2 \simeq y^2
+ 1$. The general 2D metric which solves this equation is
\be
    \exd s^2 = a \Bigl[ (\exd y^1)^2 + 2b \; \exd y^1 \, \exd y^2
    + c \, (\exd y^2)^2 \Bigr] \,,
\ee
where $a$, $b$ and $c$ are arbitrary constants, and so are the
three moduli of a 2-torus. One of these, $a$, describes overall
re-scalings of the size of the metric (the so-called {\sl
breathing mode}), and is generically a modulus because of a scale
invariance of the supergravity equations in higher dimensions. The
other two moduli describe changes to the geometry at fixed volume
(specifically changes to what is called its complex structure). A
Calabi-Yau geometry can have hundreds of similar moduli, which can
be divided into two categories: those describing modifications to
its complex structure, and the rest -- including the breathing
mode -- that are known as K\"ahler moduli.

Moduli are of particular interest when studying compactifications
because the classical field equations guarantee the existence of a
massless 4D scalar field for each modulus of the extra-dimensional
metric. To see how this works, first recall how to compactify a
fluctuation in a 10D scalar field, $\delta \phi(x,y)$, whose 10D
field equation is $\Box_{10} \delta \phi = g^{MN} D_M D_N \delta
\phi = 0$. Evaluated for a product metric like
eq.~\pref{ProductMetricEqn}, this becomes $(\Box_4 + \Box_6)
\delta \phi = 0$, where $\Box_6 = g^{mn} D_m D_n$ and $\Box_4 =
\eta^{\mu\nu} \partial_\mu \partial_\nu$. If we decompose $\delta
\phi(x,y)$ in terms of eigenfunctions, $u_k(y)$, of $\Box_6$
--- {\it i.e.} where $\Box_6 u_k = -\mu^2_k \,u_k$ --- we have
\be \label{KKScalarExpansion}
    \delta \phi(x,y) = \sum_k \varphi_k(x) \, u_k(y) \,,
\ee
and the equations of motion for $\phi$ imply $(\Box_4 - \mu^2_k)
\varphi_k = 0$. The 10D field decomposes as an infinite number of
4D Kaluza-Klein fields, each of whose 4D mass is given by the
corresponding eigenvalue, $\mu_k$. In particular a massless mode
in 4D corresponds to a zero eigenvalue: $\Box_6 u_k = 0$.

A similar analysis also applies for the fluctuations, $\delta
g_{MN}(x,y)$, in the 10D metric about a specific background
geometry such as eq.~\pref{ProductMetricEqn}. Focussing on metric
components in the extra dimensions, $\delta g_{mn}(x,y)$, allows
an expansion similar to eq.~\pref{KKScalarExpansion}
\be
    \delta g_{mn}(x,y) = \sum_k \varphi_k(x) \, h_{mn}^k(y) \,,
\ee
where $h_{mn}(y)$ are tensor eigenfunctions for a particular 6D
differential operator (the {\sl Lichnerowitz operator}) obtained
by linearizing the Einstein equations, $\Delta_6 h^k_{mn} = -
\mu^2_k \, h^k_{mn}$. Again the 10D equation of motion,
$\Delta_{10} \delta g_{mn} = 0$, implies each 4D mode,
$\varphi_k(x)$, satisfies $(\Box_4 - \mu^2_k) \varphi_k = 0$, and
so has mass $\mu_k$.

The significance of moduli is that they provide zero
eigenfunctions for $\Delta_6$, and so identify massless 4D scalar
fields within the KK reduction of the extra-dimensional metric.
The zero eigenfunction is given by the variation of the background
metric in the direction of the moduli. Schematically, if
$\omega_a$ are the moduli of the background metric,
$g_{mn}(y;\omega)$, and if $h^a_{mn} = \partial g_{mn}/\partial
\omega_a$, then $\Delta_6 h^a_{mn} = 0$. Physically, these are
zero eigenfunctions because varying a modulus in a given solution
to the Einstein equations gives (by definition) a new solution to
the same equations, and so in particular an infinitesimal
variation in this direction is a zero mode of the linearized
equations.

Because the 4D moduli fields, $\varphi_a(x)$, are massless they
necessarily appear in the low-energy 4D effective action which
governs the dynamics at scales below the KK scale, $M_c$. If we
focus purely on the moduli and the 4D metric (and ignore other
fields), then the low-energy part of this action must take the
general form of eq.~\pref{MultiScalarActionEqn}, but with a
potential, $V$, which is independent of the moduli,
$\varphi^a(x)$.

\subsubsection*{Moduli and inflation}

Moduli (and any other classically massless scalars) are a mixed
blessing for inflationary models. The Good News is they provide a
large number of candidate scalar fields in the 4D effective
theory, any of which might play the role of the inflaton.
Furthermore, a slow roll could be possible because their potential
is often very shallow, being required to be flat to the accuracy
with which it is known that configurations like $g_{mn}(y;\omega)$
are solutions for all $\omega_a$. Typically the field in question
is only approximately a modulus, although some can be exactly
massless if one of the supersymmetries is unbroken. Even in
supersymmetric cases it often happens that moduli remain massless
to all orders in perturbation theory, but appear in the 4D scalar
potential once non-perturbative effects are considered.

Indeed a number of these scalar fields have been proposed as
possible inflatons \cite{earlystringinflation}, however before the
discovery of branes within string theory all of the proposed
inflationary scenarios had difficulties. One difficulty for the
moduli of supersymmetric vacua was the need to compute
non-perturbative contributions, which made the calculation of the
inflaton potential difficult. Branes provide a way forward on two
fronts: they allow supersymmetry-breaking effects to be more
simply computed, such as with the use of brane-antibrane dynamics;
and they play a central role in the geometries arising in the
modulus-stabilization programme. The ability to compute explicitly
led to an explosion of inflaton proposals, including metric moduli
\cite{ModuliInfl}, massless modes arising from extra-dimensional
gauge fields \cite{GaugeInfl}, inter-brane separations
\cite{DvaliTye,BBI1,BBI2,angles,MoreBraneInfl}, more stringy
modes, \cite{TachyonInfl} and so on.

On the other hand, the Bad News is that it is usually impossible
to know for sure whether a given light scalar can be the inflaton
until the full potential is understood which governs the dynamics
of {\it all} of the low-energy moduli. This is because a slow roll
requires the potential to be shallow in its {\it steepest}
downward direction. If one finds an inflaton potential that is
shallow enough to obtain inflation before understanding the
corrections which stabilize some of the moduli, one must worry
that these corrections ruin the inflationary solution by providing
steeper directions along which the inflaton could roll without
inflating. Unfortunately, progress on understanding modulus
stabilization was a long time coming in string theory, and the
lack of this understanding proved to be a long-standing obstacle
to identifying how inflation might arise within a stringy context.

\subsubsection*{Modulus Stabilization: Branes and Fluxes}

Major progress on string inflation became possible with the
development of tools for understanding how to stabilize most of
the moduli for a few kinds of stringy vacua. This progress started
with the identification of how to generalize \cite{GKP} the 4D
supersymmetric compactifications of the field equations of Type
IIB supergravity in 10 dimensions to include the presence of
parallel D3, D7 branes (plus 7+1-dimensional surfaces having
negative tension, called orientifold planes).

These branes complicate the dynamics of the internal dimensions in
several important ways. First, they do so through the
gravitational fields they create, which have the effect of
modifying the metric of eq.~\pref{ProductMetricEqn} into the
following form,
\be \label{GKPMetricEqn}
    \exd s^2 = h^{-1/2}(y) \eta_{\mu\nu} \, \exd x^\mu \, \exd
    x^\nu + h^{1/2}(y) g_{mn}(y) \, \exd y^m \, \exd y^n \,,
\ee
with the {\sl warp factor}, $h(y)$, depending on the positions of
the various branes. The metric $g_{mn}(y)$ appearing here is a
Ricci-flat Calabi-Yau type metric, of the form which arose in the
absence of the branes.

A second important difference to the dynamics of the internal
dimensions which arises once branes are present is the presence of
nontrivial configurations of the various antisymmetric tensor
fields, for which they act as sources. The total flux of these
fields through topologically nontrivial surfaces in the extra
dimensions is quantized, such as
\be
    M_s^2 \int_S F \propto n_1 \quad \hbox{and} \quad
    M_s^2 \int_S H \propto n_2 \,,
\ee
where $S$ is a 3-cycle, $F = \exd C$ and $H = \exd B$ are 3-form
fluxes, and $n_1$ and $n_2$ are integers that depend on which
3-surface $S$ is considered. The presence of such fluxes has two
important consequences: ($i$) they can (but need not) break the
remaining 4D supersymmetry, and ($ii$) they can remove some of the
moduli of the extra-dimensional geometry, such as changes to the
area of these surfaces $S$. These are no longer moduli when fluxes
are present because flux quantization implies the value of fields
like $C_{MN}$ must grow as the areas of these surfaces shrink,
ensuring such changes come with an energy cost.

A third potential contribution of branes to extra-dimensional
dynamics is the tension of the branes themselves. In particular,
since D7 branes fill 7 spatial dimensions, and only 3 of these are
the noncompact ones we see, they must also extend into 4 of the
compact 6 dimensions. Typically they do so by `wrapping'
themselves around a non-contractable surface, or 4-cycle, in these
extra dimensions. But D7 branes have a fixed tension, $T_7 \propto
M_s^8$, and so such wrappings provide an energy cost for
increasing the moduli describing the volume of the cycles about
which branes wrap. Precisely what this energy cost is depends on
the relative number of different kinds of branes (positive tension
$D7$ branes, or negative tension orientifold $O7$ and $O3$ planes)
wrapping any given cycle, a number which is itself subject to the
topological constraint (`tadpole condition') that the net $D3$ and
$D7$ charges must vanish (much in the same way that Gauss' Law
requires the net electric charge in any compact volume to vanish).

In the end one expects such geometries having both branes and
fluxes to have fewer moduli than do those without branes and
fluxes, and this is indeed what is found. In particular, for the
supersymmetric Type IIB compactifications described here, the
fluxes and branes turn out to remove all of the complex structure
moduli that are associated with the Calabi-Yau metric, $g_{mn}$,
appearing in eq.~\pref{GKPMetricEqn}. But not all of the moduli of
$g_{mn}$ are lifted in this way, with the K\"ahler moduli
(including the breathing mode) remaining at the classical level,
even in the presence of branes and fluxes.

\begin{figure}
\begin{center}\epsfig{file=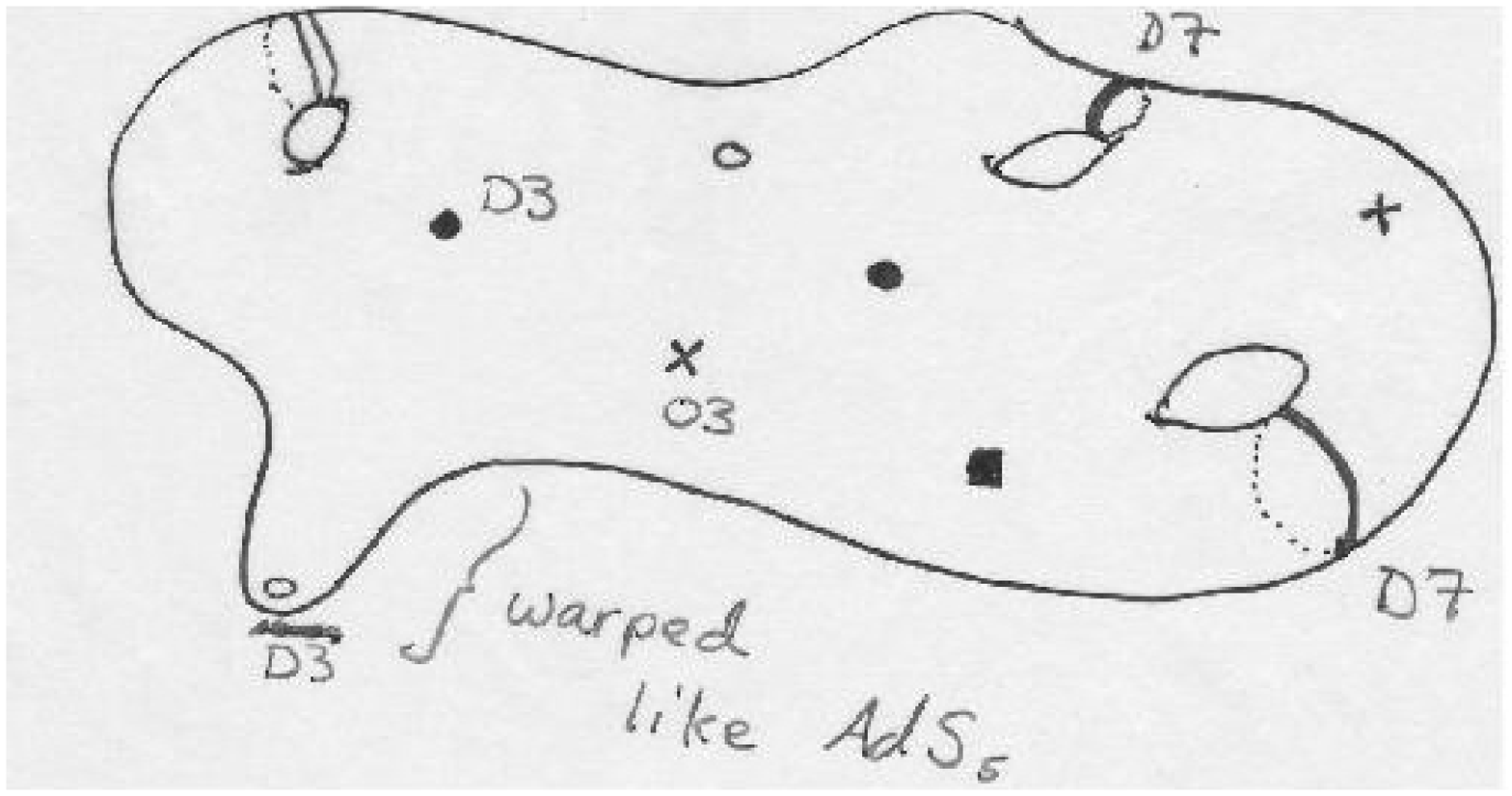,width=4in}
\caption{A cartoon of a Type IIB extra-dimensional
configuration.}\label{Fig:4.5}
\end{center}
\end{figure}

\subsubsection*{Warped Throats}

The extra dimensions which result in this way can have a
complicated and rich geometry, including the possibility of warped
throats along which the warp factor, $h(y)$, varies strongly. The
6D geometry in such a throat is well approximated by the following
polar-coordinate-like form
\be \label{ThroatMetricEqn}
    \exd s^2 = h^{-1/2} \eta_{\mu\nu} \, \exd x^\mu \, \exd x^\nu
    + h^{1/2} \Bigl[ \exd \rho^2 + \rho^2 \exd s^2_5 \Bigr]
    \quad \hbox{with} \quad
    h \simeq a^4 + b^4/\rho^4 \,,
\ee
where $\rho$ denotes proper distance along the throat (measured
with the metric $g_{mn}$) and $\exd s^2_5$ is a known metric
describing the 5 other `angular' directions. These approximations
work well away from the throat's `base' ({\it i.e.} $\rho \gg
b/a$, where $h$ becomes more slowly varying and joins into the
bulk of the internal dimensions). They also apply not too close to
its `tip' ($\rho \to 0$, where the conical singularity generically
present in the metric, $g_{mn}$, becomes smoothed out).

Notice that for $\rho \ll b/a$ we have $h \propto \rho^{-4}$ and
so the metric, eq.~\pref{ThroatMetricEqn}, takes the approximate
form
\bea \label{ThroatMetricApproxEqn}
    \exd s^2 &\simeq& \frac{\rho^2}{b^2} \, \eta_{\mu\nu}
    \,\exd x^\mu \,\exd x^\nu
    + \frac{b^2 \, \exd \rho^2}{\rho^2} + \exd s^2_5 \nn\\
    &=& e^{2\xi/b} \, \eta_{\mu\nu} \,\exd x^\mu \,\exd x^\nu
    + \exd \xi^2 + \exd s^2_5  \,,
\eea
where we change variables using $\rho = \rho_0 \, e^{\xi/b}$, and
absorb a factor of $\rho_0$ into the 4D coordinates, $x^\mu$. The
restriction of this metric to the 5 dimensions spanned by the
coordinates $\{x^\mu,\xi\}$ is the 5D de Sitter metric, and so
eq.~\pref{ThroatMetricApproxEqn} shows that the 4D warp factor
varies exponentially quickly with proper distance, $\xi$, along
the throat. (Once corrections to the geometry near the throat's
tip are included one finds $h_{\rm tip} = h(\rho \to 0)$ does not
diverge.) This is precisely the kind of fast variation of 4D scale
in the extra dimensions which could play a role in the hierarchy
problem, \`a la Randall and Sundrum \cite{RS}.

\subsubsection*{The 4D Point of View}

The Type IIB compactifications to 4 dimensions of ref.~\cite{GKP}
generically all share two properties: ($i$) they are either $N=1$
supersymmetric in 4D, or break this supersymmetry by a small
amount compared to $M_c$; and ($ii$) they preserve at least one
(but usually many) massless moduli at the classical level.
Consequently they can have an interesting dynamics at energies
well below $M_c$, which it should be possible to capture with an
$N=1$ supersymmetric 4D effective field theory.

The field content of any such a 4D supergravity generically
consists of: ($i$) chiral matter multiplets, whose bosonic
components are complex scalar fields, $\varphi^i$; ($ii$) gauge
multiplets, whose bosonic components are gauge potentials,
$A_\mu^a$; and ($iii$) the supergravity multiplet, whose bosonic
component is the massless KK mode of the 4D metric itself,
$g_{\mu\nu}$. (If more than one 4D supersymmetry were to survive
to energies below $M_c$ then a fourth kind of multiplet,
consisting of a gravitino and a gauge boson, would also be
required.) Since the surviving moduli are 4D scalars, we expect
these to fall into 4D chiral multiplets, and so be represented by
complex scalar fields, $\varphi^i$.

Once expressed in the 4D Einstein frame ({\it i.e.} with the
metric chosen so that the 4D gravity lagrangian density is $\cL =
-\frac12 M_p^2 \sqrt{-g} g^{\mu\nu} R_{\mu\nu}$) the interactions
amongst these fields are described by 4D $N=1$ supergravity
\cite{Cremmer}, which (at low energies, where the lowest
derivatives dominate) is completely characterized by three
functions of the chiral scalars: ($i$) the holomorphic
superpotential, $W(\varphi)$; ($ii$) the holomorphic gauge
coupling function, $f_{ab}(\varphi)$; and ($iii$) the K\"ahler
potential, $K(\varphi,\overline\varphi)$. In particular, the
kinetic terms for the gauge potentials, $A^a_\mu$, are given in
terms of $f_{ab}$ by
\be \label{4DGaugeKineticSUSYAction}
    \frac{\cL_{g\,kin}}{\sqrt{-g}} = - \frac14 \,
    \Bigl( \hbox{Re}\,f_{ab} \Bigr) \, F^a_{\mu\nu} F_a^{\mu\nu} \,,
\ee
and so if $f_{ab} = f_a \delta_{ab}$ then the gauge coupling is
given by $1/g_a^2 = \hbox{Re}\, f_a$. The scalar-field kinetic
terms and self-interactions are similarly given by
\be \label{4DScalarSUSYAction}
    \frac{\cL_{s}}{\sqrt{-g}} = - G_{i\overline\jmath}
    (\varphi,\overline\varphi)\,
    \partial^\mu \varphi^i \, \partial_\mu \overline\varphi^{\,
    \overline\jmath} - V(\varphi,\overline\varphi)\,,
\ee
with target space metric for the scalars given by
$G_{i\overline\jmath} = K_{,i\overline\jmath}$, which is a
K\"ahler metric, and we adopt Planck units for which $M_p = (8\pi
G)^{-1/2} = 1$.

The scalar potential is $V = V_F + V_D$, where
\be \label{VDDefinition}
    V_D = \frac12 f^{ab} D_a D_b \quad\hbox{with}\quad
    D_a = K_{,i} \delta_a \varphi^i \,,
\ee
$f^{ab}$ is the matrix inverse of the gauge coupling matrix,
$\hbox{Re}\,f_{ab}$, and $\delta_a \varphi^i$ denotes the
variation of the scalar fields under a gauge transformation (so
$V_D$ arises only when there are low-energy gauge multiplets
present, coupled to the scalars). The remaining term in $V$ is
\be \label{VFDefinition}
    V_F = e^{K} \Bigl[ G^{\overline\imath j}
    \overline{D_i W} \, D_j W - 3 |W|^2 \Bigr] \,,
\ee
where, as usual, $G^{\overline\imath j}$ is the inverse metric to
$G_{i\overline\jmath}$, and the quantity $D_iW$ denotes the
K\"ahler covariant derivative of $W$, defined by
\be
    D_i W = W_{,i} + K_{,i} \, W \,.
\ee
It turns out that $D_i W$ is the order parameter for supersymmetry
breaking, and so must vanish for stationary points of this
potential to preserve supersymmetry.

\begin{quote}{\bf Exercise 11:}
Show that any solution to $D_i W = 0$ (for all $i$) is also a
stationary point for $V_F$. Show also that gauge invariance of the
superpotential, $W_{,i} \, \delta_a \varphi^i = 0$, ensures that
$D_i W = 0$ implies $V_D = 0$. \label{Ex:11}
\end{quote}

The functions $K$ and $W$ can be computed semiclassically for the
remaining moduli in the Type IIB compactifications of
ref.~\cite{GKP} by directly dimensionally reducing the
higher-dimensional action, and this gives
\be \label{DimReductionGeneralKW}
    K = -2 \ln \Bigl( M_s^6 \cV_6 \Bigr) \quad \hbox{and} \quad
    W = W_0 \,,
\ee
where $\cV_6$ denotes the volume of the internal 6 dimensions, as
measured using the metric $g_{mn}$ and expressed as a function of
its complex moduli, $\varphi^i$. $W_0$, on the other hand, is a
$\varphi^i$-independent constant, which can be computed in terms
of the extra-dimensional fluxes which have been turned on
\cite{GVW}. If the fluxes involved do not break supersymmetry,
then $W_0$ vanishes, but $W_0$ is typically nonzero if these
fluxes break the remaining 4D supersymmetry.

\medskip\noindent{\it Example with one modulus:}
For example, one modulus which always survives at the classical
level (due to a classical scale invariance of the
higher-dimensional supergravity equations) is the field
corresponding to the overall breathing mode of the extra
dimensions. Writing the internal metric as $g_{mn}(y) = r^2 \hat
g_{mn}(y)$, with, say, $M_s^6 \int \exd^6y \sqrt{\hat g} = 1$,
then we first seek the complex field, $\varphi$, which contains
the 4D modulus $r(x)$. In principle this can be obtained by
examining the supersymmetry transformation laws, to see which
fields transform in the standard form for a 4D multiplet
\cite{BFQ}, but a shortcut to the result can be found by examining
the dependence on $r$ of the gauge kinetic terms for a gauge field
on one of the D7 branes wrapped about some 4-cycle $\Sigma$. The
result obtained by dimensional reduction is
\bea
    \cL_{g\,kin} &=& -\frac14 \int_\Sigma \exd^4y \sqrt{-g} \, g^{\mu\nu}
    g^{\lambda\rho} F_{\mu\lambda} F_{\nu \rho} + \cdots \nn\\
    &=& - \frac{r^4}{4} \, \eta^{\mu\nu} \eta^{\lambda\rho}
    F_{\mu\lambda} F_{\nu\rho} \int \exd^4y \sqrt{-\hat g} \, h(y)
    + \cdots \,,
\eea
which, when compared with the supersymmetric 4D gauge kinetic
function shows that $\hbox{Re} \, f = k r^4$, with $k \propto
\int_\Sigma \exd^4y \sqrt{\hat g} \, h$. Since 4D supersymmetry
requires $f$ to be a holomorphic function of the complex modulus
$\varphi$, it follows that we can define $\varphi$ such that $f =
\varphi$, with $\hbox{Re} \, \varphi = k r^4$.

Given this relation between $\varphi$ and $r$ we may compute the
K\"ahler potential $K$, using the known $r$-dependence of the 6D
volume: $M_s^6 \cV_6 = M_s^6 r^6 \int \exd^6y \sqrt{\hat g} =
r^6$. This shows that $\cV_6 \propto (\hbox{Re} \,
\varphi)^{3/2}$, and so
\be \label{KSingleModulus}
    K(\varphi,\overline\varphi) = - 2 \ln \Bigl( M_s^6 \cV_6
    \Bigr) = -3 \ln \Bigl( \varphi + \overline\varphi \Bigr) \,,
\ee
up to an irrelevant additive constant. The fact that $K$ depends
only on Re $\varphi$ can also be deduced on symmetry grounds once
the supersymmetry transformations are used to identify which
fields appear in Im $\varphi$. $K$ cannot depend on Im $\varphi$
at the classical level because the theory turns out to be
invariant under constant shifts of Im $\varphi$.

\begin{quote}{\bf Exercise 12:}
Verify that using the K\"ahler potential of
eq.~\pref{KSingleModulus} in eq.~\pref{4DScalarSUSYAction} gives
the correct kinetic terms for $r(x)$, by comparing the result with
what you obtain by directly dimensionally reducing the
higher-dimensional Einstein-Hilbert action, $\cL = - \frac12 M_s^8
\sqrt{-g} \, R$, using the metric, eq.~\pref{GKPMetricEqn}, with
$g_{mn} = r^2(x) \hat{g}_{mn}(y)$. Do not forget to go to the 4D
Einstein frame by also re-scaling the 4D metric, $g_{\mu\nu} \to
r^{-6} g_{\mu\nu}$. \label{Ex:12}
\end{quote}

A check on the whole picture comes when the above results for $W$
and $K$ are used to compute the scalar potential for $\varphi$,
using the general expression, eq.~\pref{VFDefinition}. Consistency
requires the result must vanish, $V = 0$, since $\varphi$ is a
modulus and so cannot have a scalar potential (to the accuracy
used to derive $W$ and $K$). Notice first that $\varphi$ does not
transform under gauge transformations (so long as none of the D7
gauge groups are anomalous), so $V_D = 0$ and $V = V_F$.
Specializing eq.~\pref{VFDefinition} for $V_F$ to a constant
superpotential, $W = W_0$, then gives
\be \label{VFConstantW0}
    V = e^{K} \Bigl[ G^{\overline\imath j}
    K_{,\overline\imath} \, K_{,j} - 3 \Bigr]
    |W_0|^2 \,.
\ee
Finally, using eq.~\pref{KSingleModulus} in this expression gives
$V_F \equiv 0$ for all $\varphi$, because the K\"ahler potential
satisfies the remarkable identity
\be \label{NoScaleIdentity}
    G^{\overline\imath j} K_{,\overline\imath} \, K_{,j} \equiv 3
    \,.
\ee
Models whose K\"ahler potential satisfies this identity are known
as {\sl no-scale} models \cite{Cremmer,NoScale}. They play an
important role in low-energy string theory because they capture
the property that the low-energy 4D potential cannot depend on
moduli fields.

Since $V$ vanishes, any value of $\varphi$ provides an equally
good classical vacuum for the low-energy 4D theory. Notice,
however, that if $W_0 \ne 0$ then supersymmetry is typically
broken for most of these values, since the order parameter for
supersymmetry breaking is $D_\varphi W = K_{,\varphi} W_0$. This
ensures the effective 4D picture agrees with the
higher-dimensional point of view, because $W_0$ is only nonzero if
the higher-dimensional fluxes break 4D supersymmetry.

\medskip \noindent{\it Examples with several moduli:}
A second example of practical later interest is to
compactifications for which more than one modulus survives at the
classical level, corresponding to a collection of complex moduli,
$\varphi^i$. For many of these the K\"ahler potential, $K$, of the
moduli has been explicitly computed, with some having the form
\be \label{KMultiModulus}
    K(\varphi,\overline\varphi) = - 2 \ln \left[ (\tau^1)^{3/2}
    - \sum_{i\ne1} k_i (\tau^i)^{3/2} \right] \,,
\ee
where $\tau^i = \hbox{Re}\, \varphi^i$ and $k_i$ are calculable
constants for a given Calabi-Yau geometry. In these models $V =
V_F$, and the superpotential is constant, $W = W_0$, so we are
again led to eq.~\pref{VFConstantW0} as the scalar potential.
Remarkably, we again obtain $V_F \equiv 0$ in this case, because
the K\"ahler potential, eq.~\pref{KMultiModulus}, also satisfies
the no-scale identity $G^{\overline\imath j} K_{,\overline\imath}
\, K_{,j} \equiv 3$.

\begin{quote}{\bf Exercise 13:}
Explicitly show that the K\"ahler potential given in
eq.~\pref{KMultiModulus} satisfies the no-scale identity,
eq.~\pref{NoScaleIdentity}. \label{Ex:13}
\end{quote}

\subsubsection*{Corrections to the Semi-classical Picture}

A consistent low-energy 4D picture for the dynamics of moduli
exists for Type IIB string vacua, but so far the resulting scalar
dynamics does not inflate because the scalar potentials are
precisely flat. However the functions $K$ and $W$ used to this
point are computed by direct dimensional reduction using the
higher-dimensional classical action, and the potential can become
more complicated once corrections are included which introduce an
energy cost to changing the value of the low-energy fields,
$\varphi^i$.

There are two important kinds of corrections of this sort which
are known to arise: ($i$) string loop corrections, involving
powers of $g_s \sim e^\phi$; and ($ii$) $\alpha'$ corrections, to
do with the higher-dimensional supergravity equations themselves
only being low-energy approximations to the full string theory.
(The notation $\alpha' \propto M_s^{-2}$ is defined for historical
reasons, and controls the second type of corrections because they
are typically suppressed by powers of a low-energy scale (like
$M_c$) compared with $1/M_s^2 = \alpha'$.)

Some of the effects of these corrections on $K$, $W$ and $f_{ab}$
are known. It is known that the holomorphic superpotential, $W$,
does not receive either of these kinds of corrections, to all
orders in perturbation theory, a result called the
non-renormalization theorem \cite{NonRenorm}. It can, however, be
corrected once non-perturbative contributions are included. The
K\"ahler potential, $K$, is not similarly protected, however, with
the contribution of higher-curvature $\alpha'$ corrections in the
extra-dimensional action correcting $K$ to become
\cite{KahlerCorr}
\be \label{KalphaCorrected}
    K = -2 \ln \left( M_s^6 \cV_6 + \frac{\xi}{2} \right) \,,
\ee
where $\xi = - \chi(\cM)/[2(2\pi)^3]$ being a calculable
coefficient depending on the Euler number, $\chi(\cM)$, of the
extra-dimensional geometry, $\cM$. Notice that the new term inside
the logarithm is suppressed relative to the first one by powers of
$1/\cV_6$, as is typical for $\alpha'$ corrections. Notice also
that the corrected K\"ahler potential no longer satisfies the
no-scale identity, eq.~\pref{NoScaleIdentity}.

\subsubsection*{The KKLT Framework}

The first approach to fix all of the moduli within the Type IIB
framework --- by Kachru, Kallosh, Linde and Trivedi, or KKLT
\cite{KKLT} --- starts with the assumption that only one modulus,
$\varphi$, survives the flux compactification, leading to a
constant superpotential, $W=W_0$, and the K\"ahler potential of
eq.~\pref{KSingleModulus}. The remaining modulus is then imagined
to be fixed through a non-perturbative correction to the
superpotential, of the form
\be \label{WNPSingleModulus}
    W(\varphi) = W_0 + A \, \exp\Bigl[-a \varphi \Bigr] \,,
\ee
where $A$ and $a$ are both constants. This functional form for the
non-perturbative correction to $W$ is known to arise in two kinds
of situations: in the presence of some brane-related instantons
\cite{D3Instantons}, or if the low-energy gauge group associated
with some of the D7 branes contains an asymptotically-free
non-abelian gauge group, $G$. (For instance, since the gauge
coupling function is $f_{ab}(\varphi) = \varphi \, \delta_{ab}$
for such a gauge group, if $G = SU(N)$ and there are no matter
multiplets carrying $SU(N)$ quantum numbers, then condensation of
gauginos \cite{GC,WBGIII} in the vacuum leads to a superpotential
of the above form, with $A$ nonzero and $a = 2\pi/N$. In this case
the exponential dependence of $W$ on $\varphi$ reflects a vacuum
energy which depends non-perturbatively on the gauge coupling
constant, $g^{-2} \propto \hbox{Re}\, \varphi$.)

KKLT analyze the potential generated using the non-perturbative
superpotential of eq.~\pref{WNPSingleModulus} together with the
uncorrected K\"ahler potential of eq.~\pref{KSingleModulus}. Is it
consistent to use non-perturbative corrections to $W$ when not
keeping perturbative contributions to $K$? It can be, depending on
the size of $W_0$. To see this imagine that $K = K_0 + K_p$ and $W
= W_0 + W_{np}$, where $K_p$ denotes the perturbative corrections
to $K$ and $W_{np}$ is the (much smaller) non-perturbative
contribution to $W$. The corresponding contributions to $V_F$ then
have the schematic form $V_F = V_0 + V_p + V_{np}$ where $V_0 = 0$
because of the no-scale form of the K\"ahler potential, while
\bea
    V_p &=& O(K_p |W_0|^2) + O(K_p^2 |W_0|^2) + \cdots \nn\\
    V_{np} &=& O(W_0 W_{np}) + O(K_p W_0 W_{np}) + O(|W_{np}|^2)
    + \cdots \,,
\eea
and the ellipses contain further subdominant terms. For generic
values of $W_0$ the perturbative contributions to $V_F$ dominate
the non-perturbative ones, but if $W_0$ should be anomalously
small, {\it e.g.} $W_0 \sim W_{np}$, then the terms involving
$K_p$ become subdominant even when $W_{np}$ cannot be neglected.

Using the leading-order K\"ahler potential,
eq.~\pref{KSingleModulus}, and including the non-perturbative
superpotential, eq.~\pref{WNPSingleModulus}, gives a potential
which depends nontrivially on $\varphi$, with $V \to 0$ as
$|\varphi| \to \infty$, falling to a nontrivial minimum for
nonzero $\varphi = \varphi_m$ \cite{KKLT}. Furthermore, although
the domain of validity of the $\alpha'$ expansion is large
$\hbox{Re}\,\varphi$, this domain can extend down to small enough
values to trust the position of this minimum provided we choose
$W_0 \sim W_{np}(\varphi_m)$.

The resulting minimum turns out to be supersymmetric, since
\be
    \Bigl. D_\varphi W \Bigr|_{\varphi_m} = -aA e^{-a\varphi_m} -
    \frac{3\left[W_0 + A e^{-a\varphi_m} \right]}{
    \varphi_m + \overline\varphi_m}  = 0
\ee
there, and so
\be \label{KKLTVacuumEnergy}
    V(\varphi_m,\overline\varphi_m)
    = - \, \frac{3 \left|W_0 + A e^{-a\varphi_m} \right|^2 }{
    (\varphi_m + \overline\varphi_m)^3 }
    = - \frac{\left|a A e^{-a\varphi_m} \right|^2}{3 (\varphi_m +
    \overline \varphi_m)}
    < 0\,.
\ee

\subsubsection*{Uplifting}

Although this successfully fixes the last of the moduli, it does
so in a way which does not break supersymmetry, and with the
geometry of the noncompact 4 dimensions being given by anti-de
Sitter space due to the negative vacuum energy density,
eq.~\pref{KKLTVacuumEnergy}. For this reason it is useful to
modify the system slightly, both to break supersymmetry and to
raise the vacuum energy to zero (or positive) values. The idea is
to do so in a way which does not ruin the success of the modulus
stabilization just discussed.

KKLT suggested doing so by adding an anti-D3 brane to the system.
The problem is that such a $\overline{\hbox{D3}}$ breaks all of
the supersymmetries that are preserved by the Calabi-Yau geometry,
and so need not appear within the effective 4D theory in a way
that is captured by 4D $N=1$ supergravity. Although this gives
much less control over the corrections to the calculation, the
damage can be kept small if the contribution of the antibrane to
the low-energy action can be made parametrically weak. This can
plausibly be done in the case that there is a strongly warped
throat, because in this case the antibrane can minimize its energy
by moving to the throat's tip. It can do so because at the tip the
dimensional reduction of the anti-brane tension (starting in the
10D Einstein frame) is small, with
\be
    \cL_{\overline{D3}} = - T_3 \int \exd^4x \; \sqrt{-g}
    = - T_3 \int \exd^4x \; \frac{\sqrt{-\hat g}}{h_{\rm tip} r^{12}}
    = - k^3 T_3 \int \exd^4x \; \frac{\sqrt{-\hat g}}{h_{\rm tip}
    (\hbox{Re}\,\varphi)^3} \,.
\ee
Here the second equality uses $g_{\mu\nu} = r^{-6} \hat
g_{\mu\nu}$, as is required to go to the 4D Einstein frame once we
re-scale the internal metric by $g_{mn} = r^2 \hat g_{mn}$, and
the third equality uses the connection $\hbox{Re} \, \varphi = k
r^4$. Since the value of the warp factor at the throat's tip turns
out to depend on $r$ like $h_{\rm tip} = h_0 r^{-4} = k h_0
(\hbox{Re}\,\varphi)^{-1}$, we see that the antibrane contribution
to the potential becomes
\be \label{UpliftingPotentialD3bar}
    V_{\overline{D3}} = \frac{\cE}{(\hbox{Re}\, \varphi)^2} \,,
\ee
where $\cE \simeq k^2 T_3/h_0 > 0$.

The point of this exercise is that the value of the parameter,
$h_0$, can be tuned over an extremely wide range of values because
it is given in Type IIB compactifications as an exponential of the
various integers which label the quantized fluxes within the extra
dimensions. Consequently, it is possible to adjust these integers
to ensure that $h_0$ is sufficiently large that the contribution
of the antibrane to the low-energy action can be computed
perturbatively in $\cE$, which to leading order means simply
adding eqs.~\pref{VFDefinition} and
\pref{UpliftingPotentialD3bar}. Once this is done, the resulting
potential can be adjusted to continue having a local minimum at
$\varphi \simeq \varphi_m$ for which $V$ vanishes or is positive.
The asymptotic region at $|\varphi| \to \infty$, where $V \to 0$,
is then separated from this minimum by a potential barrier, making
the local minimum unstable to tunnelling. However the barrier
width can easily be wide enough to make the lifetime of this
tunnelling long enough to be stable for all practical purposes.

An alternative tack on uplifting is to try to do so using physics
which does not itself badly break supersymmetry (unlike the
D3-bar) and so which can be described purely within the framework
of 4D $N=1$ supergravity. One way to do so is to turn on magnetic
fluxes on some of the D7 branes, since this allows supersymmetry
to be broken in a parametrically small way. The resulting energy
is positive, and appears within the low-energy supergravity as a
contribution to the positive potential, $V_D$, of
eq.~\pref{VDDefinition} \cite{BKQ}. It can be tricky to realize
this mechanism explicitly in brane constructions, due to the need
to ensure that the low-energy theory does not acquire new light
fields, and so modify the KKLT stabilization argument
\cite{BKQplus}. (See also \cite{Banff1} for a different uplifting
proposal.)

\subsection{Some Inflationary Models}

With this lengthy preamble it is now possible to describe briefly
some of the inflationary proposals that have been made to date.
The examples presented here are not meant to be exhaustive, but
instead are chosen to illustrate some of the insights which stand
to be gained by making a connection between inflation and string
theory.

\begin{figure}
\begin{center}\epsfig{file=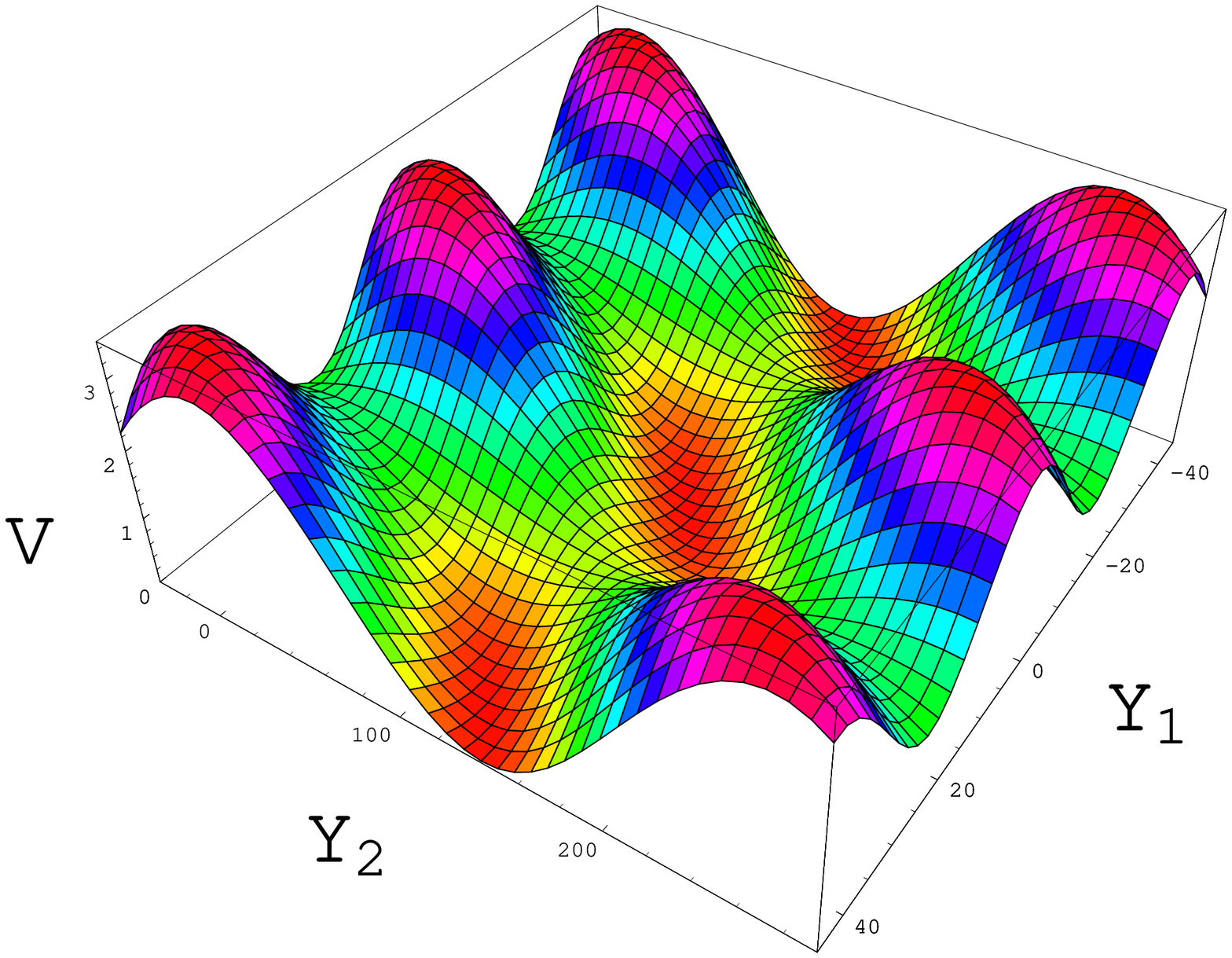,width=3in}
\caption{A sketch of the scalar potential as a function of the
imaginary parts of the two moduli once the real parts are
minimized, for the $\IP^4_{[1,1,1,6,9]}$ model of
ref.~\cite{racetrack2}.}\label{Fig:8}
\end{center}
\end{figure}

\subsubsection*{Racetrack Models}

The simplest approach is to ask if moduli themselves can play the
role of the inflaton \cite{ModuliInfl,racetrack}. More precisely,
do the 4D effective potentials for those vacua having a small
number of moduli have regions for which the slow-roll conditions
are satisfied? Although this appears not to be possible for the
simplest single-modulus example examined by KKLT, it does seem to
be possible for only marginally more complicated cases having two
complex moduli, $\varphi^1$ and $\varphi^2$ \cite{racetrack2}.

The simplest such an example is based on the Calabi-Yau manifold
$\IP^4_{[1,1,1,6,9]}$, which has a K\"ahler potential of the form
of eq.~\pref{KMultiModulus} \cite{DenefDouglas}, with $k_2 = 1$.
The non-perturbative superpotential for this case may also be
computed, and is given by
\be
    W = W_0 + A e^{-a \varphi^1} + B e^{-b \varphi^2} \,,
\ee
for calculable constants $A$, $B$, $a$ and $b$. Finally, motivated
by what would arise in the presence of a D3, the uplifting
potential can be taken to be $V_{\overline{D3}} = \cE/\cV_6^2$. As
may be seen from Figure \ref{Fig:8}, the scalar potential which
results has a complicated form as a function of the four real
fields, $\hbox{Re}\,\varphi^i$ and $\hbox{Im}\, \varphi^i$.
Although inflation is not generic for this potential, a numerical
search shows that it can occur for specific choices for the
various parameters appearing within the superpotential
\cite{racetrack2}. It is not yet known whether the precise values
required can plausibly arise from explicit choices for the
underlying Calabi-Yau geometry.

\begin{figure}
\begin{center}\epsfig{file=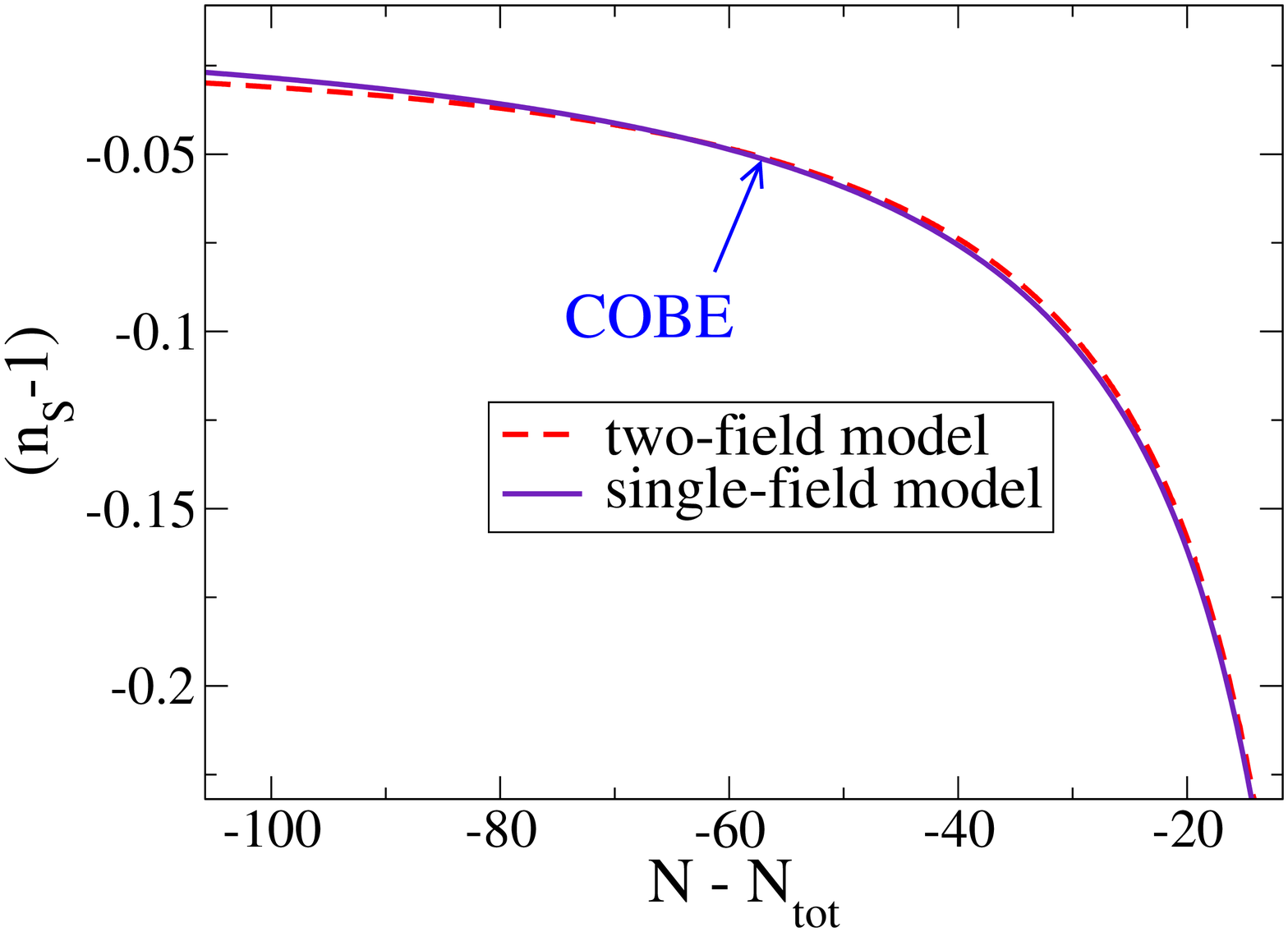,width=3in}
\caption{A comparison of a single-field calculation of the scalar
spectral index with the full result for the $\IP^4_{[1,1,1,6,9]}$
model of ref.~\cite{racetrack2}.}\label{Fig:9}
\end{center}
\end{figure}

This example --- called {\sl `Better' Racetrack Inflation} ---
already teaches us a number of things about string inflation.
First, the inflationary trajectories generically involve
complicated motions in the 4-dimensional field space, which are
not well described by having only the imaginary or real part of
one of the moduli $\varphi^i$ evolving with all of the others held
fixed. However, as Figure \ref{Fig:9} shows, because these fields
are typically rolling roughly in a fixed direction over the
comparatively short interval of horizon exit, its observational
predictions (such as a scalar spectral index $n_s \simeq 0.95$)
are nonetheless well captured by a single-field estimate. Because
inflation occurs near the top of a saddle point for $V$, the
relevant single-field model is in this case of the small-field
form. This, together with the generic decoupling of high-energy
modes which is a feature of the effective field theories during
inflation \cite{TPIUS,Shenker}, gives confidence that string
modifications do not undermine the basic observational evidence
that inflation may have taken place.

Another important feature of the Racetrack models is their strong
sensitivity to the parameters chosen for the superpotential. The
very existence of a slow roll can be destroyed merely by varying
these parameters by a percent or less \cite{racetrack2}. This is
similar to what is encountered in simple single-field models,
where potential parameters must be adjusted with similar accuracy
in order to ensure both $\epsilon$ and $\eta$ are small enough to
provide sufficient inflation. The slightly more complicated model
described in the next section may be more successful on this
particular score.

\subsubsection*{K\"ahler Moduli Inflation}

{\sl K\"ahler Moduli Inflation} \cite{KMI,Lev} works within a
class of Type IIB string vacua that are interesting in their own
right, which differ from the KKLT minima by not assuming $W_0$ to
be anomalously small. In this case the perturbative $\alpha'$
corrections to $K$ are no longer negligible, and their presence
gives rise to new minima for the potential. In order to trust
these new minima within the context of the $\alpha'$ expansion, it
is necessary to work with Calabi-Yau vacua having more than one
modulus \cite{newminima}. Among their attractive phenomenological
features are the enormous range of volumes, $\cV_6$, which are
possible for the underlying Calabi-Yau space (due to the
exponential dependence of $\cV_6$ on the parameters of the
compactification), as well as the fact that supersymmetry is not
preserved at the minimum (even before uplifting by a
$\overline{\hbox{D3}}$ brane) since $D_iW(\varphi_m) \ne 0$.

The simplest models of this class known to have scalar potentials
that inflate involve three moduli, $\varphi^i$ with $i = 1,2,3$.
Their K\"ahler potential is as given in eq.~\pref{KMultiModulus},
supplemented by the perturbative correction of
eq.~\pref{KalphaCorrected}, and their superpotential is
\be
    W = W_0 + \sum_i A_i e^{-a_i \varphi^i} \,.
\ee
The full scalar potential is then obtained by combining the
resulting supersymmetric expression, $V_F$, with an uplifting term
of the form $V_{\overline{D3}} \propto 1/\cV_6^2$.

Denoting $\tau^i = \hbox{Re}\, \varphi^i$, this potential can lead
to inflation in the regime where $\tau^3$ is much larger than the
others, with $e^{-a_3 \tau^3} \sim O(\cV_6^{-2}) \ll 1$. In this
case the motion largely involves only $\tau^3$, with $V$
approximated by the expression
\be \label{KMIPotential}
    V \simeq V_0 - C (\tau^3_c)^{4/3} \exp \Bigl[ -
    c (\tau^3_c)^{4/3}
    \Bigr] \,,
\ee
where $\tau_c^3$ denotes the canonically normalized variable along
the $\tau^3$ direction \cite{KMI}. Slow roll in this case requires
only $\tau^3_c$ to be sufficiently large, which lies within the
domain of the approximations used to compute $V$. Furthermore,
since the roll is towards smaller values of $\tau^3_c$, eventually
this condition fails and corrections to eq.~\pref{KMIPotential}
become important, providing an exit from inflation.

The attractive new feature of this model is the insensitivity of
the slow-roll conditions from specific choices for the parameters
(like $c$ and $C$) that are explicitly given in the potential.
Whether it is similarly independent of other implicit choices of
parameters, such as those possibly arising once string loop
corrections are incorporated into the potential, is not yet known.

\subsubsection*{Inflation due to Brane Motion}

Another broad class of inflationary constructions within string
theory relies on using the positions of various branes as the
inflaton \cite{DvaliTye}. In particular, using the separation
between an antibrane and a brane (or configuration of other
branes) as the inflaton turns these models into useful tools for
exploring inflationary possibilities in string theory, by allowing
supersymmetry breaking to be incorporated in a calculable way
\cite{BBI1}.

Within this framework inflaton dynamics is governed by the
potential describing the various forces acting between the various
branes. Finding inflation is difficult for these models because
although inter-brane forces typically fall off like a power of the
inter-brane separation, branes can never get far enough apart from
one another within the extra dimensions to allow this falloff to
become shallow enough for a slow roll to occur \cite{BBI1}. This
observation has led to the proposal of a variety of mechanisms for
achieving sufficiently weak inter-brane forces, involving the
interactions of branes oriented at angles to one another
\cite{angles}, dual formulations of branes at angles
\cite{GaugeInfl}, D3 branes falling towards D7 branes \cite{D3D7},
and so on \cite{moremodels}. These models usually resemble Hybrid
Inflation in their predictions, because of the appearance of an
open-string tachyon (expressing their instability towards
annihilation) once the branes approach to within the string length
of one another.

Since brane positions, $z^i$, appear in the low-energy effective
theory together with other moduli, real progress has become
possible once these ideas were embedded into a framework which
stabilizes the various moduli \cite{KKLMMT}. The simplest proposal
starts with the basic one-modulus model defined with extra
dimensions having a strongly warped throat \`a la KKLT. Brane
dynamics is then added by including a mobile D3 brane which is
free to move, and is drawn down the throat by its attraction
towards the anti-D3 which sits at its tip. The trick to make this
precise is to cast both the modulus-stabilizing and inter-brane
forces in terms of an effective 4D supergravity, since this gives
control over the corrections which are possible to the leading
semiclassical approximations.

A D3 brane added to a Type IIB vacuum in this way changes both the
K\"ahler potential and superpotential of the low-energy 4D
supergravity, and each of these changes describes a different kind
of inter-brane force. Modifications making the K\"ahler function
depend on the presence of the 3-brane position, $z^i$, modifies
eq.~\pref{KSingleModulus} to take the form
\be \label{KforKKLMMT}
    K(\varphi,z,\overline\varphi,\overline{z})
    = -3 \ln \Bigl[ \varphi + \overline\varphi - \kappa \,
    k(z,\overline z) \Bigr] \,,
\ee
where $\kappa$ is a constant and $k(z,\overline z)$ is the
K\"ahler potential for the Calabi-Yau metric, $g_{mn}(y)$, itself,
in the sense that $g_{i\overline\jmath}(z,\overline z) =
\partial_i \partial_{\overline\jmath} k$ for an appropriate choice
of coordinates. The correctness of this form for the K\"ahler
potential may be inferred by requiring agreement with the
dimensionally-reduced kinetic term, eq.~\pref{BraneActionEqn} for
the D3-brane \cite{giddingsdewolfe}, and requiring that the
supersymmetric potential for the modulus vanishes identically when
$W = W_0$ (see Exercise 14).

\begin{quote}{\bf Exercise 14:}
Show that the K\"ahler potential, $K$, of eq.~\pref{KforKKLMMT}
satisfies the no-scale identity, eq.~\pref{NoScaleIdentity}, and
so $V_F = 0$ when the superpotential is constant, $W = W_0$.
\label{Ex:14}
\end{quote}

The potential, eq.~\pref{KforKKLMMT}, describes a force on the D3
brane once the moduli get stabilized because once $W$ depends on
$\varphi$, $V_F$ acquires nontrivial dependence on $z^i$.
Physically, the absence of such a potential when $W=W_0$ expresses
the absence of a net static force between the D3 and the other
branes present in the extra dimensions. However this absence of a
net force happens due to the cancelling (due to the supersymmetry
of the background geometry) of a variety of inter-brane forces
having their origin in the exchange of massless bulk states
(gravitons, dilatons, and so on). However, if the D3 is moved
within the extra dimensions the distribution of forces acting on
the branes adjusts, as they try to maintain their cancellation at
the new position of the D3. This adjustment in turn causes the
volume modulus, $\varphi$, to change, as the internal geometry
responds to new distribution of forces. The change of the
extra-dimensional volume costs no energy so long as the breathing
mode is a modulus. But once this modulus has been stabilized (by
having $W$ depend on $\varphi$) the energy cost associated with
this adjustment induces a force (expressed by the interactions
between $\varphi$ and $z^i$ in $K$) which tends to localize the D3
at a specific position within the extra dimensions.

Modifications that introduce a $z^i$ dependence directly into $W$
describe a second kind of force experienced by the D3. This force
arises due to the back-reaction of the D3 onto the background
extra-dimensional geometry, since this changes the volume of the
cycle wrapped by any D7 branes, and thereby changes the gauge
couplings of the interactions on these branes (such as those which
generate $W_{np}$). In the low-energy supergravity this effect
appears as a calculable $z$-dependence to the constant $A = A(z)$
appearing in eq.~\pref{WNPSingleModulus} \cite{baumann1}.

\subsubsection*{KKLMMT-type Models}

Ref.~\cite{KKLMMT} performed the first search for inflation, using
eq.~\pref{KforKKLMMT} with the non-perturbative superpotential,
eq.~\pref{WNPSingleModulus}, together with the uplifting term,
eq.~\pref{UpliftingPotentialD3bar}. They found that although the
strong warping in the throat tends to favor a slow D3 roll, the
coupling between $z^i$ and $\varphi$ embodied by
eq.~\pref{KforKKLMMT} generically steepens this potential
sufficiently to prevent inflation's occurrence.

Inflation within this context requires a more detailed balancing
of the forces acting on the D3 brane. One way this might occur
would arise if the above-mentioned volume-stabilization force were
to localize the brane at a position removed from the tip of the
throat, because in this case the pull of the mobile D3 towards
this point can be balanced against its Coulomb attraction towards
the anti-D3 which is situated at the throat's tip. In this case a
slow roll can occur when the D3 is close to where these forces
balance, and ends if the D3 slowly rolls off as it succumbs to its
attraction to the anti-D3 brane \cite{realistic}. As mentioned
earlier, the observational predictions for this inflation fall
into the category of Hybrid Inflation, with the two fields
physically corresponding to the interplay between the inter-brane
separation and an open-string tachyon which describes the
instability towards mutual annihilation. As a result models of
this form exist for which both $n_s > 1$ \cite{realistic} and $n_s
< 1$ \cite{jim}.

However, from the point of view of providing a string embedding of
inflation, this kind of picture suffers from two drawbacks. First,
it assumes the forces on the D3 brane stabilize it away from the
throat's tip, without providing an explicit extra-dimensional
construction which does so. Secondly, by relying on the
brane-antibrane Coulomb force, it steps outside of the low-energy
4D supergravity approximation, and so makes difficult the
quantification of the possible corrections to the semiclassical
approximation which might arise.\footnote{Of course, this
objection also applies to most of the other proposed brane-based
inflationary mechanisms.}

\begin{figure}
\begin{center}\epsfig{file=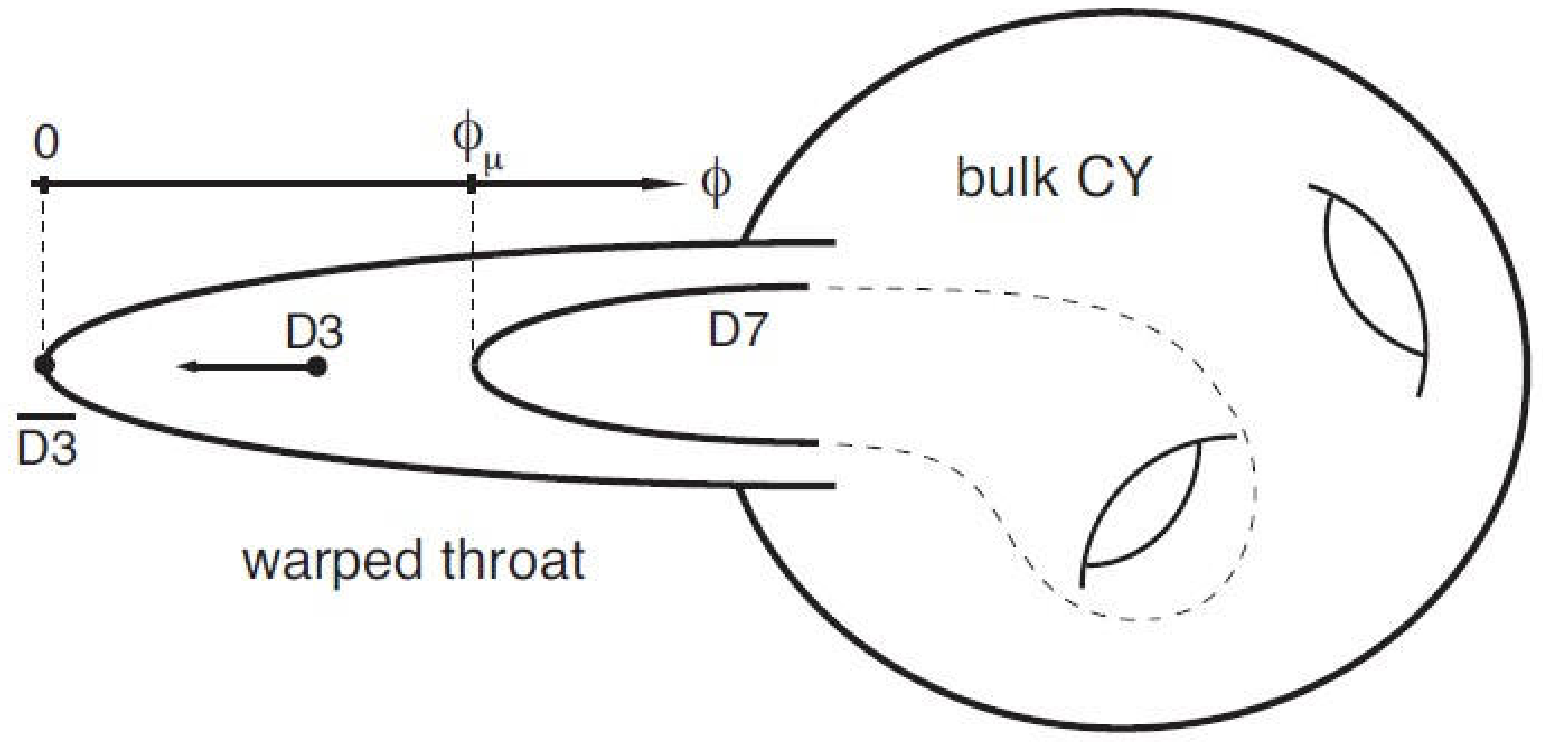,width=4in}
\caption{A sketch of a D7 descending partially into a warped
throat, as assumed in the inflationary scenario of
ref.~\cite{delicate}.}\label{Fig:10}
\end{center}
\end{figure}

A more convincing stringy grounding of this type of inflation in
string theory instead requires a description of all forces in
terms of the low-energy supergravity. This has recently become
possible using the $z$-dependence of the superpotential
\cite{baumann1} which arises when a D7 extends partially down into
a warped throat along particular kinds of cycles (see Figure
\ref{Fig:10}). In this case, the resulting $z$-dependence of $W$
shows that D3 branes in the same throat can experience a balance
of forces towards the tip and towards the D7 brane, allowing
slow-roll inflation to occur for some choices of the various
parameters describing the underlying vacuum \cite{delicate}.

\subsubsection*{Brane Annihilation and Reheating}

Once the D3 brane and the anti D3-brane come to within the string
length of one another, stringy physics intervenes and the two
branes annihilate one another. This annihilation process has two
potentially important observational implications. First,
annihilation takes place through having their world sheets
fragment into pairs a D1 and $\overline{\hbox{D1}}$ branes (or,
D-strings), which then find one another and continue to annihilate
in a cascade towards the vacuum state \cite{BBI1}. The competition
of this annihilation rate with the expansion of the universe can
be described in a manner very similar to the Kibble process
describing phase transitions, familiar to cosmologists. This
allows a quantitative estimate of the number of D1 and
$\overline{\hbox{D1}}$ that fail to find their anti-branes to
annihilate, with the result that they can be abundant enough to be
detectable as cosmic strings in the present universe
\cite{henrycosmic}. Furthermore (although this depends more on the
details of the underlying Calabi-Yau geometry) these strings can
be stable enough to avoid having decayed during the intervening
epochs \cite{CopelandMyersPolchinski}. The observation of such
cosmic strings together with inflation would provide compelling
circumstantial evidence for brane-based inflation.

The second implication of annihilation is the mechanism it
provides for reheating the later universe \cite{BBI1}, by
liberating the brane tensions which provide the underlying
inflationary energy density. Once liberated, one must ask whether
this energy can get funnelled efficiently enough into observable
low-energy degrees of freedom to provide sufficient reheating.
Since the observable degrees of freedom in these models tend to
reside on other, spectator, branes, a potential danger here is
that the released energy is dumped too efficiently into invisible,
bulk degrees of freedom rather than into observable modes.
However, an important observation \cite{WarpedRH} is that strong
warping can help with the efficiency of energy transfer into the
observed sector, provided that this observed sector resides at the
tip of a strongly warped region (as tends to be required in any
case by particle physics issues, like the Hierarchy Problem). This
low-energy mechanism is supported, with some caveats, by the
subsequent more detailed string calculations \cite{OtherWRH}.

\subsubsection*{DBI Inflation: Beyond Slow Rolls}

A related string-based inflationary proposal, again based on brane
motion, differs from all of the others by not relying on the usual
slow-roll approximation, and so also has a somewhat different
observational signature. In this model --- known as {\sl DBI
Inflation} --- a D3 brane is again envisioned to roll down a
strongly warped throat, attracted to an anti-D3 at the tip, but
the motion is taken to be relativistically rapid rather than slow.
Paradoxically, the energy of such a system can produce accelerated
inflationary expansion, despite the motion being the opposite of a
slow roll \cite{DBI}.

The starting point for this proposal is the action for a
relativistically moving D3 brane moving through a throat, and with
a cosmological 4D metric,
\be
    \exd s^2 = h^{-1/2}(y) \Bigl[ - \exd t^2 + a^2(t) \, \exd
    \vec{x}^2 \Bigr] + h^{1/2}(y) \, g_{mn}(y) \, \exd y^m \exd
    y^n \,.
\ee
Denoting the distance to the brane from the throat's tip by
$q(t)$, the brane action takes the form
\be \label{DBIAction}
    S = -\int \exd^4x \, a^3 \left[ \frac{T_3}{h(q)} \Bigl(
    \sqrt{1 - h(q) \dot q^2 /T_3} - 1 \Bigr) + V(q) \right] \,,
\ee
where $h(q) \simeq b^4/q^4$ in the throat. The square-root term in
the square brackets represents the contribution of the first ({\sl
Dirac-Born-Infeld}, or DBI) term of eq.~\pref{BraneActionEqn},
while the second ($-1$) term is due to the Chern-Simons coupling
({\it i.e.} to $C$ of eq.~\pref{BraneActionEqn}). Notice that
these cancel when $\dot q^2 = 0$, showing the above-mentioned
absence of a static force on the D3. In the potential, $V(q) = V_0
+ \frac12 m^2 q^2 - k/q^4$, $V_0$ describes the tension of other
branes, $\frac12 m^2 q^2$ phenomenologically describes the forces,
discussed above, which act to localize the brane at the throat's
tip, and $k/q^4$ describes the Coulomb attraction towards the
antibrane, also located at the tip.

Notice that in the limit of a slow roll, when $\dot q^2$ is small,
the lagrangian density of eq.~\pref{DBIAction} reduces to a
standard non-relativistic point-particle action, $\cL \simeq a^3
\left[ \frac12 \dot q^2 - V \right]$. The full action provides the
relativistic generalization, and takes the form of the action for
a relativistic point particle, but with a speed, $v^2/c^2 = h(q)
\dot q^2/T_3$. Some comment is required about the validity of
using the full form of eq.~\pref{DBIAction}, including the full
structure of the square root, given that this cannot be regarded
as a standard expansion in derivatives as typically arises at low
energies. Is it consistent to drop all higher derivatives (like
$\ddot q$) in $S$ while keeping all powers of $\dot q^2$ all
higher derivatives?

The relativistic particle action is one of the few cases where it
can be a consistent approximation to trust the entire square-root
action while neglecting higher time derivatives. It is
self-consistent to do so because as the motion becomes more and
more relativistic, $v^2/c^2$ asymptotes to 1 and the equations of
motion imply the higher derivatives go to zero. When $h$ is a
constant the same is true for the DBI action,
eq.~\pref{DBIAction}, since its equations of motion imply that
$\ddot q$ and higher derivatives become suppressed in the
ultra-relativistic limit. The same should also hold if the spatial
variation of $h(q)$ is sufficiently slow.

How can this kind of relativistic motion be consistent with a
lengthy period of inflation and the equation of state, $p < -
\frac13 \rho$ (and so potential-energy domination) which inflation
requires? The answer is in the warping: ($i$) when passing through
a strongly warped region $h \gg 1$, and so $\dot q^2$ can be small
(so inflation last a long time) even if $h \dot q^2/T_3$ is
$O(1)$; and ($ii$) because the kinetic energy's pre-factor of
$1/h$ suppresses it relative to $V$ in strongly-warped regions,
even if the motion is relativistic.

Because the motion is not slow, the predictions of DBI inflation
cannot be inferred using the slow-roll expressions of the previous
sections, which are entirely expressed in terms of the derivatives
of the scalar potential. Instead we must generalize to define
slow-roll parameters that rely only on what is important: the
approximate constancy of $H$ during inflation. To this end define
the generalized slow-roll parameters $\tilde\epsilon$ and
$\tilde\eta$ by \cite{WMAPInflation}
\be
    \tilde \epsilon \equiv - \frac{\dot H}{H^2}
    \quad\hbox{and}\quad
    \tilde \eta \equiv \frac{\dot {\tilde\epsilon}}{\tilde \epsilon
    H} \,,
\ee
and so on, for successively higher derivatives.

To make contact between these definitions and the action, consider
the general situation \cite{Kinflation} where
\be
    S = \int \exd^4x \; a^3 \, p(q,\cX) \,,
\ee
where $\cX = \frac12 \dot q^2/T_3$. The action of interest,
eq.~\pref{DBIAction}, corresponds to the special case where
\be
    p(q,\cX) = - \frac{T_3}{h(q)} \Bigl[ 1 - 2h(q) \cX \Bigr]^{1/2}
    + \frac{T_3}{h(q)} - V(q) \,.
\ee
The energy density computed from this action is then
\be
    \rho(q,\cX) = 2\cX p_{,\cX} - p \,,
\ee
and it is useful to define the `speed of sound',
\be
    c_s^2 = \frac{p_{,\cX}}{\rho_{,\cX}} =
    \frac{p_{,\cX}}{p_{,\cX} + 2 \cX p_{,\cX\cX}} \,,
\ee
which when specialized to the action, eq.~\pref{DBIAction},
becomes
\be
    c_s^2 = 1 - 2h\cX = \frac{1}{\gamma^2} \,,
\ee
where the relativistic $\gamma$ factor is defined, as usual, by
$\gamma \equiv \Bigl[ 1 - 2h \cX \Bigr]^{-1/2} \ge 1$, with
relativistic motion characterized by $\gamma \gg 1$. Using these
expressions in the Friedmann and Raychaudhuri equations,
eqs.~\pref{FriedmannEqn} and \pref{RaychaudhuriEqn}, to evaluate
$H$ and its derivatives, then gives, for instance
\be
    \tilde \epsilon = \frac{\cX p_{,\cX}}{M_p^2 H^2}
    = \frac{3 \cX p_{,\cX}}{2\cX p_{,\cX} - p} \,,
\ee
which reduces in the non-relativistic case, $p \simeq T_3 \cX -
V$, to the usual slow-roll result $\tilde\epsilon \simeq \frac32
\dot q^2/V \simeq \epsilon$.

The expressions for the amplitude of primordial fluctuations then
generalize from the usual slow-roll results,
eqs.~\pref{CurvaturePowerSpectrum} and \pref{DeltaSqTSlowRoll}, to
\be
    \Delta^2_\Phi = \frac{H^2}{8 \pi^2 M_p^2 \tilde\epsilon \,c_s}
    \quad \hbox{and} \quad
    \Delta^2_T = \frac{2H^2 }{\pi^2 M_p^2} \,.
\ee
From these the following formula for the spectral index are
obtained
\be
    n_s - 1 = -2\tilde\epsilon - \tilde \eta - s \,,
    \quad
    n_T = -2 \tilde\epsilon \quad\hbox{and} \quad
    r = - 16\, \tilde\epsilon \, c_s \,,
\ee
where the new contributions come from the appearance of $c_s$, and
the parameter $s$ is defined by
\be
    s \equiv \frac{\dot c_s}{c_s H} \,.
\ee
The previous slow-roll formulae are obtained in the limit $c_s =
1$, and so $s = 0$.

There is an important observational way to distinguish between
inflation of this type and that arising from an honest-to-God slow
roll \cite{BoundsonDBI}. This is because the fluctuations
predicted by DBI inflation are not Gaussian when the brane motion
is in the ultra-relativistic limit, $\gamma \gg 1$. Although it
goes beyond the scope of these lectures, the deviation from
Gaussianity can be quantified by a dimensionless parameter
$f_{NL}$, which vanishes for purely Gaussian fluctuations.
Observations of the microwave background are consistent with
Gaussian fluctuations, and currently constrain $-256 < f_{NL} <
332$. For comparison, the prediction of DBI inflation is $f_{NL}
\simeq 0.32 \,\gamma^2$, implying the observational bound $\gamma
\lsim 32$.

\subsubsection*{What We've Learned}

Recent years have seen some progress in trying to embed inflation
into a string theoretic framework, recently stimulated by strides
taken in understanding how moduli are fixed for Type IIB string
vacua, and rapid progress continues to be made. Although it is
still early days, string theory has already offered some insights
into how inflation might work within a fundamental context. Some
of these are, in a nutshell:

\begin{itemize}

\item {\bf Single-Field Slow Roll Models:} Single-field slow-roll
models (and simple multi-field models, like Hybrid Inflation)
capture most of the predictions of the known string-inflationary
scenarios. Partly this is because the tools available only allow
the exploration of string dynamics when it is described by an
effective 4D theory. But it is also true that these low-energy
field theories typically involve many light scalars during the
inflationary epoch, and although it is necessary to properly
follow the dynamics of these extra scalars when finding inflation,
their presence often does not crucially alter the observational
predictions for the spectrum of primordial fluctuations. This
gives some assurance that we are not being led far astray when
analyzing cosmological data using simple single-field models.

\item {\bf Decoupling and Robustness:} Even though there are many
heavy fields in addition to the inflationary sector, all the
evidence is that in string theory those with masses much greater
than $H_I$ decouple and so have a negligible effect during horizon
exit \cite{Shenker}. As a result it suffices to describe inflation
purely in terms of the relevant inflaton physics at the
inflationary scale. It can be possible to have decoupling break
down, such as by having nominally heavy particles become light; by
having some fields evolve non-adiabatically; or by having
inflation start just before horizon exit. But the current evidence
is that when it does so, it does so in the usual way that
time-dependent effective field theories do \cite{TPIUS}.

\item {\bf New Signatures:} Although inflation, where found so far
in string theory, is well-described by a 4D effective field
theory, several inflationary scenarios do differ in their
implications from simple slow-roll models. Brane-antbrane
inflationary mechanisms can also give rise to relic cosmic strings
\cite{BBI1,henrycosmic,CopelandMyersPolchinski}, and the detection
of these would provide considerable circumstantial evidence for
this kind of mechanism. DBI inflationary models can predict
non-Gaussian primordial fluctuations, and their detection would
definitively rule out inflation due to a single-field slow-roll
mechanism \cite{DBI}.

\item {\bf Naturalness:} For most stringy scenarios parameters in
the potentials must be adjusted in order to ensure a slow roll, at
a level which is consistent with the adjustments that are required
in simple single-field models. But two approaches may prove to be
more promising in this regard: K\"ahler Moduli Inflation
\cite{KMI}, and DBI inflation \cite{DBI}, since these may produce
inflation more robustly than other models. Whether these models
definitively emerge as more natural than others remains the
subject of active current study.

\item {\bf Reheating:} It is a bit premature to fully address
reheating issues, since no string model has yet been constructed
which provides both a convincing inflationary picture as well as a
properly formulated Standard Model sector to describe particle
physics, including a proper understanding of the Hierarchy Problem
\cite{realistic,BLK}. Both are required to address reheating after
inflation, but the first indications are that stringy inflationary
scenarios provide a number of novel challenges and opportunities
for reheating \cite{WarpedRH,OtherWRH}.

\end{itemize}

Further insights are certain to emerge as the inflationary options
become better investigated.

\section*{Acknowledgements}
I would like to thank the organizers of these schools for their
kind invitation to present these lectures, and to thank my
collaborators for their help in understanding the many thorny
inflationary issues. My research has been supported by the Natural
Sciences and Engineering Research Council of Canada, by McMaster
University and by the Killam Foundation. Research at Perimeter
Institute is supported in part by the Government of Canada through
NSERC and by the Province of Ontario through MEDT.

\end{document}